\documentclass[fleqn, usenatbib]{mnras}

\usepackage{newtxtext, newtxmath}

\usepackage[T1]{fontenc}

\DeclareRobustCommand{\VAN}[3]{#2}
\let\VANthebibliography\thebibliography
\def\thebibliography{\DeclareRobustCommand{\VAN}[3]{##3}\VANthebibliography}

\usepackage{graphicx}	
\usepackage{amsmath}	
\usepackage{float}
\usepackage[export]{adjustbox}
\usepackage{natbib}
\usepackage{xcolor}
\usepackage{pdfcomment} 
\usepackage{orcidlink}
\usepackage{lineno}
\usepackage{hyperref}
\usepackage{appendix} 
\usepackage{threeparttable}
\usepackage{booktabs}

\DeclareRobustCommand{\Eqref}[1]{equation~\ref{#1}}
\DeclareRobustCommand{\Figref}[1]{Fig.~\ref{#1}}
\DeclareRobustCommand{\Tabref}[1]{Table~\ref{#1}}
\DeclareRobustCommand{\Secref}[1]{Section~\ref{#1}}
\DeclareRobustCommand{\Appref}[1]{Appendix~\ref{#1}}
\mathchardef\mhyphen="2D

\newcommand{\ppisn}{PPISN}
\newcommand{\ppisne}{PPISNe}
\newcommand{\ppisni}{PPISN-I}
\newcommand{\ppisnei}{PPISNe-I}

\newcommand{\pisn}{PISN}
\newcommand{\pisne}{PISNe}
\newcommand{\pisni}{PISN-I}
\newcommand{\pisnei}{PISNe-I}

\newcommand{\slsn}{SLSN}
\newcommand{\slsne}{SLSNe}
\newcommand{\slsni}{SLSN-I}
\newcommand{\slsnei}{SLSNe-I}

\newcommand{\ccsn}{CCSN}
\newcommand{\ccsne}{CCSNe}
\newcommand{\ccsni}{CCSN-I}

\newcommand{\binaryc}{\textsc{binary\_c}}
\newcommand{\compas}{\textsc{compas}}
\newcommand{\startrack}{\textsc{startrack}}

\newcommand{\bpass}{\textsc{bpass}}
\newcommand{\mobse}{\textsc{mobse}}

\newcommand{\solarmass}{\mathrm{M}_{\odot}}
\newcommand{\eventratedensity}{\mathrm{Gpc}^{-3}\,\mathrm{yr}^{-1}}

\newcommand{\euclid}{\textit{EUCLID}}
\newcommand{\jwst}{\textit{JWST}}
\newcommand{\lsst}{\textit{LSST}}
\newcommand{\ROMAN}{\textit{ROMAN}}

\newcommand{\extraML}{\Delta M_{\mathrm{PPI,\,extra\,ML}}}
\newcommand{\COshift}{\Delta M_{\mathrm{PPI,\,CO\,shift}}}
\newcommand{\upwardvariation}{$\COshift = +10\,\solarmass$}
\newcommand{\downwardvariation}{$\COshift = -15\,\solarmass$}

\newcommand{\fryerprescription}{delayed}
\newcommand{\slsnredshift}{0.028}
\newcommand{\ligoredshift}{0.2}
\newcommand{\powerlawpeak}{\textsc{power law + peak}}
\newcommand{\mprimary}{M_{\mathrm{primary}}}
\newcommand{\triplealpha}{3\alpha}
\newcommand{\carbonoxygen}{{}^{12}{\rm{C}}{(\alpha ,\gamma )}^{16}{\rm{O}}}
\newcommand{\mppisncutoff}{M_{\mathrm{{\ppisn}\,cutoff}}}

\newcommand{\bhbh}{BBH}
\newcommand{\overdensity}{over-density}
\newcommand{\apostrophes}[1]{`#1'}
\newcommand{\ofourrun}{O4}

\defcitealias{farmerMindGapLocation2019}{F19}

\newcommand{\mytitle}{Pulsational pair-instability supernovae in gravitational-wave and
   electromagnetic transients}
\newcommand{\myshorttitle}{{\ppisne} and in GW and EM transients}
\title[\myshorttitle]{\mytitle}

\hypersetup{
    colorlinks=true,
    linkcolor=blue,
    filecolor=magenta,
    urlcolor=cyan,
    pdftitle={\mytitle},
    bookmarksdepth=3
}

\author[D.~D.~Hendriks et al.]{
D.~D.~Hendriks $^{1}$\thanks{E-mail: \href{mailto:dh00601@surrey.ac.uk}{dh00601@surrey.ac.uk}\ (DDH)} \orcidlink{0000-0002-8717-6046},
  L.~A.~C.~van Son $^{2,\ 3,\ 4}$ \orcidlink{0000-0001-5484-4987},
  M.~Renzo $^{5}$ \orcidlink{0000-0002-6718-9472},
  R.~G.~Izzard $^{1}$ \orcidlink{0000-0003-0378-4843},
  \newauthor
  R.~Farmer $^{4}$ \orcidlink{0000-0003-3441-7624}
\\
  $^{1}$Department of Physics, University of Surrey, Guildford, GU2 7XH, Surrey, UK\\
  $^{2}$Center for Astrophysics, Harvard \& Smithsonian, 60 Garden St., Cambridge, MA 02138, USA\\
  $^{3}$Anton Pannekoek Institute for Astronomy, University of Amsterdam, Science Park 904, 1098XH Amsterdam, The Netherlands\\
  $^{4}$Max-Planck-Institut für Astrophysik, Karl-Schwarzschild-Straße 1, D-85741 Garching, Germany\\
  $^{5}$Center for Computational Astrophysics, Flatiron Institute, New York, NY 10010, US\\
}

\date{Accepted 2023 September 14. Received 2023 September 13; in original form 2023 August 2}

\pubyear{2023}

\begin{document}
\label{firstpage}
\pagerange{\pageref{firstpage}--\pageref{lastpage}}
\maketitle

\begin{abstract}
  Current observations of binary black-hole ({\bhbh}) merger events show support
  for a feature in the primary BH-mass distribution at $\sim\,35\,\solarmass$,
  previously interpreted as a signature of pulsational pair-instability
  ({\ppisn}) supernovae. Such supernovae are expected to map a wide range of
  pre-supernova carbon-oxygen (CO) core masses to a narrow range of BH masses,
  producing a peak in the BH mass distribution. However, recent numerical
  simulations place the mass location of this peak above
  $50\,\solarmass$. Motivated by uncertainties in the progenitor's evolution and
  explosion mechanism, we explore how modifying the distribution of BH masses
  resulting from {\ppisn} affects the populations of gravitational-wave (GW) and
  electromagnetic (EM) transients. To this end, we simulate populations of
  isolated {\bhbh} systems and combine them with cosmic star-formation
  rates. Our results are the first cosmological {\bhbh}-merger predictions made
  using the {\binaryc} rapid population synthesis framework. We find that our
  fiducial model does not match the observed GW peak. We can only explain the
  $35\,\solarmass$ peak with {\ppisne} by shifting the expected CO core-mass
  range for {\ppisn} downwards by $\sim{}15\,\solarmass$. Apart from being in
  tension with state-of-the art stellar models, we also find that this is likely
  in tension with the observed rate of hydrogen-less super-luminous
  supernovae. Conversely, shifting the mass range upward, based on recent
  stellar models, leads to a predicted third peak in the BH mass function at
  $\sim{}64\,\solarmass$. Thus we conclude that the $\sim{}35\,\solarmass$
  feature is unlikely to be related to {\ppisn}.
\end{abstract}



\section{Introduction}
\label{sec:intro}
The gravitational-wave (GW) observatories operated by the LIGO VIRGO KAGRA (LVK)
collaboration have started to measure signals from GW mergers
\citep{ligoscientificcollaborationandvirgocollaborationGWTC1GravitationalWaveTransient2019,
  abbottGWTC2CompactBinary2021}, and with the recent release of the GWTC-3 there
are now $\sim{}90$ confirmed compact object merger observations
\citep{abbottPopulationMergingCompact2023,
  theligoscientificcollaborationOpenDataThird2023}, the majority of which are
binary black hole (\bhbh) mergers.
These observations show structure in the distribution of primary masses,
$\mprimary$, i.e., the most massive object in the binary at the time of {\bhbh}
merger. Parameteric models of the observations
\citep[e.g.,][]{abbottPopulationPropertiesCompact2021,
  abbottPopulationMergingCompact2023, farahThingsThatMight2023}, as well as
non-parametric models \citep[e.g.,][]{sadiqFlexibleFastEstimation2022,
  callisterParameterFreeTourBinary2023}, consistently infer a feature, e.g., a
change in power-law slope, or the presence of a Gaussian peak, between $32$ and
$38\,\solarmass$, suggesting that this feature is robust. The exact nature and
origin of this feature is unclear, but the often-proposed explanation is that it
originates from a pile up of BH masses due to {\ppisne}
\citep[][]{talbotMeasuringBinaryBlack2018,
  stevensonImpactPairinstabilityMass2019,
  belczynskiEvolutionaryRoadsLeading2020, karathanasisBinaryBlackHoles2023}.
However, several alternative explanations for such a feature have also been
proposed \citep[e.g.,][]{liDivergenceMassRatio2022,
  antoniniCoalescingBlackHole2023, brielUnderstandingHighmassBinary2023}.

{\ppisne} occur when a very massive star is dynamically unstable due to runaway
electron-positron pair-formation in their cores which remove high-energy
photons. This leads to a decrease in the radiation pressure and an increase of
the mass-density, and causes a softening of the equation of state, i.e., a
decrease of the adiabatic index.
The process results in an initial collapse, the explosive ignition of oxygen in
the core, and a subsequent pulsating behaviour through which the star loses
mass, or a single violent explosion that leaves behind no remnant
\citep{barkatDynamicsSupernovaExplosion1967,
  rakavyInstabilitiesHighlyEvolved1967, woosleyPulsationalPairInstability2007,
  woosleyPulsationalPairInstabilitySupernovae2017,
  marchantPulsationalPairinstabilitySupernovae2019,
  renzoPredictionsHydrogenfreeEjecta2020, renzoSensitivityLowerEdge2020,
  farmerConstraintsGravitationalWave2020,
  faragResolvingPeakBlack2022}. {\ppisne} cause a wide range of pre-supernova
core masses to form BHs in a narrow range of remnant masses, which leads to an
{\overdensity} and a subsequent mass-gap at higher BH masses. The magnitude of
this {\overdensity}, or pile-up, depends on the width of the pre-supernova
core-mass range that undergoes {\ppisn} and consequently on how sensitive the
{\ppisn} mass-loss is to the pre-supernova core mass. A broader pre-supernova
core-mass range, i.e., a shallower {\ppisn} remnant mass curve, leads to a
larger pile-up.

Detailed models of single stars allow estimates of the mass lost during the
pulsations at a given helium (He) or CO core mass
\citep{renzoPairinstabilityMassLoss2022}. These are used to calculate an upper
limit on the BH remnant mass that can form after subsequent core-collapse (CC)
supernovae (\citealt{farmerMindGapLocation2019}, from hereon
\citetalias{farmerMindGapLocation2019}; \citealt{faragResolvingPeakBlack2022})
and the location of a feature in the primary-mass distribution due to {\ppisne}
which is consistently predicted at masses $\gtrsim{}40-45\,\solarmass$
\citep{talbotMeasuringBinaryBlack2018, stevensonImpactPairinstabilityMass2019,
  belczynskiEvolutionaryRoadsLeading2020}.
This mass is significantly ($\gtrsim{}5\,\solarmass$) greater than the
$\sim{}35\,\solarmass$ location of the feature inferred from GW data.
Moreover, it is remarkably robust against the most common uncertainties in
massive stellar evolution, such as metallicity, mixing, and neutrino physics
\citep{farmerMindGapLocation2019}.

However, there are uncertainties that lead to larger variations in the mass
range and remnant mass of stars that undergo {\ppisn}.
Several processes have been suggested that lead to shifts in the CO-core masses
that undergo PPISN. These include uncertainties in the nuclear burning rates
that affect the carbon-to-oxygen (C/O) ratio in the core
\citep{deboer12C16OReaction2017, farmerMindGapLocation2019,
  farmerConstraintsGravitationalWave2020, costaFormationGW190521Stellar2021,
  woosleyPairinstabilityMassGap2021, mehtaObservingIntermediatemassBlack2022,
  faragResolvingPeakBlack2022, shenNewDetermination12C2023}, rotation, which
provides both more massive cores and enhanced dynamical stability
\citep{glatzelFateRotatingPairunstable1985,maederEvolutionRotatingStars2000,
  chatzopoulosEffectsRotationMinimum2012, marchantImpactStellarRotation2020},
beyond-Standard-Model physics which can either affect the C/O ratio
\citep{croonMissingActionNew2020}, or lead to reduced dynamical stability
\citep{croonMissingActionNew2020, saksteinAxionInstabilitySupernovae2022,
  moriLightCurvesEvent2023} at lower masses, and lastly dark-matter annihilation
which acts like an additional heating source \citep{zieglerFillingBlackHole2021,
  zieglerGapNoMore2022}. These processes could lower the CO core masses that
undergo PPISN by up to $-20\,\solarmass$ (axion instability) or increase them by
up to $+10\,\solarmass$ (reaction rates, rotation).
Moreover, some theoretical studies predict additional mass loss, either in the
post-PPI CC due to changes in core structure of PPI stars affecting the
propagation of the core-bounce shock
\citep{marchantPulsationalPairinstabilitySupernovae2019,
  renzoPredictionsHydrogenfreeEjecta2020, powellFinalCoreCollapse2021,
  rahmanPulsationalPairinstabilitySupernovae2022}, or by how convection
transports energy during the PPI \citep{renzoSensitivityLowerEdge2020}.
Furthermore, recent SN observations are well modelled by post-PPI mass loss
\citep{ben-amiSN2010mbDirect2014, kuncarayaktiBactrianBroadlinedTypeIc2023,
  linSuperluminousSupernovaLightened2023}. Both theoretical studies and
observational estimates find, at most, $10\,\solarmass$ in additional mass loss.

The location of the {\pisn} mass-gap has broad implications. If the high-mass
feature at $\sim{}35\,\solarmass$ is indeed caused by {\ppisn}, it
observationally constrains the maximum BH mass that stars can form below the
full disruption by PISN, and thus the lower bound of the so-called {\pisn}
mass-gap \citep{woosleyEvolutionExplosionMassive2002,
  renzoPredictionsHydrogenfreeEjecta2020, woosleyPairinstabilityMassGap2021}.
Only stars with He-core masses $\gtrsim{}120\,\solarmass$ at the onset of
collapse, which experience photo-dissociation during the pair-instability, can
directly collapse into more-massive BHs \citep{bondEvolutionFateVery1984,
  renzoPredictionsHydrogenfreeEjecta2020,
  siegelSuperkilonovaeMassiveCollapsars2022}. The {\pisn} mass-gap further helps
determine the fractional contribution of different gravitational-wave source
channels to the overall population of {\bhbh} mergers, because systems with
masses in the gap originate from channels other than isolated-binary evolution
\citep{arcaseddaFingerprintsBinaryBlack2020, baibhavMassGapSpin2020,
  safarzadehBranchingRatioLIGO2020, wongJointConstraintsFieldcluster2021}. The
location of the mass gap also constrains stellar physics, like the
aforementioned uncertain nuclear reaction rates $\carbonoxygen$ and
$\triplealpha$ \citep{farmerMindGapLocation2019,
  mehtaObservingIntermediatemassBlack2022, faragResolvingPeakBlack2022,
  shenNewDetermination12C2023}. Lastly, the location of the mass-gap and the
pile-up may be redshift independent sign-posts for cosmological applications
\citep[][and references therein]{farrFuturePercentLevelMeasurement2019}.

One way to constrain the physics of {\ppisne} is to compare our simulations
directly to the observed rate of electromagnetic transients. Unfortunately,
unambiguous transient observations of {\ppisn} are currently not available.
Theoretical modelling of {\pisne} light-curves show that their light-curves
generally rise slowly and that some are very luminous at peak luminosity
\citep[][]{kozyrevaObservationalPropertiesLowredshift2014},
while those caused by {\ppisne} and the subsequent interaction of their ejecta
with the circumstellar medium or previously ejected mass-shells
\citep{moriyaPulsationsRedSupergiant2015,
  woosleyPulsationalPairInstabilitySupernovae2017,
  renzoPredictionsHydrogenfreeEjecta2020} have shorter rise-times but
equally-high peak luminosities \citep{woosleyPulsationalPairInstability2007}.
Some super-luminous supernovae ({\slsne}) could be powered by either {\pisne} or
{\ppisne}, and indeed some of these are suggested observations of {\pisne} or
{\ppisne} \citep[e.g.,][]{lunnanUVResonanceLine2018,
  linSuperluminousSupernovaLightened2023, schulze1100DaysLife2023,
  aamerPrecursorPlateauPreMaximum2023}, although as of yet none of them have
been confirmed to be caused by either {\pisne} or {\ppisne}. Moreover, there is
growing evidence from light curves, spectra and rates that not all {\slsne} are
powered by {\pisne} or {\ppisne} \citep{nichollSlowlyFadingSuperluminous2013,
  kozyrevaCanPairinstabilitySupernova2015, perleyHostgalaxyProperties322016,
  gilmerPairinstabilitySupernovaSimulations2017}. Estimates of (P){\pisn}
event-rate densities are useful to study stellar evolution
\citep[e.g.][]{dubuissonCosmicRatesBlack2020, brielEstimatingTransientRates2022,
  tanikawaEuclidDetectabilityPair2023}, and could be compared to {\slsn} rates
to determine whether these rates are in tension
\citep{nichollSlowlyFadingSuperluminous2013}. Although uncontroversial
detections are lacking, there are debated candidates for both {\pisne} and
{\ppisne}, e.g., \textit{SN 1961V} \citep{woosleySN1961VPulsational2022},
\textit{SN 1000+0216} \citep{cookeSuperluminousSupernovaeRedshifts2012},
\textit{SN 2010mb} \citep{ben-amiSN2010mbDirect2014}, \textit{PTF10mnm}
\citep{kozyrevaObservationalPropertiesLowredshift2014}, \textit{iPTF14hls}
\citep{wangIPTF14hlsCircumstellarMedium2022}, \textit{iPTF16eh}
\citep{lunnanUVResonanceLine2018}, \textit{SN 2016iet}
\citep{gomezSN2016ietPulsational2019}, \textit{SN 2017egm}
\citep{linSuperluminousSupernovaLightened2023}, \textit{SN 2018ibb}
\citep{schulze1100DaysLife2023} and \textit{SN 2019szu}
\citep{aamerPrecursorPlateauPreMaximum2023}. However, their interpretation is
sufficiently uncertain that estimating a rate from these observations is still a
challenge (however, see also \citealt{nichollSlowlyFadingSuperluminous2013}).

In this study we explore how the remnants of {\ppisne} affect the distribution
of $\mprimary$ for the {\bhbh} systems merging at redshift
$z{}\sim{}\ligoredshift$, and compare this primary-mass distribution to the
current observations. We focus our results on redshift $z{}\sim{}\ligoredshift$
because this is where current observations provide the strongest constraints
\citep{abbottPopulationMergingCompact2023}. We evolve isolated binary systems
and convolve the resulting {\bhbh} systems with recent star-formation rate
prescriptions \citep{sonRedshiftEvolutionBinary2022}, combined with a new
{\ppisne} remnant-mass prescription \citep{renzoPairinstabilityMassLoss2022}. We
introduce variations in this prescription to capture the effects of uncertain or
new physics. Moreover, we estimate the rates of {\ppisne} and {\pisne} and
compare them to the observed {\slsne} to constrain our variations. We aim to
evaluate whether the peak at $35\,\solarmass$ is explained by BHs formed through
{\ppisne}.

The layout of this paper is as follows. In \Secref{sec:Method} we explain our
method to simulate populations of {\bhbh}s through population synthesis
(\Secref{sec:input-physics} and~\ref{sec:simulated-populations}) and describe
our approach to convolving our binary populations with star-formation rates
(\Secref{sec:cosmo-convo}). In \Secref{sec:updated-routines} we explain our
variations of the {\ppisn} mechanism. In \Secref{sec:results} we show the
primary-mass distributions at $z=\ligoredshift$, and {\bhbh} merger and EM
transient event-rate densities as a function of redshift in our fiducial
populations and the populations with variations on the {\ppisne} mechanism. We
discuss our findings and conclude in Sections~\ref{sec:discussion}
and~\ref{sec:conclusion}.

\section{Method}
\label{sec:Method}
We simulate populations of binary-star systems using {\binaryc}, a binary
population-synthesis framework based on the stellar-evolution algorithm of
\citet{hurleyComprehensiveAnalyticFormulae2000, hurleyEvolutionBinaryStars2002},
which makes use of the single-star models of
\citet{polsStellarEvolutionModels1998} and provides analytical fits to their
evolution as in \citet{toutRapidBinaryStar1997} with updates in
\citet{izzardNewSyntheticModel2004, izzardPopulationNucleosynthesisSingle2006,
  izzardPopulationSynthesisBinary2009, claeysTheoreticalUncertaintiesType2014,
  izzardBinaryStarsGalactic2018, izzardCircumbinaryDiscsStellar2022,
  hendriksBinaryCpythonPythonbased2023}.

We combine the results of these populations with cosmological star-formation
rates, similar to, e.g., \citet[][]{dominikDOUBLECOMPACTOBJECTS2013,
  dominikDoubleCompactObjects2015, belczynskiCOMPACTBINARYMERGER2016,
  mandelMergingBinaryBlack2016, chruslinskaInfluenceDistributionCosmic2019,
  neijsselEffectMetallicityspecificStar2019, sonRedshiftEvolutionBinary2022,
  tanakaOptimalEnvelopeEjection2023}, to estimate the rate and mass distribution
of merging {\bhbh} systems as a function of redshift.

\subsection{Population synthesis and input physics}
\label{sec:input-physics}
For an in-depth review of the relevant physical processes in binary stellar
physics, see \citet{langerPresupernovaEvolutionMassive2012,
  postnovEvolutionCompactBinary2014, demarcoImpactCompanionsStellar2017} and
\citet{petrovicEvolutionMassiveBinary2020}. We highlight our choices of physics
prescriptions for the processes relevant to this study in the following
sections.

\subsubsection{Mass transfer, stability, and common-envelope evolution}
\label{sec:mass-transf-stab}
During their evolution, stars in binary systems interact with their companion by
expanding and overflowing their Roche-Lobe (RL), resulting in mass flowing from
the donor star to its companion.
We take the mass-transfer rate of the donor from
\citet{claeysTheoreticalUncertaintiesType2014}. When the accretor has a
radiative envelope, we limit the mass-accretion rate to $10$ times its thermal
limit,
$\dot{M}_{\mathrm{acc\ thermal\ limit}} = 10\,\dot{M}_{\mathrm{KH,\ acc}}$,
where
$\dot{M}_{\mathrm{KH,\ acc}} = M_{\mathrm{acc}}/\tau_{\mathrm{KH}}\
\solarmass\,\mathrm{yr}^{-1}$ 
with $\tau_{\mathrm{KH}}$ the global Kelvin-Helmholtz timescale of the accretor
and the factor of 10 roughly accounts for the fact that initially only the outer
envelope, which has a shorter timescale than the global $\tau_\mathrm{KH}$,
responds to mass accretion. We do not similarly limit the accretion rate of
giant-type stars with convective envelopes because we assume that they shrink in
response to mass accretion \citep{hurleyEvolutionBinaryStars2002}. We do not
expect this assumption to have a dominant impact because, over all redshifts,
only up to $10$ per cent of the merger rate of our {\bhbh} mergers consists of
systems that undergo any episode of mass transfer onto a giant-like star. We
further limit the accretion rate onto compact objects by the Eddington accretion
rate limit.
We assume any mass transfer exceeding the accretion rate limits is lost from the
system. Moreover, we assume that that mass carries a specific angular momentum
equal to the specific orbital angular momentum of the accretor, the so-called
isotropic re-emission mass loss \citep{sobermanStabilityCriteriaMass1997}.
We calculate the stability of mass transfer based on the critical mass ratio,
$q_{\mathrm{crit}} = M_{\mathrm{accretor}}/M_{\mathrm{donor}}$, at the onset of
mass transfer. For stars on the main sequence, Hertzsprung gap, giant branch,
early AGB and thermally pulsing AGB we use the $q_{\mathrm{crit}}$ of
\citet{geAdiabaticMassLoss2015, geAdiabaticMassLoss2020}. For the remaining
stellar types we use the $q_{\mathrm{crit}}$ of
\citet{claeysTheoreticalUncertaintiesType2014}.

Recent studies suggest that the rate of {\bhbh} mergers that experience and
survive common-envelope (CE) evolution might be overestimated
\citep{marchantNewRouteMerging2016, klenckiItHasBe2021,
  gallegos-garciaBinaryBlackHole2021, olejakImpactCommonEnvelope2021}, and they
argue that mass transfer should either generally be more stable or the ejection
of the envelope much more difficult and hence the stars merge (see, however,
\citealt{renzoRejuvenatedAccretorsHave2023}). Independently, both
\citet{sonNoPeaksValleys2022} and \citet{brielUnderstandingHighmassBinary2023}
showed that the CE channel is not necessary to explain the rate of {\bhbh}
mergers, while the converse is true for binary neutron-star mergers
\citep{chruslinskaDoubleNeutronStars2018, tanakaOptimalEnvelopeEjection2023}.
More importantly, \citet{sonRedshiftEvolutionBinary2022} shows that high-mass
{\bhbh} systems are almost exclusively formed through the stable mass-transfer
channel, and that the CE channel is inefficient for the formation of systems
with $\mprimary > 25\,\solarmass$. Other population synthesis studies like
\citet[][\startrack]{belczynskiBlackHoleBlack2022},
\citet[][\mobse]{mapelliPropertiesMergingBlack2019,mapelliCosmicEvolutionBinary2022}
and \citet[][\bpass]{brielUnderstandingHighmassBinary2023}, come to the same
conclusion. In this work we test this with {\binaryc} and find the same results
(\Secref{sec:primary-mass-distribution-variations} and
\Appref{app:primary-additional}). Therefore, we focus on the stable
mass-transfer channel and generally exclude merging systems that survive a CE
(indicated with \apostrophes{\textit{excluding CE}}) from our primary mass
distribution results and our merger rate densities unless explicitly indicated
with \apostrophes{\textit{including CE}} (see also
Fig.~\ref{app:fiducial_distribution_highlight}).

\subsubsection{Wind mass loss}
\label{sec:wind-mass-loss}
We follow \citet{schneiderPresupernovaEvolutionCompactobject2021} in our choice
of wind mass loss prescriptions, with the exception of their LBV-wind
prescription.
For hot-star ($T_{\mathrm{eff}} > 1.1\,\times\,10^{4}$K) winds we use
the prescriptions from \citet{vinkNewTheoreticalMassloss2000,
  vinkMasslossPredictionsStars2001}. For Wolf-Rayet star wind mass
loss we use the prescription of
\citet{yoonBetterUnderstandingEvolution2017}.
For low-temperature ($T_{\mathrm{eff}} < 10^{4}$K) stellar winds we use
\citet{reimersCircumstellarAbsorptionLines1975} mass loss on the first giant
branch, with $\eta = 0.4$, and
\citet{vassiliadisEvolutionLowIntermediatemass1993} on the asymptotic giant
branch (AGB).
At intermediate temperatures we linearly interpolate.
Beyond the Humphreys-Davidson limit \citep{humphreysLuminousBlueVariables1994} we
use the prescription for LBV-winds as described in
\citet[][]{hurleyComprehensiveAnalyticFormulae2000}. We do not include the
effects of rotationally-enhanced mass loss.

\subsubsection{Neutrino loss during compact object formation}
\label{sec:neutrino-loss-during}
For stars that only experience a {\ccsn} we calculate the baryonic remnant mass,
$M_{\mathrm{rem,\ bary}}$, using the {\fryerprescription} prescription of
\citet{fryerCompactRemnantMass2012}.
We calculate the gravitational remnant mass, $M_{\mathrm{rem,\ grav}}$, of BHs
formed through {\ppisne} and {\ccsne} from
\begin{equation}
  \label{eq:neutrino_loss}
  M_{\mathrm{rem,\ grav}} = M_{\mathrm{rem,\ bary}} -
  \min\left(0.5\,\solarmass, 0.1\,\times\,M_{\mathrm{rem,\ bary}}\right)
\end{equation}
\citep{zevinExploringLowerMass2020}.
Equation~\ref{eq:neutrino_loss} reduces the compact-object mass because of loss
of neutrinos during the collapse of the star. Because even in extremely massive
stars the CC releases a few $10^{53}-10^{54}$\,erg in neutrinos, we limit this
correction to $0.5\,\solarmass{}\simeq{}10^{54}\,\mathrm{erg}/c^2$
\citep{aksenovNeutronizationMatterStellar2016, zevinExploringLowerMass2020,
  rahmanPulsationalPairinstabilitySupernovae2022}.

\subsubsection{Envelope ejection following neutrino losses}
\label{sec:envel-eject-foll}
During CC, rapid changes in core mass because of neutrino emission change the
potential energy of a star, and lead to a pressure wave travelling outward. This
pressure wave, in some cases, evolves into a shock wave. In stars with low
envelope binding energy ($> -10^{48}\ \mathrm{erg}$), like red super giants,
this leads to a loss of (part of) the outer envelope
\citep{nadezhinSecondaryIndicationsGravitational1980,
  lovegroveVeryLowEnergy2013, piroTakingOutUnnovae2013}. Because the expected
mass loss depends on the structure of the core and the binding energy of the
envelope, most mass is lost from red (super) giants. Stars with compact
envelopes, such as blue and yellow super giants or Wolf-Rayet stars, are not
expected to lose much mass \citep{fernandezMassEjectionFailed2018,
  ivanovMassEjectionFailed2021}. We thus apply this effect only to red (super)
giants when the explosion is expected to fail (i.e.
  $f_{\mathrm{fallback}} = 1$), and assume that everything outside the He core
is lost,
\begin{equation}
  \label{eq:lovegrove_woosley_nadezhin}
  \Delta M_{\mathrm{\nu,\ env}} = M_{\mathrm{tot}} - M_{\mathrm{He}},
\end{equation}
where $\Delta M_{\mathrm{\nu,\ env}}$ is the ejected mass due to neutrino loss,
$M_{\mathrm{tot}}$ is the total mass of the star and $M_{\mathrm{He}}$ is the
mass of its He core. We assume this mass is ejected symmetrically and does not
introduce a natal kick to the star, other than a {\apostrophes{Blaauw}} kick
\citep{blaauwOriginBtypeStars1961}, due to the change in centre of mass. We do
not apply this mass loss term to blue and yellow supergiants and Wolf-Rayet
progenitors. In cases where the explosion is successful, the matter that may be
ejected because of the neutrino losses would anyway be easily removed by the SN
shock (as accounted for by the {\fryerprescription}), therefore we do not need
to apply Eq.~\ref{eq:lovegrove_woosley_nadezhin} when
$f_{\mathrm{fallback}} < 1$.

\subsubsection{Supernova natal kick}
\label{sec:supernova-natal-kick}
Stars that undergo CC may receive a natal momentum kick due to asymmetries in
the resulting explosion \citep[][]{shklovskiiPossibleCausesSecular1970,
  fryerNeutronStarKicks2004, jankaNatalKicksStellar2013,
  grefenstetteDISTRIBUTIONRADIOACTIVE44Ti2016,
  holland-ashfordComparingNeutronStar2017,
  katsudaIntermediatemassElementsYoung2018}. We calculate the supernova kick by
sampling a kick speed, $V_{\mathrm{sampled\,kick}}$ from a Maxwellian
distribution with dispersion of
$\sigma_{\mathrm{kick}} = 265\,\mathrm{km\ s^{-1}}$ and sampling a direction
isotropically on a sphere \citep{hobbsStatisticalStudy2332005}. We scale the
natal kick speed with the fallback fraction,
$f_{\mathrm{fallback}} = M_{\mathrm{fallback}}/M_{\mathrm{SN ejecta}}$, where
$M_{\mathrm{fallback}}$ is the total mass that falls back onto the remnant and
$M_{\mathrm{SN ejecta}}$ is the initial total supernova ejecta, as
\begin{equation}
  \label{eq:scaled_kick_speed}
  V_{\mathrm{scaled\,kick}} = V_{\mathrm{sampled\,kick}}(1-f_{\mathrm{fallback}}).
\end{equation}
We calculate this fraction through the {\fryerprescription} {\ccsn} prescription
of \citet{fryerCompactRemnantMass2012}. In \Appref{app:supern-kick-scal} we
discuss a different scaling.
Moreover, even if the supernova ejecta do not impart a natal kick, as long there
is any mass ejected, the system still experiences a Blaauw kick.
In {\ppisne} we assume spherically-symmetric ejecta, and no natal kick other
than the Blaauw kick \citep{chenTWODIMENSIONALSIMULATIONSPULSATIONAL2014,
  chenGasDynamicsNickel562020,
  chenRadiationHydrodynamicalSimulationsPulsational2022}.

\subsection{Simulated populations}
\label{sec:simulated-populations}

Binary-star systems are characterised by their initial primary mass, $M_{1}$,
secondary mass, $M_{2}$, orbital period, $P$, eccentricity, $e$, and
metallicity, $Z$. To evolve a population of binary systems, we vary each of
these initial properties by sampling from their probability distributions. In
this study we assume all the probability distributions are separable and can be
calculated independently.

For the initially more massive star (primary mass, $M_{1}$) we assume an initial
mass function (IMF) of \citet{kroupaVariationInitialMass2001}. We sample
$N_{M1}$ stars between $7.5$ and $300\,\solarmass$. Stars of an initially lower
mass do not form BHs and we do not include these in our populations.
We sample the initially less-massive star from a flat distribution in
$q = M_{2}/M_{1}$ \citep{sanaBinaryInteractionDominates2012} between
$0.08/M_{1}$ and 1, with a resolution $N_{q}$.
We sample the orbital period $P_{\mathrm{\mathrm{orb}}}$ of the binary systems
from a logarithmically spaced distribution between $0.15$ and $5.5$
$\mathrm{log}_{10}\left(\,P_{\mathrm{orb}}/\mathrm{d}\right)$, with the distribution
function from \citet{kobulnickyNewLookBinary2007} for systems with a primary
mass below $15\,\solarmass$ and the power-law in
$\mathrm{log}_{10}\left(\,P_{\mathrm{orb}}/\mathrm{d}\right)$ distribution
function with exponent 0.55 from \citet{sanaBinaryInteractionDominates2012} for
systems with $M_{1} > 15\,\solarmass$.
We neglect the possibility of initially eccentric binaries because with the
tidal circularisation model of \citet{hurleyEvolutionBinaryStars2002} that we
employ, they all circularise before interaction
\citep[][]{minkMERGERRATESDOUBLE2015}.

We assign a probability, $p_{i}$, to each system, $i$, which is a product of the
probability density functions of each variable and the step size in phase space,
see \citet{izzardPopulationSynthesisBinary2018} for a detailed explanation of
the method. \Appref{app:merger-rate-calculation} shows how we use the
probabilities $p_{i}$ and the binary fraction $f_{\mathrm{bin}}$ in our
merger-rate calculation. Throughout this study we assume a constant binary
fraction $f_{\mathrm{bin}} = 0.7$.

\newcommand\singleres{750}
\newcommand\binaryres{75}
We use a resolution of $N_{M1} = \singleres$ for our single-star system
parameter distributions.
We use a resolution of $N_{M1} = \binaryres$, $N_{q} = \binaryres$,
$N_{P} = \binaryres$ for our binary-system parameter distributions. We simulate
$N_{Z} = 12$ populations of single and binary systems, with metallicity equally
spaced in $\mathrm{log_{10}}(Z)$ between $\mathrm{log_{10}}(Z)=-4$
(corresponding to very metal-poor stars with negligible wind mass-loss) and
$\mathrm{log_{10}}(Z)=-1.6$ (corresponding to super-solar stars with strong wind
mass loss). At each supernova we sample the natal kick direction and magnitude
(\Secref{sec:supernova-natal-kick}) $N_{\mathrm{SN\ kick}} = 4$ times, and
divide the probability fraction of the system as $p_{i}/N_{\mathrm{SN\
	kick}}$. This amounts to an initial total of $\sim{}4\,\times\,10^{5}$
binaries at each metallicity, of which a subset splits due to multiple kick
samples.

\subsection{Cosmological star formation history}
\label{sec:cosmo-convo}
We calculate the intrinsic redshift-dependent merger-rate density of {\bhbh}
systems, $\mathcal{R}_{\mathrm{\bhbh}}(z_{\mathrm{merge}},\ \zeta)$, merging at
redshift $z_{\mathrm{merge}}$, or the corresponding merging lookback time,
$t^{*}_{\mathrm{merge}}$, with a set of system properties, $\zeta$, (e.g.,\
orbital period, primary mass, metallicity) similarly to the {\compas} code
\citep{neijsselEffectMetallicityspecificStar2019,
  broekgaardenImpactMassiveBinary2021, sonRedshiftEvolutionBinary2022,
  rileyRapidStellarBinary2022}. We define the intrinsic redshift-dependent
merger-rate density as,
\begin{equation}
  \label{eq:merger_rate_density}
  \begin{aligned}
    \mathcal{R}_{\mathrm{\bhbh}}(z_{\mathrm{merge}},\ \zeta) =
    \int_{Z_{\mathrm{min}}}^{Z_{\mathrm{max}}} dZ
    \int_{0}^{t^{*}_{\mathrm{first\, SFR}}-t^{*}_{\mathrm{merge}}}dt_{\mathrm{delay}}\\
    \mathcal{N}_{\mathrm{form}}(Z,\,t_{\mathrm{delay}},\,\zeta)
    \times \mathrm{SFR}(Z,\,z_{\mathrm{birth}}).
  \end{aligned}
\end{equation}
The integrand consists of the number of {\bhbh} systems per formed solar mass,
$\mathcal{N}_{\mathrm{form}}(Z,\,t_{\mathrm{delay}},\,\zeta)$, as a function of
metallicity $Z$, delay time, $t_{\mathrm{delay}}$, and system properties,
$\zeta$, and the star-formation rate density,
$\mathrm{SFR}(Z, z_{\mathrm{birth}})$, as a function of $Z$ and the birth
redshift, $z_{\mathrm{birth}}$. The delay time, $t_{\mathrm{delay}}$, is the sum
of the time it takes from the systems birth to the moment the second BH forms
(DCO formation), $t_{\mathrm{form}}$, and the time it takes the DCO to inspiral
and merge due to emission of gravitational wave radiation,
$t_{\mathrm{inspiral}}$. The inspiral time $t_{\mathrm{inspiral}}$ of the
{\bhbh} system is computed from
\citet{petersGravitationalRadiationMotion1964}. The birth redshift corresponds
the birth lookback time of the system,
$t^{*}_{\mathrm{birth}} = t^{*}_{\mathrm{merge}} +
t_{\mathrm{delay}}$. Generally, times with a * superscript are lookback times
and those without are durations.
We integrate this over metallicity between the metallicity bounds
$Z_{\mathrm{min}}$ and $Z_{\mathrm{max}}$ (\Secref{sec:simulated-populations}),
and over the delay time $t_{\mathrm{delay}}$ between $0$ and
$t^{*}_{\mathrm{first\, SFR}}-t^{*}_{\mathrm{merge}}$ to avoid integrating
beyond $t^{*}_{\mathrm{first\, SFR}}$, which is the lookback time of first
star-formation and has the corresponding first star-formation redshift
$z_{\mathrm{first\, SFR}}$.

We determine $\mathcal{N}_{\mathrm{form}}(Z,\,t_{\mathrm{delay}},\,\zeta)$ by
simulating populations of binary stars (\Secref{sec:simulated-populations} and
\Appref{app:merger-rate-calculation}) with primary stars between $7.5$ and
$300\,\solarmass$, and we convolve {\bhbh} systems with the star formation rate,
$\mathrm{SFR}(z,Z)$, of \citet{sonRedshiftEvolutionBinary2022}, with redshifts
between 0 and $z_{\mathrm{First\ SFR}} = 10$ and a step size of $dz = 0.025$
through the discretized version of \Eqref{eq:merger_rate_density}.
We use the \textit{PLANCK13} \citep{adePlanck2013Results2014} cosmology to
calculate redshift as a function of the age of the Universe and the volume
spanned by the redshift shells.

To calculate the total merger-rate density at a given redshift we integrate
\Eqref{eq:merger_rate_density} over all system properties $\zeta$,
\begin{equation}
  \label{eq:total_merger_rate_at_given_redshift}
  \mathcal{R}_{\mathrm{\bhbh}}(z_{\mathrm{merge}}) = \int \mathcal{R}_{\mathrm{\bhbh}}(z_{\mathrm{merge}},\ \zeta)\ d\zeta.
\end{equation}

While the total merger rate is degenerate in both the adopted
cosmology/star-formation rate prescription and the adopted stellar physics
\citep[e.g.][]{broekgaardenImpactMassiveBinary2022}, the locations of the
features in the mass distribution of merging {\bhbh}s are robust under the
uncertainties of the star-formation rates
\citep{sonLocationsFeaturesMass2023}. In this study we therefore fix the
star-formation rate prescription and only vary the prescription for {\ppisne}.

\section{The PPISN remnant mass prescription and its variations}
\label{sec:updated-routines}
\begin{figure*}
  \centering
  \includegraphics[width=0.75\textwidth]{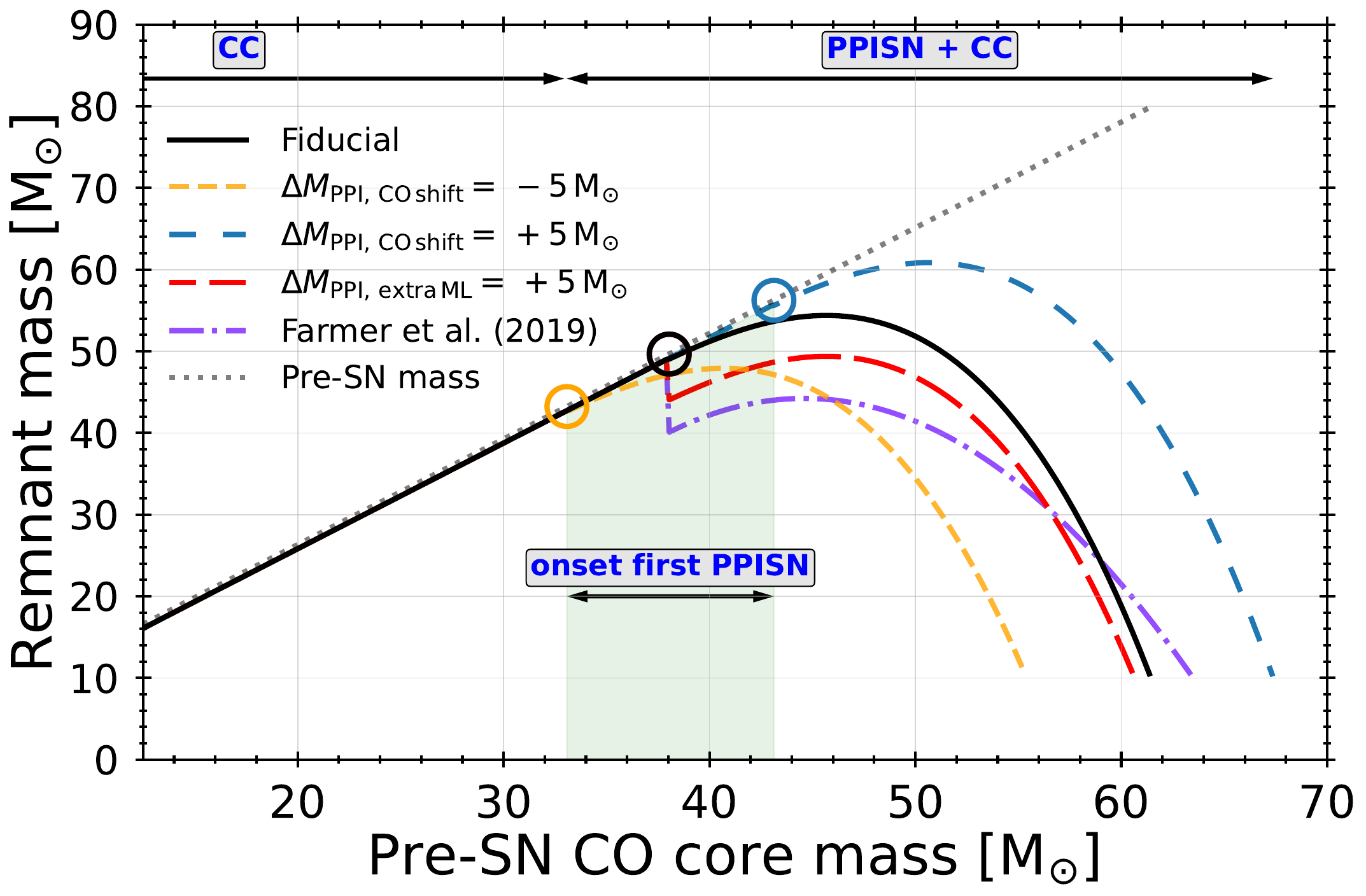}
  \caption[Remnant mass vs. pre-SN CO core mass in single stars at metallicity
  $Z = 0.001$ for our fiducial, \citetalias{farmerMindGapLocation2019},
  $\COshift = -5\,\solarmass$, $\extraML = +5\,\solarmass$ and
  $\COshift = +5\,\solarmass$ models.]{Remnant mass vs. pre-SN CO core mass in
	single stars at metallicity $Z = 0.001$. The grey, dotted line shows the
	pre-SN mass, the black line shows the remnant mass with our fiducial
	implementation \citep{renzoPairinstabilityMassLoss2022}, and the purple
	dash-dotted line shows the remnant mass as given by the prescription of
	\citetalias{farmerMindGapLocation2019}. The dashed-coloured lines indicate
	example variations on the {\ppisne} prescription. The orange-dashed line
	indicates a downward CO core-mass shift of $\COshift = -5\,\solarmass$, the
	red long-dashed line indicates an additional mass loss of
	$\extraML = +5\,\solarmass$ and the blue loosely-dashed line indicates an
	upward CO core-mass shift of $\COshift = +5\,\solarmass$. The corresponding
	circles indicate the CO core mass of the {\ppisn} onset for each
	variation. The green shaded region indicates the range of {\ppisne} onset CO
	core masses spanned by the example variations.}
  \label{fig:schematic}
\end{figure*}

We model the mass loss of (P){\pisne} with the prescription of
\citet{renzoPairinstabilityMassLoss2022}, which is based on the detailed models
of \citetalias{farmerMindGapLocation2019}. This prescription takes a
{\apostrophes{top-down}} approach, that is it prescribes the total mass lost for
a given CO core mass, rather than directly prescribes a remnant mass. This
allows us to incorporate all possible mass-loss mechanisms when compact objects
with masses above and below the {\ppisn} regime form without introducing
artificial jumps in the remnant mass function. We show an example pre-SN core to
remnant-mass relation for our fiducial model at metallicity $Z=0.001$ in
\Figref{fig:schematic}.
\citetalias{farmerMindGapLocation2019} also provide a remnant-mass prescription
based on their detailed models which we include in our study. We call this the
\citetalias{farmerMindGapLocation2019} model.

We assume that stars with a minimum CO core mass
$M_{\mathrm{CO,\ Min,\ PPISN}} = 38\,\solarmass$ after carbon burning undergo
{\ppisne}. If pulsations lead to a remnant mass below $10\,\solarmass$ we regard
the supernova as a {\pisn} which leaves no remnant behind
\citep{marchantPulsationalPairinstabilitySupernovae2019}. For CO core masses
greater than $114\,\solarmass$ we assume direct collapse to a BH following the
photodisintegration instability \citep{bondEvolutionFateVery1984,
  renzoPredictionsHydrogenfreeEjecta2020}. If, at the onset of pulsations, the
star still has a hydrogen-envelope, we assume this is always expelled, and the
hydrogen envelope mass is added to the prescribed mass loss due to pulsations
\citep[appendix B, ][]{renzoPredictionsHydrogenfreeEjecta2020}.

%

In \Secref{sec:intro} we mention several processes that introduce a large
uncertainty in CO core masses that undergo PPI compared to our fiducial
model. This motivates us to consider introducing a parameter to shift the CO
core-mass range that undergoes PPI in our prescription.
Moreover, the observational \citep{ben-amiSN2010mbDirect2014,
  kuncarayaktiBactrianBroadlinedTypeIc2023,
  linSuperluminousSupernovaLightened2023} and theoretical
\citep{powellFinalCoreCollapse2021,
  rahmanPulsationalPairinstabilitySupernovae2022} indications of additional
post-PPI mass-loss motivates us to consider this in our prescription as well.

We capture these effects by modifying our prescription from
\citet{renzoPairinstabilityMassLoss2022} to allow such variations and hence our
predicted {\ppisn }mass loss is,
\begin{equation}
  \begin{aligned}
  \label{eq:modified-top-down}
  \Delta M_{\mathrm{PPI}} &= (0.0006 \log_{10}Z + 0.0054)\,\times\,\\
  &(M_{\mathrm{CO}} - \COshift - 34.8)^{3}\\ &- 0.0013\,\times\,(M_{\mathrm{CO}} -
  \COshift - 34.8)^{2} \\ & + \extraML,
  \end{aligned}
\end{equation}
where $\COshift$ is the mass by which we shift the CO core-mass requirement for
{\ppisne}. Negative $\COshift$ shifts the core-mass range to lower masses, and
vice-versa. $\extraML$ represents additional, post-pulsation, mass loss, and $Z$
is the metallicity of the star.

We vary $\COshift$ between $-20\,\solarmass$ and $+10\,\solarmass$, and
$\extraML$ between $0\,\solarmass$ and $+20\,\solarmass$, in steps of
$5\,\solarmass$. We note that $\COshift = +10\,\solarmass$ qualitatively behaves
like \citet{faragResolvingPeakBlack2022} and $\COshift = -15\,\solarmass$
behaves qualitatively like the $f=0.5$ model of
\citet{moriLightCurvesEvent2023}, corresponding to an axion mass of half the
electron mass.

Varying {$\COshift$} and {$\extraML$} allows us to determine how a shift of the
{\ppisn} CO core-mass range or additional mass loss from {\ppisne} affects the
remnant-mass distribution, and specifically whether these changes lead to a
feature in the primary-mass distribution at $\sim{}35\,\solarmass$.

\Figref{fig:schematic} shows an example of the remnant-mass distribution as a
function of the pre-SN CO core mass at $Z = 0.001$ with a CO core mass shift of
$\COshift = \pm5\,\solarmass$.
The shift of the range of CO core masses that undergo {\ppisn} both affects the
(P){\pisne} as well as {\ccsne}. By shifting the range to lower or higher
masses, the CO core masses that undergo CC decrease or increase respectively. It
is important to note that a translation in the CO core mass that undergoes
{\ppisn} does not translate directly to the same shift in ZAMS masses. This
difference is caused by a non-linear relation between the ZAMS mass and the
pre-SN CO core mass
\citep[e.g.,][]{limongiPresupernovaEvolutionExplosive2018}. In
\Appref{sec:effect-single-stars} we show examples of the dependence of the
remnant mass as a function of ZAMS mass and metallicity.

We further note that additional mass loss ($\extraML$) only affects stars that
already undergo {\ppisne}. Hence it does not affect the rate of {\ccsne} but
only the rate of, and ratio between, {\ppisne} and {\pisne}, because too much
additional mass loss turns a {\ppisn} into a {\pisn}.
Another effect of our implementation is that at high additional mass loss
($\extraML > 10$), the most massive BH formed through single-star evolution is
from direct CC, not through {\ppisn}+CC.
We show how $\extraML = 10\,\solarmass$ affects the initial-final-mass relation
in a grid of masses and metallicities \Appref{sec:effect-single-stars}.

\section{Results}
\label{sec:results}
In this section we present the results of our simulations. We present the
primary BH-mass distributions of our models with varying {\ppisne} mechanism
properties in \Secref{sec:primary-mass-distribution-variations} and the
event-rate densities of EM-transient events as well as GW-merger events in
\Secref{sec:rate-dens-supern}.
We emphasise that in \Secref{sec:primary-mass-distribution-variations} we
exclude systems that undergo CE evolution. See \Secref{sec:mass-transf-stab} for
our motivation and~\Appref{app:primary-additional} for results that include CE
evolution.

\subsection{Primary-mass distributions}
\label{sec:primary-mass-distribution-variations}

\begin{figure*}
  \centering
 \includegraphics[width=\textwidth]{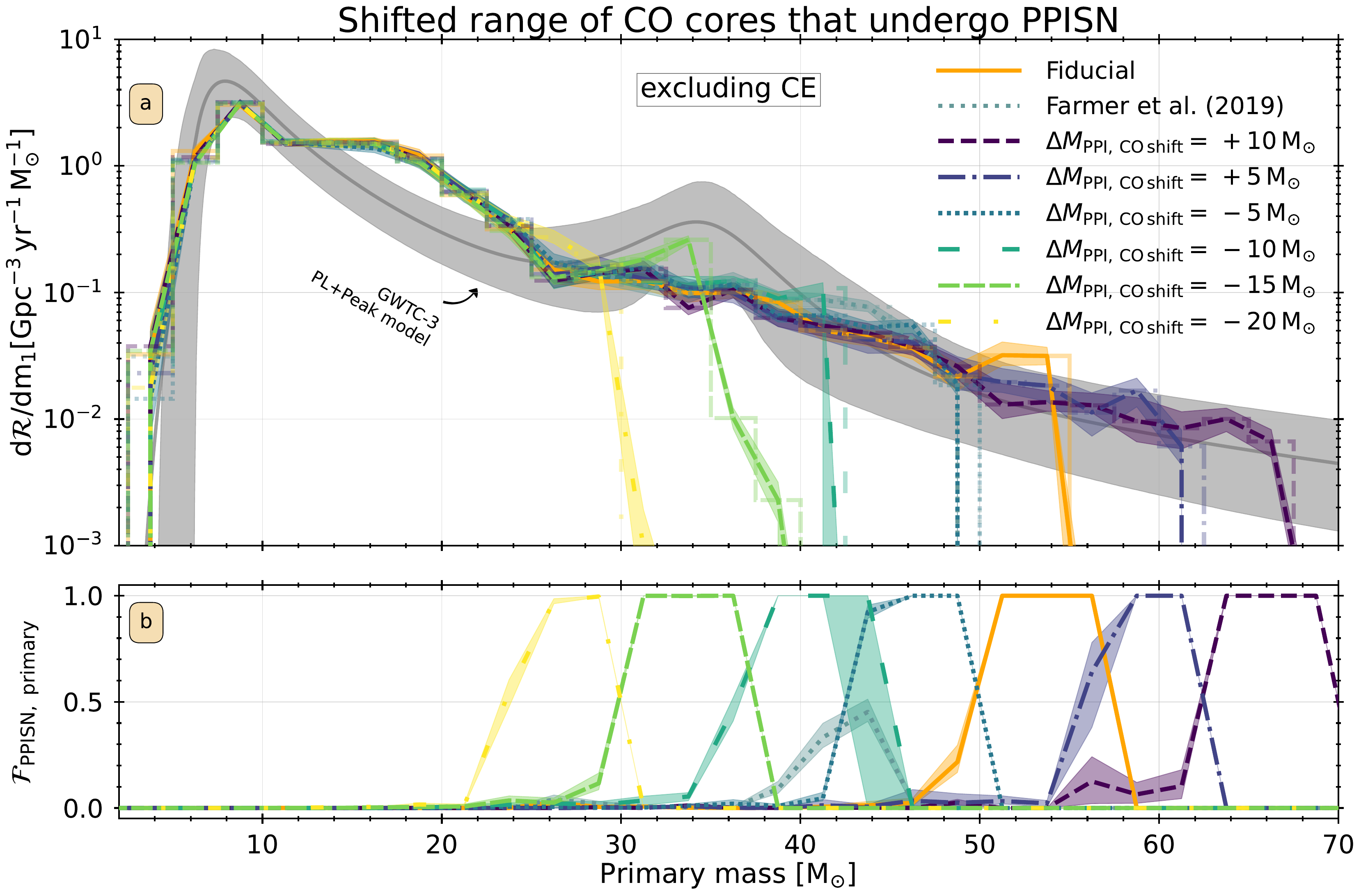}
 \caption[Merger rate density and fraction of primary mass black holes that form
 through \ppisne as a function of primary mass for our fiducial,
 \citetalias{farmerMindGapLocation2019}, $\COshift = +10\,\solarmass$,
 $\COshift = +5\,\solarmass$, $\COshift = -5\,\solarmass$,
 $\COshift = -10\,\solarmass$, $\COshift = -15\,\solarmass$ and
 $\COshift = -20\,\solarmass$ models.]{Panel (a): Merger rate density as a
   function of primary mass for {\bhbh} mergers at $z = \ligoredshift$, for our
   fiducial model (orange solid), the \citetalias{farmerMindGapLocation2019}
   model (grey dotted), and our CO core-mass shift variations
   $\COshift = +10\,\solarmass$ (dark-purple dashed),
   $\COshift = +5\,\solarmass$ (blue dashed-dotted), $\COshift = -5\,\solarmass$
   (light-blue dotted), $\COshift = -10\,\solarmass$ (dark-green long-dashed)
   $\COshift = -15\,\solarmass$ (green dashed), $\COshift = -20\,\solarmass$
   (yellow long dash-dotted). The lines connect the centres of the bins
   (stepped) of width $2.5\,\solarmass$. The translucent bands around the lines
   are the regions of $90$ per cent confidence-intervals obtained by 50
   bootstrap samples of the DCO populations. The dark-grey line indicates the
   mean of the {\powerlawpeak} model of
   \citet{abbottPopulationMergingCompact2023} at $z = \ligoredshift$ and the
   grey shaded region indicates the $90$ per cent confidence interval.
   These models indicate that the fiducial model does not peak at the
   observed location, and we need to introduce a shift of as much as
   $\COshift = -15\,\solarmass$ to make the distribution peak at the observed
   mass. The upward shift of $\COshift = +10\,\solarmass$ that matches
   \citet{faragResolvingPeakBlack2022} forms a slight {\overdensity} at
   $\sim 64\,\solarmass$.
   Panel (b): fraction of systems in that bin where the primary BH formed
   through
   {\ppisn}.  \label{fig:primary-mass_distribution-at-redshift-zero-for-core-mass-shift-variation}}
\end{figure*}

We show the primary-mass distribution of merging {\bhbh}s for our CO core-mass
shift models, $\COshift$ and our additional mass loss models, $\extraML$, in
Figures
\ref{fig:primary-mass_distribution-at-redshift-zero-for-core-mass-shift-variation}
and
\ref{fig:primary-mass_distribution-at-redshift-zero-for-extra-massloss-variation}. Panels
\ref{fig:primary-mass_distribution-at-redshift-zero-for-core-mass-shift-variation}a
and
\ref{fig:primary-mass_distribution-at-redshift-zero-for-extra-massloss-variation}a
show the merger-rate density of {\bhbh} systems as a function of primary-BH
mass. Panels
\ref{fig:primary-mass_distribution-at-redshift-zero-for-core-mass-shift-variation}
(b) and
\ref{fig:primary-mass_distribution-at-redshift-zero-for-extra-massloss-variation}
(b) show the fraction of primary-mass BHs that are formed through {\ppisne}.

Our fiducial primary-mass distribution at redshift $z = \ligoredshift$
(\Figref{fig:primary-mass_distribution-at-redshift-zero-for-core-mass-shift-variation}a,
orange line) peaks at about $10\,\solarmass$ in good agreement with the LVK
observations \citep{abbottPopulationMergingCompact2023}.
Moreover, in the intermediate range of $12 < \mprimary/\solarmass < 25$ we
predict more mergers than observed, which is seen in several BSE-based rapid
population-synthesis codes \citep[e.g.,][]{mapelliCosmicEvolutionBinary2022,
  sonLocationsFeaturesMass2023}, though the origin of this over-production is
unknown. This region contains systems that undergo at least one stable
mass-transfer episode and we do find indications that the mass-transfer
stability prescriptions affect the width and height of this over-density.
Additionally, we find that among systems with $\mprimary > 12\,\solarmass$ some
merge without undergoing any mass transfer but are able to merge because they
form with very high eccentricity ($e > 0.9$) upon DCO formation. The fraction of
systems that merge through this channel is low ($<10$ per cent) at low
($12\,\solarmass$) primary mass but slowly increases with primary mass to about
$30$ per cent above $45\,\solarmass$.
We note that we find that at a primary mass of $15\,\solarmass$, 50 per cent of
the merging systems undergo CE, down to 10 per cent at $25\,\solarmass$ and 0
per cent at $30\,\solarmass$ (\Appref{app:primary-additional}). This justifies
our exclusion of systems that go through CE and our choice to focus on the
high-mass end of the primary-mass distribution.

Related to {\ppisne}, we find the following results.
First, in our fiducial model, we find a {\ppisn} pile-up between
$50-52\,\solarmass$ and the rate in this pile-up is double the rate of systems
with primary masses just below the peak ($\mprimary =
48-50\,\solarmass$). {\ppisne} also lead to a maximum primary mass
$\mppisncutoff \sim{}55\,\solarmass$, which sets the lower edge of the
{\pisn}-mass gap.
Secondly, within the $51\mhyphen55\,\solarmass$ region associated with the
pile-up, $100$ per cent of the BHs form through {\ppisne}. We find an extended
region between $49 \mhyphen57\,\solarmass$ where, for a given primary mass, at
least $25$ per cent of BHs are formed through {\ppisn}
(\Figref{fig:primary-mass_distribution-at-redshift-zero-for-core-mass-shift-variation}
(b), orange line).
Thirdly, we find that systems in the mass range where the primary BHs are
predominantly formed through {\ppisne} show very high ($\geq0.9$) eccentricity
upon DCO formation. Systems in this region do not gain much eccentricity due to
the low mass ejecta of the {\ppisne}. We find, however, that the eccentricity is
mostly a result of the supernova of the initially lower mass-companions.
More generally, we note that from $30\,\solarmass$ upward, our merging systems
almost exclusively have high ($\geq0.9$) eccentricity at DCO formation,
i.e.~after the second SN. This indicates that they merge primarily because of
their eccentricity which strongly reduces their inspiral time
\citep{petersGravitationalRadiationMotion1964}. Without this eccentricity, the
majority of these systems are too wide to merge in a Hubble time.
We find that the systems that undergo no mass transfer also form with high
eccentricity and merge because their inspiral time is reduced because of
this. Especially in the mass range where the primary BHs are predominantly
created through {\ppisne} ($50 < M_{\mathrm{primary}}/\solarmass < 55$, many
($40$ per cent) never undergo mass transfer, but are formed with large ($>0.9$)
eccentricities upon DCO formation.

The distribution of primary masses in our \citetalias{farmerMindGapLocation2019}
model shows similar behaviour to our fiducial model for masses below
$\mprimary < 36\,\solarmass$, but differs at the high-mass end
($\mprimary \geq 36\,\solarmass$). The most massive primary mass is
$\mppisncutoff = 49\,\solarmass$ and there is a slight {\overdensity} at
$42\,\solarmass$. Moreover, around the over-dense region
($36 < \mprimary/\solarmass < 46$), the fraction of primary masses that undergo
{\ppisn} is at most $\sim0.5$, meaning that a large fraction of systems in that
{\overdensity} have primary BHs that undergo no {\ppisn}, but are rather formed
directly through a {\ccsn} (\Figref{fig:schematic}
and~\Secref{sec:updated-routines}).

\subsubsection{Shift in CO core mass for pair-instability}
\label{sec:res_shift}
To reflect uncertainties in the CO core masses that undergo {\ppisne}, we vary
$\COshift$. We show our primary-mass distributions from our $\COshift$ models in
\Figref{fig:primary-mass_distribution-at-redshift-zero-for-core-mass-shift-variation}. While
our fiducial model is based on the detailed stellar models of
\citetalias{farmerMindGapLocation2019}, the $\COshift = +10\,\solarmass$
variation behaves like the more-recent results of
\citet{mehtaObservingIntermediatemassBlack2022} and
\citet[][]{faragResolvingPeakBlack2022} with more densely sampled
$\carbonoxygen$ reaction rates, and improved spatial and temporal resolution.
Below $20\,\solarmass$, the distribution of primary BH masses is not strongly
affected by these variations.
Reducing the CO core-mass threshold for {\ppisne} decreases the most massive BH
mass, and shifts the location of pile-up from {\ppisn} downwards. We have to
shift the range of CO core masses that undergo {\ppisn} down by
$10-15\,\solarmass$ to move the {\ppisn} pile-up near the observed feature at
$35\,\solarmass$.
Our upward-shift variation models show an increase in maximum BH mass, and
generally a less pronounced, but not absent, pile-up of BHs formed through
{\ppisn}.
All primary-mass distributions in our CO core-mass shift models show that the
mass-range around the pile-up is entirely populated by primary BHs that are
formed through {\ppisne}.

In summary, we find that varying $\COshift$ shifts the location of the {\ppisn}
pile-up. To have the {\ppisn} feature appear near the observed
$32-38\,\solarmass$ peak, we need a shift of $\COshift \simeq
-15\,\solarmass$. The variations motivated by the models of
\citet{faragResolvingPeakBlack2022}, i.e.~an upward shift of
$\COshift \simeq 10\,\solarmass$, create a shallow {\overdensity} at
$\simeq{}64\,\solarmass$. Current observations show no structure in this region,
but the current (\ofourrun) and planned (O5) observing runs will help unveil any
existing structure in the primary BH mass distribution in this mass range.

\subsubsection{Extra mass loss during, or after, pulsational pair-instability}
\label{sec:extraML}
\begin{figure*}
  \centering \includegraphics[width=\textwidth]{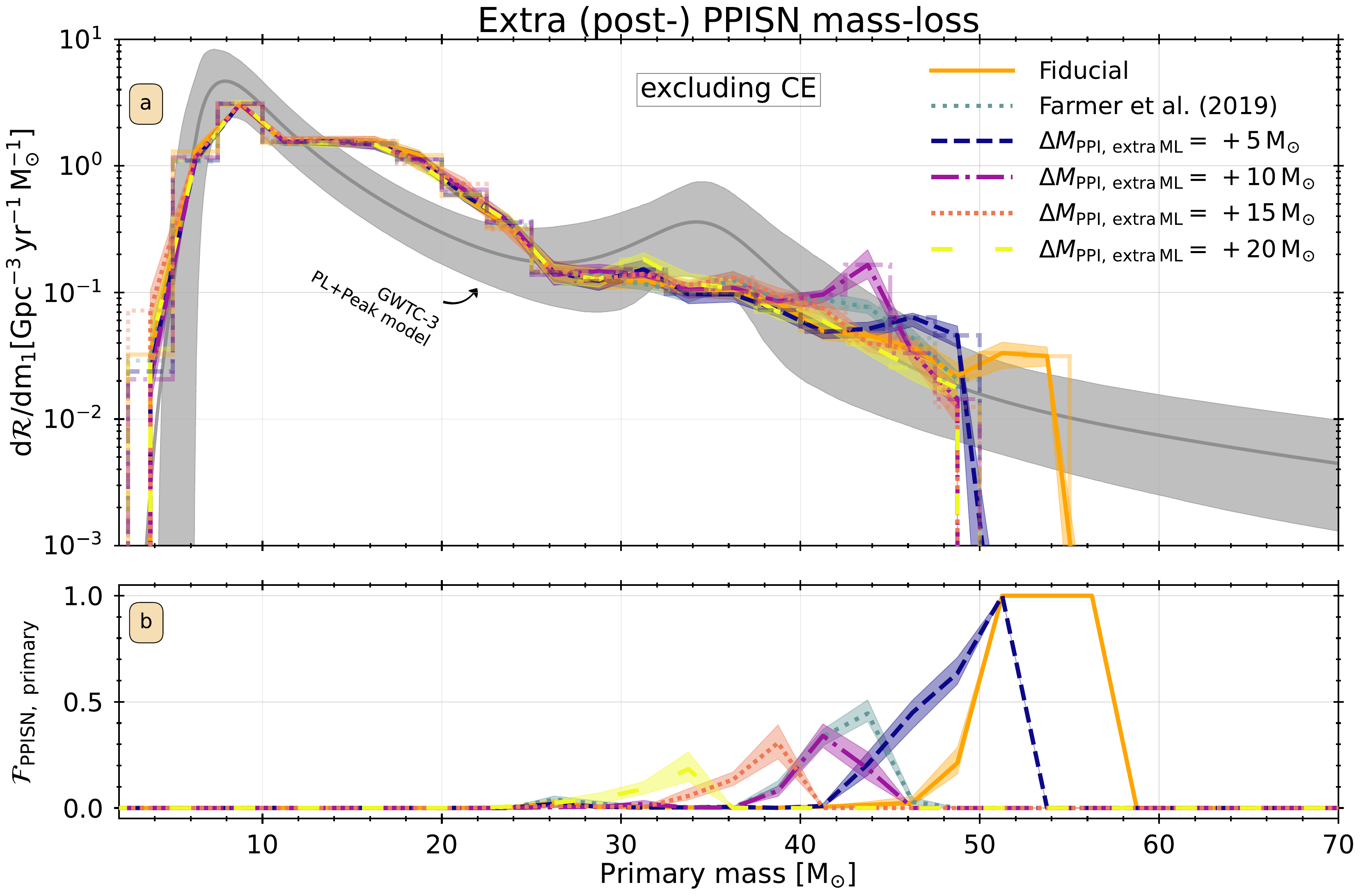}
  \caption[As
  \Figref{fig:primary-mass_distribution-at-redshift-zero-for-core-mass-shift-variation}
  for our fiducial, \citetalias{farmerMindGapLocation2019},
  $\extraML = 5\,\solarmass$, $\extraML = 10\,\solarmass$,
  $\extraML = 15\,\solarmass$, and $\extraML = 20\,\solarmass$ models.]{As
	\Figref{fig:primary-mass_distribution-at-redshift-zero-for-core-mass-shift-variation}
	for our fiducial model (orange solid), the
	\citetalias{farmerMindGapLocation2019} model (grey dotted), and our
	additional mass loss models $\extraML = 5\,\solarmass$ (dark-blue dashed),
	$\extraML = 10\,\solarmass$ (purple dashed-dotted), and
	$\extraML = 15\,\solarmass$ (red dotted), $\extraML = 20\,\solarmass$
	(yellow long-dashed).
    These models indicate that additional mass loss models lower the merger rate
    density of systems with $\mprimary > 48\,\solarmass$, and for some models
    ($\extraML = 5\,\solarmass$ and $\extraML = 10\,\solarmass$) increases the
    rate between $\mprimary = 40$ and $48\,\solarmass$, forming a peak. Below
    $40\,\solarmass$ the primary mass distribution is not affected by any amount
    of additional mass loss.
  } \label{fig:primary-mass_distribution-at-redshift-zero-for-extra-massloss-variation}
\end{figure*}

Both theory and observations suggest that some amount of additional mass loss
occurs post-PPI, which we model with $\extraML$.
We show our $\extraML$ variation simulations in
\Figref{fig:primary-mass_distribution-at-redshift-zero-for-extra-massloss-variation}. Our
results show that introducing additional mass loss to the {\ppisne} affects the
distribution of primary-BH masses in several ways.
First, removing extra mass lowers $\mppisncutoff$, and affects the location and
magnitude of the pile-up. Our additional mass loss models,
$\extraML = 5\,\solarmass$ and $\extraML = 10\,\solarmass$, shift
$\mppisncutoff$ down by up to $10\,\solarmass$. This is associated with an
increased magnitude of a pile-up of up to an order of magnitude.
The \citetalias{farmerMindGapLocation2019} model peaks at the same mass as our
$\extraML = 10\,\solarmass$ models, and shows a similar fraction of primaries
that are formed through {\ppisne} in the region of their pile-up. The
$\extraML = 10\,\solarmass$ model, however, shows a pile-up with double the
magnitude of the \citetalias{farmerMindGapLocation2019} model.
Though some of our $\extraML$ models increase the magnitude of the pile up
feature, these features are no longer exclusively populated by systems that
undergo {\ppisn}. This is for similar reasons as the feature in the
\citetalias{farmerMindGapLocation2019} model: the additional mass loss
introduces a jump in the remnant-mass function such that the most massive BH
comes from a {\ccsn} (\Figref{fig:schematic}).
Removing more than $10\,\solarmass$ does not affect $\mppisncutoff$ because any
BH formed by a {\ppisn} is of lower mass than the most massive BH formed through
CC. Moreover, the distribution of primary BHs less massive than
$\sim{}30\,\solarmass$ is not affected by our $\extraML$ models.

In summary, we find that additional mass loss,
$0 \leq \extraML \leq 10\,\solarmass$, lowers the location of the peak by up to
$10\,\solarmass$. Moreover, the rate in the pile-up increases by almost an order
of magnitude. This mechanism does not allow us to match the observed
$32-38\,\solarmass$ peak. After applying $\extraML \geq 10\,\solarmass$ the
{\ppisne} are sub-dominant across the entire mass range
($\mathcal{F}_{\mathrm{{\ppisn},\ primary}} < 0.1$), and stop affecting the
primary-mass distribution.

\begin{figure*}
  \centering
  \includegraphics[width=\textwidth]{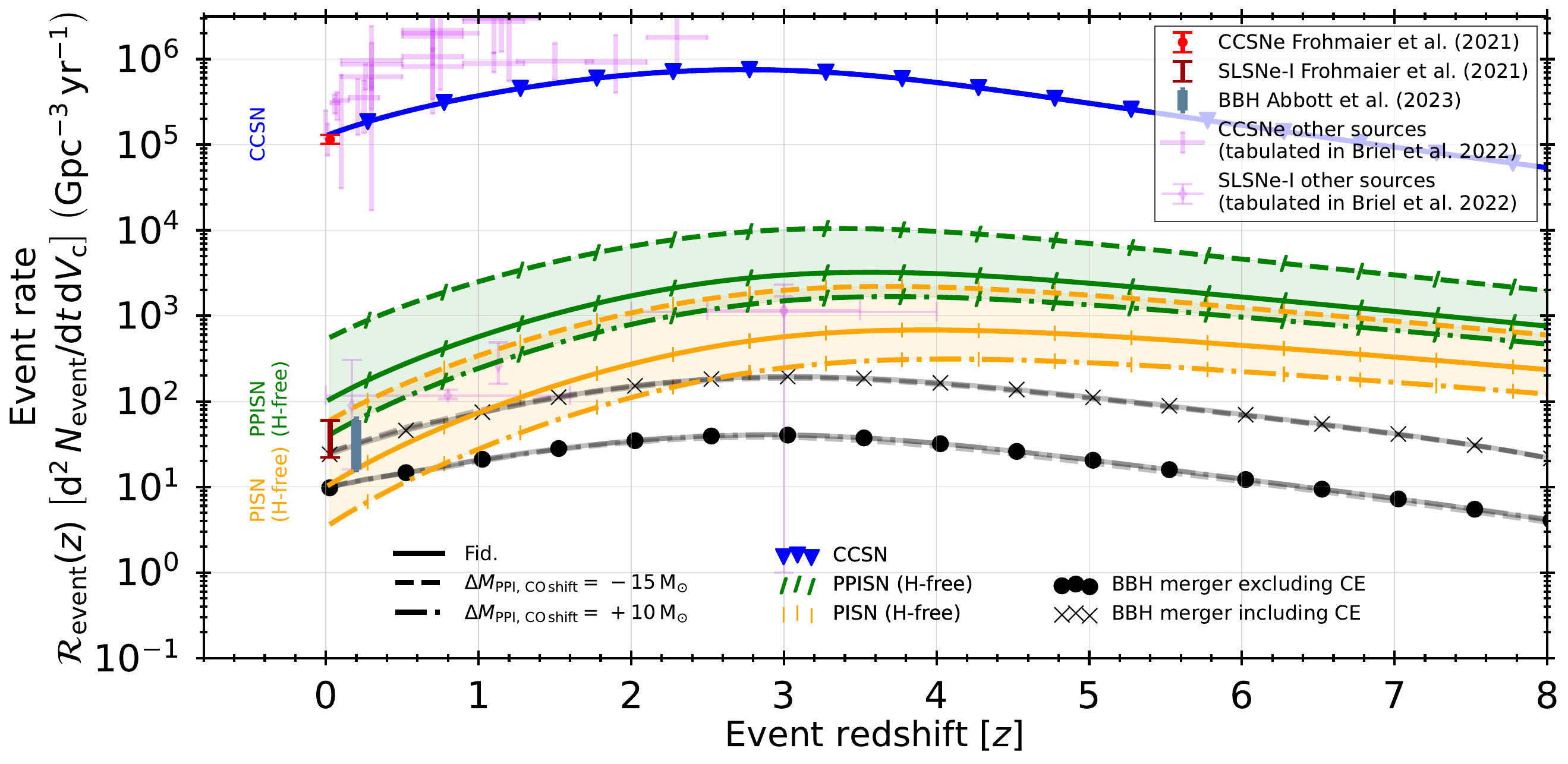}
  \caption[Intrinsic event-rate density evolution as a function of redshift for
  our fiducial, $\COshift = -15\,\solarmass$ and $\COshift = +10\,\solarmass$
  models.]{Intrinsic event-rate density, $\mathcal{R}_{\mathrm{event}}$,
	evolution in our fiducial simulations (all solid), the
	$\COshift = -15\,\solarmass$ variation (all dashed) and the
	$\COshift = +10\,\solarmass$ variation (all dash-dotted). We show our
	{\ccsne} transient-event (blue with downward triangle), {\ppisne} (green
	with forward slash) and {\pisne} (orange with vertical line), as well as the
	{\bhbh}-merger rates when we exclude CE systems (solid circles, the line
	styles match are the same as the transient event-rate variations) and those
	where we include CE systems (crosses). The translucent green and orange
	regions indicate the rate ranges spanned by the two variations for {\ppisne}
	and {\pisne} respectively. The other {$\COshift$} variations have rates that
	fall within these regions, except for $\COshift = -20\,\solarmass$. We
	indicate the observed transient event-rate densities of
	\citet{frohmaierCoreCollapseSuperluminous2021} for {\ccsne} (red) and
	{\slsne} (pink) as well as the expected GW merger event-rate densities of
	\citet{abbottPopulationMergingCompact2023} for {\bhbh} mergers (blue-gray
	bar). Moreover, we indicate the event-rate density estimates from other
	sources for {\ccsne} and {\slsne}, both tabulated in
	\citet{brielEstimatingTransientRates2022}, by pink errorbar symbols and pink
	capped-errorbar symbols with diamond centers. The (P){\pisni} rates increase
	in our {\downwardvariation} model and decrease in our {\upwardvariation}
	model, spanning about an order of magnitude in rate densities for both
	{\ppisni} and {\pisni} between the models. {\ccsne} transient- and
	{\bhbh}-merger rates are not affected in the two models compared to our
	fiducial model.}
  \label{fig:fiducial-sn-rate-density-evolution}
\end{figure*}

%
\subsection{Event-rate densities as a function of redshift}
\label{sec:rate-dens-supern}
In the following section we present our supernova-event rate density, i.e.~the
event rate in a given volume of space, and {\bhbh}-merger rate densities as a
function of redshift in our fiducial model as well as our core-mass shift models
{\downwardvariation} and {\upwardvariation}. We choose to show only these two
models because the former leads to a match of our modelled {\ppisn} pile-up
location with the observed peak, while the latter fits with the latest estimates
of stellar evolution and nuclear-reaction rates
\citep{faragResolvingPeakBlack2022}.
We calculate the intrinsic supernova rate density similar to the merger rate,
except that we use the time that the star took from birth to supernova,
$t_{\mathrm{SN}}$, as the timescale in the convolution
(\Eqref{eq:merger_rate_density}), instead of the delay time,
$t_{\mathrm{delay}}$.

Our {\ccsne} include type Ibc and type II supernovae, but exclude
\textit{failed} supernovae, i.e.~CC supernovae where the shock fails to unbind
any mass according to the {\fryerprescription} prescription of
\citet{fryerCompactRemnantMass2012}. We note that all our {\ppisne} and {\pisne}
are hydrogen-poor, i.e., {\ppisnei} and {\pisnei}, and we do not find any
hydrogen-rich {\ppisne} or {\pisne} in our simulations. In our models we find
that our stars (self-)strip and lose their hydrogen envelope before they undergo
(P){\pisn}, which may be caused by overestimated wind mass loss
\citep{beasorExtremeScarcityDustenshrouded2022}. 

In \Figref{fig:fiducial-sn-rate-density-evolution} we show the event-rate
density, $\mathcal{R}_{\mathrm{event}}$, which is the number of events,
$N_{\mathrm{event}}$, per unit time, $\mathrm{d}t$, per unit comoving volume,
$\mathrm{d}V_{\mathrm{c}}$, in units of number per $\mathrm{yr}$ per
$\mathrm{Gpc}^{3}$, of the supernova and {\bhbh}-merger events, both including
and excluding systems that undergo CE evolution from our fiducial,
{\downwardvariation} and {\upwardvariation} models. The rates are intrinsic,
i.e., are not weighted by detectability in any particular survey.

In \Figref{fig:fiducial-sn-rate-density-evolution} we also show the volumetric
rates at $z = \slsnredshift$ based on bias-corrected ZTF observations
\citep{frohmaierCoreCollapseSuperluminous2021}. These include the combined
hydrogen-rich and hydrogen-poor stripped-envelope {\ccsne}, at a rate of
$1.151^{+0.15}_{-0.13}\,\times\,10^{5}\,\eventratedensity$, {\slsnei}, at a rate
of $3.5^{+2.5}_{-1.3}\,\times\,10^{1}\,\eventratedensity$. We compare these to
our predicted {\ccsn} and (P){\pisni} rates. Moreover, we indicate estimates of
{\ccsne} and {\slsne} at higher redshifts from other sources that are tabulated
in \citet{brielEstimatingTransientRates2022}.

We show the {\slsni} rate because {\ppisnei} and {\pisnei} may be associated
with a subset of {\slsnei}, but we stress that not all {\ppisne} and {\pisne}
necessarily display {\slsne}-like transients.
We summarise the SN event-rate results at $z=\slsnredshift$ from our selected
models and compare them to the observations from
\citet{frohmaierCoreCollapseSuperluminous2021} in \Tabref{tab:rates_table}.

%
Our fiducial model shows a {\ccsn} transient-rate density of
$\sim{}10^{5}\,\eventratedensity$ at $z = 0$, increasing to
$\sim{}7\,\times\,10^{5}\,\eventratedensity$ by redshift $z \sim 3$ then
decreasing to $\sim{}6\,\times\,10^{4}\,\eventratedensity$ at $z = 8$.
Our fiducial {\ccsn} rate, as well as the rates of either variations, match
closely the rates of \citet{frohmaierCoreCollapseSuperluminous2021}. This
indicates that overall we reproduce the observed {\ccsn}-rate density and also
that variations in the {\ppisn} mechanism do not affect this rate strongly. This
is because the IMF disfavours stars massive enough to undergo {\ppisn} relative
to all {\ccsn} progenitors. Overall we find a reasonable match with the other
sources for {\ccsne} that are tabulated in
\citet{brielEstimatingTransientRates2022}, often matching the lower-bound
estimate of the rate.

We find a {\bhbh} merger rate of $\sim{}10\,\eventratedensity$ at $z \sim~0$,
excluding systems that undergo CE, which increases to
$\sim{}40\,\eventratedensity$ at $z \sim{}2.5$ then decreases to
$\sim{}4\,\eventratedensity$ at $z = 8$. These rates are not significantly
affected by changes in the CO core-mass range of {\ppisne}, because the merger
rate is dominated by systems with primary masses around $\sim{}10\,\solarmass$
(\Figref{fig:primary-mass_distribution-at-redshift-zero-for-core-mass-shift-variation},
\citealt{liFlexibleGaussianProcess2021, veskeCharacterizingObservationBias2021,
  edelmanAinNoMountain2022, tiwariExploringFeaturesBinary2022,
  abbottPopulationMergingCompact2023}). The rate of {\bhbh} mergers, if we
include those that undergo a CE event and survive, is about a factor of 3 larger
than those that exclude CE events over all redshifts. At $z = \ligoredshift$ the
rate including CE systems matches well with GW observations.
\Secref{sec:primary-mass-distribution-variations} shows that our fiducial
{\bhbh} mergers excluding CE systems match the overall shape of the observed
primary-mass distribution well, but here we find that including CE systems is
needed to match the observed rate integrated over all BH masses.

Our fiducial {\ppisni} transient-rate density at $z = 0$ is
$\sim{}10^{2}\,\eventratedensity$, which increases to
$\sim{}3\,\times\,10^{3}\,\eventratedensity$ at $z \sim 3$, and then it
decreases to $\sim{}7\,\times\,10^{2}\,\eventratedensity$ at $z = 8$.
Our fiducial {\pisni} transient-rate density at $z = 0$ is
$\sim{}10\,\eventratedensity$, which increases to
$\sim{}8\,\times\,10^{2}\,\eventratedensity$ at $z \sim 3$, and then it
decreases to $\sim{}1.5\,\times\,10^{2}\,\eventratedensity$ at $z = 8$.
Both our {\ppisn} and {\pisn} rates evolve with redshift but deviate from the
shape of the total SFR-density. Both peak at $z \sim 3$, coinciding with the
cosmic star-formation rate density peak (\Figref{fig:mssfr}), but at low
redshift ($z=0$) their event-rate density is lower, by at least a factor of 5,
than at high redshift ($z=8$). This is because (P){\pisne} occur in very massive
stars only, and thus their formation strongly depends on their metallicity. Even
if the star formation rate density at $z=0$
($\sim 2\,\times\,10^{7}\,\solarmass\,\eventratedensity$) exceeds that at $z=8$
($\sim 8\,\times\,10^{6}\,\solarmass\,\eventratedensity$), the metallicity
distribution at high $z$ trends towards lower metallicities, compensating for
their lower star-formation rates, because stars at lower metallicity lose less
mass and remain massive enough to undergo (P){\pisn}.

In \Tabref{tab:rates_table} we compare our {\ppisni} rate density estimate to
the inferred {\slsni} rate density of
\citet{frohmaierCoreCollapseSuperluminous2021}, expressed as the ratio between
our predicted and their observed rates.
The inferred rate density of {\slsnei} at $z=\slsnredshift$,
$3.5^{+2.5}_{-1.3}\,\times\,10^{1}\,\eventratedensity$, falls between our
predicted {\pisnei} and {\ppisnei} rates in our fiducial model.
With our predicted {\ppisni} rate density (at $z=\slsnredshift$),
$1.06\,\times\,10^{2}\,\eventratedensity$, we find a ratio to the {\slsni} rate
of $3.03^{+1.12}_{-2.16}$ and a ratio to the {\ccsn} rate of
$9.21^{+1.02}_{-1.25}\,\times\,10^{-4}$.
With our predicted {\pisnei} rate density,
$1.07\,\times\,10^{1}\,\eventratedensity$, we find a ratio to {\slsnei} of
$0.31^{+0.11}_{-0.22}$ and a ratio to {\ccsne} of
$9.30^{+1.03}_{-1.26}\,\times\,10^{-5}$.
This implies that in our fiducial model (P){\pisne} can contribute a significant
fraction to the {\slsn} rate. However, it is important to note that {\pisne} do
not necessarily lead to {\slsn}-like transients
\citep{gilmerPairinstabilitySupernovaSimulations2017}, and the same is likely
true in {\ppisne} \citep{woosleyPulsationalPairInstabilitySupernovae2017}. We
should thus caution making conclusions from directly from these
results. Comparison to the other sources that are tabulated in
\citet{brielEstimatingTransientRates2022} give similar ratios at higher
redshifts.

While shifting the CO core-mass range for {\ppisne} does not affect the {\ccsn}
transient-rate density nor any of the {\bhbh} merger rate densities
significantly, the {\ppisni} and {\pisni} rate densities are, however, strongly
affected.

With our {\downwardvariation} model both {\ppisni} and {\pisni} supernova rates
increase by about a factor of 5 $5.60\,\times\,10^{2}\,\eventratedensity$ and
$4.10\,\times\,10^{1}\,\eventratedensity$, respectively. This is because, in
this model, lower-mass stars explode as {\ppisnei}, and because the IMF favours
lower mass stars, this rate is higher.
With this model both the {\ppisni} and {\pisni} transient event-rate densities
at $z = \slsnredshift$ are either approximately equal to, or higher than, the
inferred {\slsni} rate density.
With our predicted {\ppisni} rate density we find a ratio to the {\slsni} rate
of $16.00^{+5.94}_{-11.43}$ and a ratio to the {\ccsn} rate of
$4.87^{+0.54}_{-0.66}\,\times\,10^{-3}$.
With our predicted {\pisni} rate density we find a ratio to the {\slsni} rate of
$1.73^{+0.64}_{-1.23}$ and a ratio to the {\ccsn} rate of
$5.26^{+0.58}_{-0.71}\,\times\,10^{-4}$.
The {\pisni} rate density is only slightly higher than the mean {\slsn} rate and
falls within its error bars, but the {\ppisni} rate density is higher by more
than an order of magnitude.

With {\upwardvariation} both {\ppisni} and {\pisni} supernova rates are
decreased by about a factor of 3 relative to our fiducial model. {\ppisnei}
decrease to $4.10\,\times\,10^{1}\,\eventratedensity$ and {\pisnei} decrease to
$3.66\,\eventratedensity$. We now see the effect of the IMF disfavouring
increasingly massive stars, decreasing the rate of both phenomena.
In this model both the {\ppisni} and {\pisni} transient event-rate densities are
either approximately equal to, or lower than, the inferred {\slsni} rate density.
With our predicted {\ppisni} rate density we find a ratio to the {\slsni} rate
of $1.17^{+0.44}_{-0.84}$ and a ratio to the {\ccsn} rate of
$3.56^{+0.39}_{-0.48}\,\times\,10^{-4}$.
With our predicted {\pisni} rate density we find a ratio to the {\slsni} rate of
$0.10^{+0.04}_{-0.07}$ and a ratio to the {\ccsn} rate of
$3.18^{+0.35}_{-0.44}\,\times\,10^{-5}$.
The {\ppisni} rate density is approximately equal to the mean inferred {\slsni}
rate density, but the {\pisni} rate density is lower by more than an order of
magnitude.

\begin{table}
  \centering
  \caption[Transient event-rate densities and their ratios to observed rate
  densities from our fiducial, {\downwardvariation} and {\upwardvariation}
  models at $z=\slsnredshift$]{Transient event-rate densities and their ratios to observed rate
  densities from our fiducial, {\downwardvariation} and {\upwardvariation}
  models.
  The ratios to the observed densities for (P){\pisne} span about an order of
  magnitude between our {\downwardvariation} and {\upwardvariation} models. All
  ratios of {\pisni} to {\ccsn} are in tension with
  \citet{frohmaierCoreCollapseSuperluminous2021}, but our {\downwardvariation}
  model most so.}
\begin{threeparttable}
\resizebox{\columnwidth}{!}{%
  {\renewcommand{\arraystretch}{1.1} 
    \setlength\tabcolsep{0.1cm}
  \begin{tabular}{ll|p{1.6cm}p{2cm}p{1.5cm}}
    \toprule
    Model & SN type & Rate density\tnote{a} $\left[\eventratedensity\right]$ & Ratio to {\ccsn} rate
                                           density\tnote{a,b}
    & Ratio to {\slsni}
      rate
      density\tnote{a,b} \\ \midrule
    Fiducial  & {\ccsn}     & $1.33\,\times\,10^{5}$ & & \\
      & {\ppisni}  & $1.06\,\times\,10^{2}$ & $9.21^{+1.02}_{-1.25}\,\times\,10^{-4}$ & $3.03^{+1.12}_{-2.16}$ \\
      & {\pisni}   & $1.07\,\times\,10^{1}$ &
                          $9.30^{+1.03}_{-1.26}\,\times\,10^{-5}$ &
                                                $0.31^{+0.11}_{-0.22}$ \\
    \hline
    {\downwardvariation} & {\ccsn} & $1.33\,\times\,10^{5}$ & & \\
          & {\ppisni}  & $5.60\,\times\,10^{2}$ & $4.87^{+0.54}_{-0.66}\,\times\,10^{-3}$ & $16.00^{+5.94}_{-11.43}$                 \\
          & {\pisni}   & $6.05\,\times\,10^{1}$ &
                             $5.26^{+0.58}_{-0.71}\,\times\,10^{-4}$ & $1.73^{+0.64}_{-1.23}$            \\
    \hline
    \upwardvariation & {\ccsn}  & $1.33\,\times\,10^{5}$                  &                &                  \\
          & {\ppisni} & $4.10\,\times\,10^{1}$   &
                             $3.56^{+0.39}_{-0.48}\,\times\,10^{-4}$
                                      & $1.17^{+0.44}_{-0.84}$                  \\
          & {\pisni}   & $3.66$                 &
                             $3.18^{+0.35}_{-0.44}\,\times\,10^{-5}$                & $0.10^{+0.04}_{-0.07}$\\
    \bottomrule
  \end{tabular}}
}
\begin{tablenotes}
\item[a] At $z=\slsnredshift$.
\item[b] Using the observed rate of~\citet{frohmaierCoreCollapseSuperluminous2021}.
\end{tablenotes}
\end{threeparttable}
  \label{tab:rates_table}
\end{table}

To summarise, we find that varying the CO core-mass range of (P){\pisne}
strongly affects the transient event-rate density of these supernovae, with
little effect on the overall rates of transients associated with {\ccsne}. In
our fiducial model, both the {\ppisne} and {\pisne} could contribute to the
{\slsn} rate. Our {\downwardvariation} model increases both rates such that the
rate of {\pisne} falls within the upper bound of the error on the observed
{\slsn} rate, and the {\ppisn} rate is about a factor of 16 times higher than
the mean {\slsn} rate. Our {\upwardvariation} model has lower (P){\pisn}
transient rates compared to our fiducial model, the {\pisn} rate is about an
order of magnitude lower than the {\slsn} rate and the {\ppisn} rate
approximately matches the {\slsn} rate. We discuss the implications of these
variations, and whether they are in tension with the observed {\slsn} rate, in
\Secref{sec:transient-rates}.

\section{Discussion}
\label{sec:discussion}
In the following section we discuss the implications of our results in
\Secref{sec:results}, some choices in our modelling approach, and whether, based
on our results, the observed peak in the primary-BH mass distribution at
$35\,\solarmass$ originates from {\ppisne}.

\subsection{{\ppisn} mechanism and the primary-mass distribution}
\label{sec:vari-p{\pisne}-mech}
Our modifications to the {\ppisn} prescription of
\cite{renzoPairinstabilityMassLoss2022} encompass both a shift in CO core masses
that undergo {\ppisne} and an additional {\ppisn} or post-{\ppisn} mass loss
(\Eqref{eq:modified-top-down}). This parametric approach allows us to
explore several physical effects proposed in the literature. We discuss the
results of this exploration in this subsection.

Motivated by processes that affect the CO core mass (\Secref{sec:intro}) we
calculate merging {\bhbh} populations with $\COshift = -20\,\solarmass$ to
$\COshift =
+10\,\solarmass$. \Figref{fig:primary-mass_distribution-at-redshift-zero-for-core-mass-shift-variation}
shows that the CO core-mass shift strongly affects the location of the {\ppisn}
pile-up in the primary-mass distribution. Moreover, the (relative) magnitude of
this pile-up varies with different CO core-mass shifts.

The location and the magnitude of the {\overdensity} in our {\downwardvariation}
model matches the observed peak at $\sim35\solarmass$. In many of processes
mentioned in \Secref{sec:intro}, however, this downward shift is too large to
explain.
The beyond Standard-Model process of axion formation, however, could lead to an
effective downward shift of as much as $15\,\solarmass$ in the model of
\citet{moriLightCurvesEvent2023} where the axion mass is about half the electron
mass. They find that supernovae from these axion-induced instabilities are
similar to standard pair-formation induced supernovae, i.e. the nickel ejecta
distribution has the same overall extent and shape. Their light curves, however,
have a shorter rise-to-peak time due to the lower total mass of star, which
could differentiate between models, but also means that they do not display the
standard long rise-to-peak characteristics used to identify {\pisnei}, making it
harder to identify them as {\pisnei} in {\slsni} observations.

Several of the processes in \Secref{sec:intro} lead to an upward shift of the
CO core-mass range of stars that undergo (P){\pisne}, but we specifically
highlight the more accurate and up-to-date reaction rates and stellar models of
\citet{faragResolvingPeakBlack2022}, and model this with our {\upwardvariation}
model.
This model results in an {\overdensity} in the primary-BH mass distribution at
$\sim{}64\,\solarmass$, suggesting that a third peak in the primary-mass
distribution exists. We find that the magnitude of the peak is less pronounced,
being only slightly higher
($0.2\,\times\,10^{-2}\eventratedensity\solarmass^{-1}$) than the merger rate at
primary masses slightly lower than the location of the peak
($58-62\,\solarmass$).
We expect that the magnitude of this peak relative to the rate at masses
slightly lower than the peak, at least in part, depends on the maximum mass of
stars that we take into account in our simulations. While our upward CO
core-mass shift leads to a larger region of pre-SN core masses that undergo
{\ppisn}, if the initial masses of our stars are insufficiently massive to
populate the entire range of pre-SN CO core masses, it lowers the rate of BH
formation with masses in the expected {\ppisne}-remnant mass range. In the case
that no star is massive enough to undergo {\ppisn}, no pile-up or {\overdensity}
is formed at all.
The range of CO core-masses that undergo {\ppisn} is also a factor that
determines the magnitude of the peak. If the range is narrow, fewer stars are in
that CO core-mass range, which effectively lowers the rate of stars that undergo
{\ppisn} and form BHs in the {\ppisn}-remnant mass range. The narrower CO
core-mass range is the result of a higher sensitivity of the {\ppisn} mass loss
to the CO core-mass. Examples of this are the
$\sigma\left[\carbonoxygen\right]=-3$ models of
\citet{faragResolvingPeakBlack2022} or the strongly coupled (high $\epsilon$)
hidden-photon models of \citet{croonMissingActionNew2020}. 

We leave the exploration of the sensitivity of the peak at $\sim 64\,\solarmass$
to the maximum considered initial primary mass and the sensitivity of the
{\ppisn} mass loss to the CO core-mass for a future study.
The {\ofourrun} observation run of the LVK-collaboration
\citep{abbottProspectsObservingLocalizing2020} probes a five times larger space
than O3 and is expected to uncover more structure in the high mass range. Thus,
this peak may already be observed in {\ofourrun}. The exact location and
magnitude of this new peak may inform us about the {\ppisne} mechanism and how
massive stars that undergo {\ppisne} are.

Additionally, we calculate merging {\bhbh} populations varying the additional
mass loss from $\extraML = +5\,\solarmass$ to $\extraML = +20\,\solarmass$.
\Figref{fig:primary-mass_distribution-at-redshift-zero-for-extra-massloss-variation}
shows that additional mass loss lowers the merger rate and moves both
$\mppisncutoff$ and the {\overdensity} caused by {\ppisne} to lower masses, but
up to $\extraML = +10\,\solarmass$. This is because removing more mass results
in primary masses that are created by {\ccsne} instead of {\ppisne}, and the BHs
that are formed through {\ppisn} lose so much mass that they end up as the
secondary BH.
The \citetalias{farmerMindGapLocation2019} model is similar to our
$\extraML = +5\,\solarmass$ and $\extraML = +10\,\solarmass$ models, although it
does not have the peak in the primary-mass distribution we find at
$\sim{}42\,\solarmass$ in the $\extraML = +10\,\solarmass$. This indicates that
a {\ppisn} mass loss prescription that has an artificial discontinuity at the
CC-{\ppisn} interface has the same qualitative effect as extra mass
removal.
While there are studies, both theoretical \citep{powellFinalCoreCollapse2021,
  rahmanPulsationalPairinstabilitySupernovae2022} and observational
\citep{ben-amiSN2010mbDirect2014, kuncarayaktiBactrianBroadlinedTypeIc2023,
  linSuperluminousSupernovaLightened2023}, that indicate additional
post-{\ppisn} mass loss, removing more than $10\,\solarmass$ over the entire
range of {\ppisne} seems hard to justify, and it makes no difference to the
primary-BH mass distribution.

\subsection{Transient rates}
\label{sec:transient-rates}


Our fiducial model agrees well with the observed {\ccsne} rate from
\citet{frohmaierCoreCollapseSuperluminous2021}. We produce roughly one {\pisni}
per $10 000$ {\ccsne} and one {\ppisni} per $1 000$ {\ccsne} when $z \leq
1$. While currently there are no unambiguous rate estimates from direct
observations of {\ppisne} or {\pisne}, there are estimates based on the
non-detection of these supernovae, e.g.,
\citet[][]{nichollSlowlyFadingSuperluminous2013}. Specifically, from light-curve
analysis and {\slsn} rates, taking into account that not all {\slsne} match
{\pisn} light-curves and that not all {\pisne} are {\slsne},
\citet{nichollSlowlyFadingSuperluminous2013} concludes that the rate of {\pisne}
cannot exceed a fraction $6\,\times\,10^{-6}$ of the {\ccsn} rate of. This rate
contradicts fiducial results, because we find a ratio of {\pisnei} to {\ccsne}
of $9.30^{+1.03}_{-1.26}\,\times\,10^{-5}$ (\Tabref{tab:rates_table}).

We find that at $z=\slsnredshift$ our predicted {\ppisni} rate is approximately
equal to the {\slsni} rate of \citet{frohmaierCoreCollapseSuperluminous2021},
and that our {\pisni} rate is approximately an order of magnitude lower.
We caution, however, that our {\ppisni} and {\pisni} rates are not directly
comparable to observed {\slsni} rates.
It is clear from {\slsni} observations
\citep{nichollSlowlyFadingSuperluminous2013, ciaLightCurvesHydrogenpoor2018,
  gal-yamMostLuminousSupernovae2019} that only a small fraction of {\slsnei}
display characteristics that fit with {\pisne}, and from detailed models of
{\pisne} \citep{kasenPAIRINSTABILITYSUPERNOVAE2011,
  nichollSlowlyFadingSuperluminous2013,
  gilmerPairinstabilitySupernovaSimulations2017} it is understood that not all
{\pisne} are super luminous or necessarily show the characteristics that make
them stand out as {\pisne} in the {\slsnei} sample. Instead, some {\pisne} may
hide in a population of transients that fall between normal and {\slsne} called
Luminous Supernovae \citep{gomezLuminousSupernovaeUnveiling2022}.
The situation with {\ppisnei} is likely similar.
Events like \textit{SN2017egm} \citep{linSuperluminousSupernovaLightened2023},
\textit{iPTF16eh} \citep{lunnanUVResonanceLine2018}, \textit{PTF12dam}
\citep{tolstovPulsationalPairinstabilityModel2017} and \textit{SN 2019szu}
\citep{aamerPrecursorPlateauPreMaximum2023} are strong candidates for {\slsnei}
caused by {\ppisnei}. However, based on detailed models, not all {\ppisne} are
super-luminous \citep[e.g.,][]{woosleyPulsationalPairInstabilitySupernovae2017},
and the fact that \textit{SN 1961V} \citep{woosleySN1961VPulsational2022} and
\textit{iPTF14hls} \citep{wangIPTF14hlsCircumstellarMedium2022} possibly have
{\ppisn}-like light-curve morphologies but are not super-luminous supports this.
Because of the theoretical and observational uncertainties, we refrain here from
making quantitative estimates of the fraction of (P){\pisne} that appear as
super-luminous, and encourage further studies of both {\ppisn} and {\pisn} light
curves building on the pioneering work of
\cite{woosleyPulsationalPairInstabilitySupernovae2017,
  woosleyEvolutionMassiveHelium2019}.

Our fiducial intrinsic transient-rate density predictions at $z{}\sim 0$ for
{\ccsne} and {\ppisnei} agree well with the rates of \citet[][using the
{\compas} population-synthesis code]{stevensonImpactPairinstabilityMass2019}. We
predict about an order of magnitude more {\pisnei}, possibly due to considering
a higher maximum stellar
mass. 
The estimates from \citet[][using the {\bpass} population-synthesis
code]{brielEstimatingTransientRates2022} for {\ccsne} agree well with ours. They
estimate a {\pisni} rate density of $1-6\,\eventratedensity$ at $z{}\sim 0$,
which is $\sim5\,\eventratedensity$ lower than our fiducial rate density. They
do not provide rate estimates of {\ppisnei}. Previous results from
\citet[][\bpass]{eldridgeConsistentEstimateGravitational2019} show a similar
agreement for {\ccsne}, and match well for the {\pisnei}. Thus, we produce
similar {\ccsne} and {\pisne} rates to other studies.

We note that with the variations we introduce in this study, specifically our
$\COshift$ models, the fractions of (P){\pisne} that are {\slsnei}, and vice
versa, are not necessarily the same as in our fiducial models.
{\slsne} from {\pisne} are characterised by long rise-to-peak times due to large
masses, and long decay time-scales. \citet{moriLightCurvesEvent2023} finds that
axion instability supernovae, which we model with a downward CO core-mass shift,
behave qualitatively similarly to normal {\pisne}. The light curves of their
{\ppisne} have a slightly shorter rise-to-peak time, but the nickel-mass ejecta
and peak luminosities span the same ranges and are comparable to {\pisne}
without an additional CO core-mass shift. If the light curves and peak
luminosities behave similarly, the fraction of {\pisni} that are {\slsni} does
not change much, and thus, as our fiducial model, our {\downwardvariation} model
is likely in tension with the observations. It is unclear how {\ppisne},
specifically the fraction that display {\slsne}-like features, are affected by
either an upward or a downward CO core-mass shift, and studies similar to
\citet{moriLightCurvesEvent2023} are necessary to obtain further insight.

To provide a more quantitative exclusion or confirmation of our models, we need
observational data from current and upcoming telescopes like
{\jwst}~\citep[][]{hummelSOURCEDENSITYOBSERVABILITY2012},
{\lsst}~\citep[][]{villarSuperluminousSupernovaeLSST2018},
{\euclid}~\citep[][]{tanikawaEuclidDetectabilityPair2023} and {\ROMAN}
\citep[][]{moriyaDiscoveringSupernovaeEpoch2022} for better estimates of the
rate densities of {\ppisne} and {\pisne}, as well as more systematic modelling
of {\ppisne} and {\pisne} light curves, including variations on stellar
evolution and the {\ppisn} mechanism, to determine the fractions of these
transients that are super luminous.

\subsection{Modelling approach}
\label{sec:population-synthesis}
We use population synthesis to evolve populations with initial primary masses up
to $300\,\solarmass$ with the \binaryc\ framework. These masses go well beyond
the maximum mass of the detailed models on which \binaryc\ is based, and are
thus an extrapolation of the fitting formulae from
\citet{hurleyComprehensiveAnalyticFormulae2000, hurleyEvolutionBinaryStars2002},
which are themselves based on models of stars with initial masses
$\leq50\,M_\odot$ from \cite{polsStellarEvolutionModels1998}. Most of the stars
in our simulation that undergo {\ppisn} require initial masses in excess of
$100\,\solarmass$, and are affected by systematics in the extrapolation. Our
results are affected by the maximum initial primary mass we consider in that the
fraction of stars that remain massive enough to undergo {\ppisn}/{\pisn} is
changed. The presence of a pile-up in primary BH mass by {\ppisne}, and probably
the magnitude of this pile-up, depend on our considered maximum primary mass.
The magnitude of the shallow `peak' of primary BH masses in our
{\upwardvariation} model could increase by considering a larger maximum mass.
This, in turn, also increases the transient event-rate density of PISNe
\citep{tanikawaEuclidDetectabilityPair2023}.

We choose to use a binary fraction $f_{\mathrm{bin}} = 0.7$. Several studies
show that the binary fraction depends on initial primary mass,
\cite[e.g.,][]{moeMindYourPs2017, offnerOriginEvolutionMultiple2022}, and in
solar-mass stars it is anti-correlated with metallicity
\citep{moeCloseBinaryFraction2019,
  thieleApplyingMetallicitydependentBinary2023}. Because we are interested in
objects formed in massive-star systems only, we assume that choosing a
mass-dependent binary fraction is currently unnecessary, as most (if not all)
massive stars come in binaries or higher-order systems
\citep{sanaBinaryInteractionDominates2012} or higher multiplicity systems
\citep{offnerOriginEvolutionMultiple2022}.
Moreover, we assume that the distributions of birth parameters of our binary
systems are separable and independent. \citet{moeMindYourPs2017} and
\citet{offnerOriginEvolutionMultiple2022} show that this is not the
case. \citet{klenckiImpactIntercorrelatedInitial2018} find, however, that this
assumption does not strongly affect the rate estimates, although it does skew
the birth mass-ratio distribution of merging {\bhbh}s to lower mass ratios.

In this study we use the prescription for {\ppisn} mass loss of
\citet{renzoPairinstabilityMassLoss2022}, which is based on the detailed stellar
models of \citetalias{farmerMindGapLocation2019}. Unlike most other existing
prescriptions for {\ppisn} mass loss, this provides the mass lost in pulses due
to the {\ppisn}, rather than a remnant mass, for a given CO core mass, which
allows for a natural transition at the {\ccsne}/{\ppisne} boundary. Whether
there really is no discontinuity at the interface is unclear
\citep{renzoSensitivityLowerEdge2020}, but a prescription that artificially
introduces discontinuities should be avoided.
Taking the top-down approach from \citet{renzoPairinstabilityMassLoss2022} makes
the final remnant-mass prediction sensitive to the mass of the He layer that
lies above the CO core. The pre-SN evolution of the star, specifically the
evolution of the mass of the He layer for a given final CO core mass, affects
the final remnant mass. Several processes influence the ratio of the He to CO
core mass, like convective overshooting \citep{tanikawaPopulationIIIBinary2021,
  vinkMaximumBlackHole2021}, or wind mass loss
\citep{renzoSystematicSurveyEffects2017, woosleyEvolutionMassiveHelium2019}, or
binary interactions \citep{laplaceDifferentCorePresupernova2021}. We find a
near-constant ratio, $M_{\mathrm{He\ core}} / M_{\mathrm{CO\ core}} = 1.3$, in
all our stars that undergo {\ppisne}.

\subsection{Can the peak in the primary-BH mass distribution at $35\,\solarmass$
  be explained by {\ppisne}?}
\label{sec:can-peak-at}
We make use of observations of both GW mergers and EM transient events to
constrain our models and to answer whether the peak in the primary-BH mass
distribution at $35\,\solarmass$ can be explained by {\ppisne}.
We find that the CO core-mass range for stars to undergo {\ppisne} must shift
down by more than $10\,\solarmass$ to line up the feature from our PPISNe to the
observed feature in the primary-mass distribution at $35\,\solarmass$. This
downward shift contradicts recent results \citep{faragResolvingPeakBlack2022}
which suggest an upward shift of about $10\,\solarmass$.
Given that the {\pisne} rate in our fiducial model is already too high according
to \citet{nichollSlowlyFadingSuperluminous2013} and that
\citet{moriLightCurvesEvent2023} indicates that the light curves of our
{\downwardvariation} model behave similarly to {\ppisne} without an additional
CO core-mass range shift, we find it likely that the downward shift variation
that is required to match the GW observations is in tension with the observed
(rate of) {\slsnei}.

Our {\ppisn}-prescription variations that behave qualitatively like more recent
detailed models of (P){\pisne} ($\COshift=+10\,\solarmass$) predict a at peak
between $58-64\,\solarmass$. The transient rates associated with this variation
relieve some of the tension with observations, given that only some {\slsni} are
{\pisni}, although we still overproduce {\pisni} compared to
\citet{nichollSlowlyFadingSuperluminous2013}.
Our models therefore suggest that the $58-64\,\solarmass$ region is a promising
mass range in which to search for a new {\overdensity} of primary BH masses, and
may well be observable in the next observation runs of the LVK collaboration.

We regard a combination of a downward $\COshift$ variation and $\extraML$ as an
unlikely explanation to the peak at $35\,\solarmass$. While additional mass loss
does shift the peak to a lower mass, and an additional $\COshift$ of e.g.
$\sim 5\,\solarmass$ may create an {\overdensity} at $35\,\solarmass$, it would
still be in tension with the {\slsne} rate according to
\citet{nichollSlowlyFadingSuperluminous2013} because our fiducial model is
already in tension with that rate and any CO core-mass shift would increase this
tension.

Current and upcoming transient surveys like {\euclid}, {\jwst}, {\lsst} and
{\ROMAN} will measure increased rates of {\slsne}, {\pisne} and {\ppisne} over a
large range of redshifts. While we cannot definitively rule out downward
variation of {\downwardvariation} based on the current observations, these
surveys will provide the observational data to confirm or reject our
transient-rate estimates, and will statistically constrain the fraction of
{\slsnei} is associated with (P){\pisnei}.

Thus, given the results of our study, the fact that transient event-rate
observations indicate a likely tension with the rates of our models that lead to
a matching peak, we find it unlikely that the observed peak is due to {\ppisne}.

\subsection{If the peak at $35\,\solarmass$ is not from {\ppisne}, then what
  causes it?}
Broadly speaking, the origin of features in the primary-mass distribution are
expected to either I) mainly reflect the remnant mass distribution, or II) they
mainly reflect (binary) evolutionary selection effects that are caused by their
formation channel.
If the feature is caused by {\ppisne}, then this would fall under the first
category \citep[e.g.][for the lower-mass
analogue]{schneiderBimodalBlackHole2023, disbergFailedSupernovaeNatural2023}.
However, it is equally likely for such a feature to arise from evolutionary
effects.

A handful of studies have tried to explain the $35\,\solarmass$ peak through
causes other than {\ppisne} and the remnant mass distribution.
For example, \citet{antoniniCoalescingBlackHole2023} suggests that the
$35\,\solarmass$ peak can be explained by cluster dynamics. They find that
dynamical interactions in globular clusters lead to features in the primary-mass
distribution around $\sim{}35\,\solarmass$, as long as massive clusters form
with a half-mass density $>10^{4}\solarmass\,\mathrm{pc}^{-3}$. This is not
populated by hierarchical mergers, but does depend on the dynamical pairing of
black holes.
Alternatively, \citet{brielUnderstandingHighmassBinary2023} suggests that
isolated binary interactions are the cause of the $35\,\solarmass$ peak. They
find that the peak is not caused by pair-instability remnants, but
rather systems that undergo only stable mass transfer, possibly multiple
times. They find that a combination of mass-transfer stability that limits the
low-end of the mass range of primary-mass BHs at $35\,\solarmass$, and
quasi-homogeneous evolution limiting the upper end, leads to an {\overdensity} at
$\sim{}35\,\solarmass$.

No alternative explanation for the peak at $35\,\solarmass$ has yet been adopted
as the solution, and further research is needed to determine the correct
channel. It may not be enough to just find an {\overdensity} at $35\,\solarmass$,
and matching other properties of systems around this mass, like mass-ratio
\citep[e.g.][]{liDivergenceMassRatio2022} and spin-orbit alignment, may be
critical in finding the actual cause of the observed peak.


\section{Conclusions}
\label{sec:conclusion}
We implement a top-down pulsational pair-instability supernova mass-loss
algorithm in the binary population-synthesis code {\binaryc} and use this to
predict the merger rate and mass distribution of {\bhbh}s merging at redshift
zero.
We explore several physically motivated variations to our {\ppisn} prescription,
and study how each variation affects the mass distribution of primary masses of
merging {\bhbh}s, with a focus on the location of a peak at high BH masses. We
combine our GW- and EM-transient predictions to study {\ppisne} and {\pisne}
phenomena, and we compare these to recent observations to constrain our model
variations.

Below we list our most notable results.
\begin{enumerate}
\item Our fiducial model has no peak in the primary-mass distribution that
  matches the observed feature at $35\,\solarmass$.
\item Our CO core-mass shift variations strongly affect the location of the
  {\ppisn} pile-up such that shifting the CO core-mass range with
  {\downwardvariation} does match the location of the observed {\overdensity} in
  the primary-mass distribution. It is hard to explain this with conventional
  physics like rotation or variations in nuclear reaction rates. The upward
  shift of {\upwardvariation}, which is based on detailed models of {\ppisne}
  \citep{faragResolvingPeakBlack2022}, moves the {\overdensity} upward in
  primary BH mass by about $8-14\,\solarmass$, predicting a (slight)
  {\overdensity} at $58-64\,\solarmass$. The current LVK {\ofourrun} observation
  run will detect {\bhbh} systems more efficiently than before and could shed
  light on whether this third peak exists.
\item Our additional mass-loss variations affect the location of the
  {\overdensity} of BHs in the primary-mass distribution by about
  $5-10\,\solarmass$. Removing more mass, however, does not lead to an
  {\overdensity} at lower masses because, at lower mass, the majority of BHs
  with those primary-masses are created through {\ccsne}.
\item The transient-rate estimates of {\ccsn} in our fiducial model match well
  with the inferred rate of
  \citet{frohmaierCoreCollapseSuperluminous2021}. Their rate for {\slsni} falls
  between both our predicted {\ppisni} and {\pisni} rates. We predict a {\ppisn}
  rate $\sim3$ times higher, and a {\pisn} rate $\sim3$ times lower. Our ratio
  of {\pisni} to {\ccsni} exceeds the estimate of
  \citet{nichollSlowlyFadingSuperluminous2013}, however, indicating that our
  fiducial model disagrees with the {\slsni} rate.
\item With the {\ppisn}-prescription variation that does produce a peak at the
  correct location ({\downwardvariation}), we find that the {\ppisni} rates
  exceed the {\slsn} rates by a factor of 16, and the {\pisni} rates are almost
  double that of the {\slsni} rates. Even taking into account that not all
  (P){\pisne} produce {\slsne}, and not all {\slsne} can be explained by
  (P){\pisne}, these rates likely are in tension with the observed {\slsn} rates
  as well.
\end{enumerate}

\noindent In summary, because the large downward shift in CO core mass required
to fit the observed GW peak is difficult to explain without exotic physics
beyond the Standard Model, and new reaction rate studies even suggest an upward
shift to $58-64\,\solarmass$, and because the transient event-rates of {\ppisne}
and {\pisne} for this variation are likely in tension with the observed {\slsne}
rate, we conclude that {\ppisne} are unlikely to be responsible for the peak
feature observed at $32 \mhyphen 38\,\solarmass$.

\section*{Acknowledgements}
\label{sec:acknowledgements}
DDH wants to thank Arman Aryaeipour, Max Briel, Payel Das, Will Farr, Giovanni
Mirouh, Bob Nichol, Natalie Rees, Karel Temmink, Rob Yates for the useful
discussions, Paula Gherghinescu and Madison Walder for their artistic advice,
and the UKRI/UoS for the funding grant H120341A.
LvS acknowledges partial financial support from the National Science Foundation
under Grant No. (NSF grant number 2009131), the Netherlands Organisation for
Scientific Research (NWO) as part of the Vidi research program BinWaves with
project number 639.042.728 and the European Union’s Horizon 2020 research and
innovation program from the European Research Council (ERC, Grant agreement
No. 715063).
RGI thanks the STFC for the funding grants
\href{https://gtr.ukri.org/projects?ref=ST%2FR000603%2F1}{ST/R000603/1} and
\href{https://gtr.ukri.org/projects?ref=ST/L003910/2}{ST/L003910/2}, and the
BRIDGCE consortium.
The authors thank Selma de Mink for providing a platform for collaboration and
communication, and for long term scientific guidance. Moreover, we thank the
anonymous reviewer for the useful feedback on the manuscript.

In this research we make use of the GWTC-3 data release provided by the LIGO,
VIRGO and KAGRA collaborations
\citep{ligo_scientific_collaboration_and_virgo_2021_5655785}. Moreover, we make
use of the following software to enable this study: The cosmology module of
\textsc{Astropy} \citep{theastropycollaborationAstropyProjectSustaining2022},
\textsc{asymmetric\_uncertainty} \citep{gobatAsymmetricUncertaintyHandling2022},
the star-formation rate prescriptions of {\compas}
\citep{rileyRapidStellarBinary2022} \textsc{h5py}
\citep{collettePythonHDF52013}, \textsc{Ipython/Jupyter}
\citep{perezIPythonSystemInteractive2007,
  kluyverJupyterNotebooksPublishing2016}, \textsc{Matplotlib}
\citep{hunterMatplotlib2DGraphics2007}, \textsc{Numpy}
\citep{harrisArrayProgrammingNumPy2020}, \textsc{numpy-indexed}
\citep{hoogendoornEelcoHoogendoornNumpyArraysetops2023}, \textsc{pandas}
\citep{wesmckinneyDataStructuresStatistical2010,
  thepandasdevelopmentteamPandasdevPandasPandas2020}, \textsc{PyCBC}
\citep{nitzGwastroPycbcV22023}, \textsc{PyPDF2}
\citep{fenniakPyPDF2Library2022}, \textsc{PyTables}
\citep{pytablesdevelopersteamPyTablesHierarchicalDatasets2002}, \textsc{Python}
\citep{vanrossumPythonReferenceManual2009} and \textsc{Scipy}
\citep{virtanenSciPyFundamentalAlgorithms2020}.

\section*{Data Availability}
\label{sec:data-availability}
We will make the DCO and EM transient data used in this study available on
\href{https://doi.org/10.5281/zenodo.8083112}{10.5281/zenodo.8083112} upon
publication, along with routines to generate these data and the figures
presented in this paper. The data is generated with
\href{https://gitlab.com/binary_c/binary_c/-/tree/david/versions/2.2.2?ref_type=heads}{a
  modified version} of \textsc{binary\_c} v2.2.2 and
\href{https://gitlab.com/binary_c/binary_c-python/-/tree/development_0.9.5/2.2.2_david?ref_type=heads}{a
  modified version} of \textsc{binary\_c-python} v0.9.5/2.2.2.

\bibliographystyle{mnras}
\bibliography{PPISN_paper.bib} 

\begin{thebibliography}{}
\makeatletter
\relax
\def\mn@urlcharsother{\let\do\@makeother \do\$\do\&\do\#\do\^\do\_\do\%\do\~}
\def\mn@doi{\begingroup\mn@urlcharsother \@ifnextchar [ {\mn@doi@}
  {\mn@doi@[]}}
\def\mn@doi@[#1]#2{\def\@tempa{#1}\ifx\@tempa\@empty \href
  {http://dx.doi.org/#2} {doi:#2}\else \href {http://dx.doi.org/#2} {#1}\fi
  \endgroup}
\def\mn@eprint#1#2{\mn@eprint@#1:#2::\@nil}
\def\mn@eprint@arXiv#1{\href {http://arxiv.org/abs/#1} {{\tt arXiv:#1}}}
\def\mn@eprint@dblp#1{\href {http://dblp.uni-trier.de/rec/bibtex/#1.xml}
  {dblp:#1}}
\def\mn@eprint@#1:#2:#3:#4\@nil{\def\@tempa {#1}\def\@tempb {#2}\def\@tempc
  {#3}\ifx \@tempc \@empty \let \@tempc \@tempb \let \@tempb \@tempa \fi \ifx
  \@tempb \@empty \def\@tempb {arXiv}\fi \@ifundefined
  {mn@eprint@\@tempb}{\@tempb:\@tempc}{\expandafter \expandafter \csname
  mn@eprint@\@tempb\endcsname \expandafter{\@tempc}}}

\bibitem[\protect\citeauthoryear{Aamer et~al.,}{Aamer
  et~al.}{2023}]{aamerPrecursorPlateauPreMaximum2023}
Aamer A.,  et~al., 2023, A {{Precursor Plateau}} and {{Pre-Maximum}} [{{O II}}]
  {{Emission}} in the {{Superluminous SN2019szu}}: {{A Pulsational
  Pair-Instability Candidate}}, \mn@doi{10.48550/arXiv.2307.02487}, \url
  {https://ui.adsabs.harvard.edu/abs/2023arXiv230702487A}

\bibitem[\protect\citeauthoryear{Abbott et~al.,}{Abbott
  et~al.}{2020}]{abbottProspectsObservingLocalizing2020}
Abbott B.~P.,  et~al., 2020, \mn@doi [Living Reviews in Relativity]
  {10.1007/s41114-020-00026-9}, 23, 3

\bibitem[\protect\citeauthoryear{Abbott et~al.,}{Abbott
  et~al.}{2021a}]{abbottGWTC2CompactBinary2021}
Abbott R.,  et~al., 2021a, \mn@doi [Physical Review X]
  {10.1103/PhysRevX.11.021053}, 11, 021053

\bibitem[\protect\citeauthoryear{Abbott et~al.,}{Abbott
  et~al.}{2021b}]{abbottPopulationPropertiesCompact2021}
Abbott R.,  et~al., 2021b, \mn@doi [The Astrophysical Journal Letters]
  {10.3847/2041-8213/abe949}, 913, L7

\bibitem[\protect\citeauthoryear{Abbott et~al.,}{Abbott
  et~al.}{2023}]{abbottPopulationMergingCompact2023}
Abbott R.,  et~al., 2023, \mn@doi [Physical Review X]
  {10.1103/PhysRevX.13.011048}, 13, 011048

\bibitem[\protect\citeauthoryear{Ade et~al.,}{Ade
  et~al.}{2014}]{adePlanck2013Results2014}
Ade P. a.~R.,  et~al., 2014, \mn@doi [Astronomy \& Astrophysics]
  {10.1051/0004-6361/201321591}, 571, A16

\bibitem[\protect\citeauthoryear{Aksenov \& Chechetkin}{Aksenov \&
  Chechetkin}{2016}]{aksenovNeutronizationMatterStellar2016}
Aksenov A.~G.,  Chechetkin V.~M.,  2016, \mn@doi [Astronomy Reports]
  {10.1134/S1063772916070015}, 60, 655

\bibitem[\protect\citeauthoryear{Antonini, Gieles, Dosopoulou  \&
  Chattopadhyay}{Antonini et~al.}{2023}]{antoniniCoalescingBlackHole2023}
Antonini F.,  Gieles M.,  Dosopoulou F.,   Chattopadhyay D.,  2023, \mn@doi
  [Monthly Notices of the Royal Astronomical Society] {10.1093/mnras/stad972},
  522, 466

\bibitem[\protect\citeauthoryear{Arca~Sedda, Mapelli, Spera, Benacquista  \&
  Giacobbo}{Arca~Sedda et~al.}{2020}]{arcaseddaFingerprintsBinaryBlack2020}
Arca~Sedda M.,  Mapelli M.,  Spera M.,  Benacquista M.,   Giacobbo N.,  2020,
  \mn@doi [The Astrophysical Journal] {10.3847/1538-4357/ab88b2}, 894, 133

\bibitem[\protect\citeauthoryear{Atri et~al.,}{Atri
  et~al.}{2019}]{atriPotentialKickVelocity2019a}
Atri P.,  et~al., 2019, \mn@doi [Monthly Notices of the Royal Astronomical
  Society] {10.1093/mnras/stz2335}, 489, 3116

\bibitem[\protect\citeauthoryear{Baibhav, Gerosa, Berti, Wong, Helfer  \&
  Mould}{Baibhav et~al.}{2020}]{baibhavMassGapSpin2020}
Baibhav V.,  Gerosa D.,  Berti E.,  Wong K. W.~K.,  Helfer T.,   Mould M.,
  2020, \mn@doi [Physical Review D] {10.1103/PhysRevD.102.043002}, 102, 043002

\bibitem[\protect\citeauthoryear{Barkat, Rakavy  \& Sack}{Barkat
  et~al.}{1967}]{barkatDynamicsSupernovaExplosion1967}
Barkat Z.,  Rakavy G.,   Sack N.,  1967, \mn@doi [Physical Review Letters]
  {10.1103/PhysRevLett.18.379}, 18, 379

\bibitem[\protect\citeauthoryear{Beasor \& Smith}{Beasor \&
  Smith}{2022}]{beasorExtremeScarcityDustenshrouded2022}
Beasor E.~R.,  Smith N.,  2022, \mn@doi [The Astrophysical Journal]
  {10.3847/1538-4357/ac6dcf}, 933, 41

\bibitem[\protect\citeauthoryear{Belczynski, Repetto, Holz, O'Shaughnessy,
  Bulik, Berti, Fryer  \& Dominik}{Belczynski
  et~al.}{2016}]{belczynskiCOMPACTBINARYMERGER2016}
Belczynski K.,  Repetto S.,  Holz D.~E.,  O'Shaughnessy R.,  Bulik T.,  Berti
  E.,  Fryer C.,   Dominik M.,  2016, \mn@doi [The Astrophysical Journal]
  {10.3847/0004-637X/819/2/108}, 819, 108

\bibitem[\protect\citeauthoryear{Belczynski et~al.,}{Belczynski
  et~al.}{2020}]{belczynskiEvolutionaryRoadsLeading2020}
Belczynski K.,  et~al., 2020, \mn@doi [Astronomy \& Astrophysics]
  {10.1051/0004-6361/201936528}, 636, A104

\bibitem[\protect\citeauthoryear{Belczynski, Doctor, Zevin, Olejak, Banerje  \&
  Chattopadhyay}{Belczynski et~al.}{2022}]{belczynskiBlackHoleBlack2022}
Belczynski K.,  Doctor Z.,  Zevin M.,  Olejak A.,  Banerje S.,   Chattopadhyay
  D.,  2022, \mn@doi [The Astrophysical Journal] {10.3847/1538-4357/ac8167},
  935, 126

\bibitem[\protect\citeauthoryear{{Ben-Ami} et~al.,}{{Ben-Ami}
  et~al.}{2014}]{ben-amiSN2010mbDirect2014}
{Ben-Ami} S.,  et~al., 2014, \mn@doi [The Astrophysical Journal]
  {10.1088/0004-637X/785/1/37}, 785, 37

\bibitem[\protect\citeauthoryear{Blaauw}{Blaauw}{1961}]{blaauwOriginBtypeStars1961}
Blaauw A.,  1961, Bulletin of the Astronomical Institutes of the Netherlands,
  \href {https://ui.adsabs.harvard.edu/abs/1961BAN....15..265B} {15, 265}

\bibitem[\protect\citeauthoryear{Bond, Arnett  \& Carr}{Bond
  et~al.}{1984}]{bondEvolutionFateVery1984}
Bond J.~R.,  Arnett W.~D.,   Carr B.~J.,  1984, \mn@doi [The Astrophysical
  Journal] {10.1086/162057}, 280, 825

\bibitem[\protect\citeauthoryear{Briel, Eldridge, Stanway, Stevance  \&
  Chrimes}{Briel et~al.}{2022}]{brielEstimatingTransientRates2022}
Briel M.~M.,  Eldridge J.~J.,  Stanway E.~R.,  Stevance H.~F.,   Chrimes A.~A.,
   2022, \mn@doi [Monthly Notices of the Royal Astronomical Society]
  {10.1093/mnras/stac1100}, 514, 1315

\bibitem[\protect\citeauthoryear{Briel, Stevance  \& Eldridge}{Briel
  et~al.}{2023}]{brielUnderstandingHighmassBinary2023}
Briel M.~M.,  Stevance H.~F.,   Eldridge J.~J.,  2023, \mn@doi [Monthly Notices
  of the Royal Astronomical Society] {10.1093/mnras/stad399}, 520, 5724

\bibitem[\protect\citeauthoryear{Broekgaarden et~al.,}{Broekgaarden
  et~al.}{2021}]{broekgaardenImpactMassiveBinary2021}
Broekgaarden F.~S.,  et~al., 2021, \mn@doi [Monthly Notices of the Royal
  Astronomical Society] {10.1093/mnras/stab2716}, 508, 5028

\bibitem[\protect\citeauthoryear{Broekgaarden et~al.,}{Broekgaarden
  et~al.}{2022}]{broekgaardenImpactMassiveBinary2022}
Broekgaarden F.~S.,  et~al., 2022, \mn@doi [Monthly Notices of the Royal
  Astronomical Society] {10.1093/mnras/stac1677}, 516, 5737

\bibitem[\protect\citeauthoryear{Callister \& Farr}{Callister \&
  Farr}{2023}]{callisterParameterFreeTourBinary2023}
Callister T.~A.,  Farr W.~M.,  2023, A {{Parameter-Free Tour}} of the {{Binary
  Black Hole Population}}, \mn@doi{10.48550/arXiv.2302.07289}, \url
  {https://ui.adsabs.harvard.edu/abs/2023arXiv230207289C}

\bibitem[\protect\citeauthoryear{Callister, Farr  \& Renzo}{Callister
  et~al.}{2021}]{callisterStateFieldBinary2021}
Callister T.~A.,  Farr W.~M.,   Renzo M.,  2021, \mn@doi [The Astrophysical
  Journal] {10.3847/1538-4357/ac1347}, 920, 157

\bibitem[\protect\citeauthoryear{Chan, M{\"u}ller  \& Heger}{Chan
  et~al.}{2020}]{chanImpactFallbackCompact2020}
Chan C.,  M{\"u}ller B.,   Heger A.,  2020, \mn@doi [Monthly Notices of the
  Royal Astronomical Society] {10.1093/mnras/staa1431}, 495, 3751

\bibitem[\protect\citeauthoryear{Chatzopoulos \& Wheeler}{Chatzopoulos \&
  Wheeler}{2012}]{chatzopoulosEffectsRotationMinimum2012}
Chatzopoulos E.,  Wheeler J.~C.,  2012, \mn@doi [The Astrophysical Journal]
  {10.1088/0004-637X/748/1/42}, 748, 42

\bibitem[\protect\citeauthoryear{Chen, Woosley, Heger, Almgren  \& Whalen}{Chen
  et~al.}{2014}]{chenTWODIMENSIONALSIMULATIONSPULSATIONAL2014}
Chen K.-J.,  Woosley S.,  Heger A.,  Almgren A.,   Whalen D.~J.,  2014, \mn@doi
  [The Astrophysical Journal] {10.1088/0004-637X/792/1/28}, 792, 28

\bibitem[\protect\citeauthoryear{Chen, Woosley  \& Whalen}{Chen
  et~al.}{2020}]{chenGasDynamicsNickel562020}
Chen K.-J.,  Woosley S.~E.,   Whalen D.~J.,  2020, \mn@doi [The Astrophysical
  Journal] {10.3847/1538-4357/ab9819}, 897, 152

\bibitem[\protect\citeauthoryear{Chen, Whalen, Zhang  \& Woosley}{Chen
  et~al.}{2022}]{chenRadiationHydrodynamicalSimulationsPulsational2022}
Chen K.-J.,  Whalen D.~J.,  Zhang W.,   Woosley S.~E.,  2022,
  Radiation-{{Hydrodynamical Simulations}} of {{Pulsational Pair-Instability
  Supernovae}} (\mn@eprint {arxiv} {1904.12873}),
  \mn@doi{10.48550/arXiv.1904.12873}, \url {http://arxiv.org/abs/1904.12873}

\bibitem[\protect\citeauthoryear{Chruslinska, Belczynski, Klencki  \&
  Benacquista}{Chruslinska et~al.}{2018}]{chruslinskaDoubleNeutronStars2018}
Chruslinska M.,  Belczynski K.,  Klencki J.,   Benacquista M.,  2018, \mn@doi
  [Monthly Notices of the Royal Astronomical Society] {10.1093/mnras/stx2923},
  474, 2937

\bibitem[\protect\citeauthoryear{Chruslinska, Nelemans  \&
  Belczynski}{Chruslinska
  et~al.}{2019}]{chruslinskaInfluenceDistributionCosmic2019}
Chruslinska M.,  Nelemans G.,   Belczynski K.,  2019, \mn@doi [Monthly Notices
  of the Royal Astronomical Society] {10.1093/mnras/sty3087}, 482, 5012

\bibitem[\protect\citeauthoryear{Cia et~al.,}{Cia
  et~al.}{2018}]{ciaLightCurvesHydrogenpoor2018}
Cia A.~D.,  et~al., 2018, \mn@doi [The Astrophysical Journal]
  {10.3847/1538-4357/aab9b6}, 860, 100

\bibitem[\protect\citeauthoryear{Claeys, Pols, Izzard, Vink  \& Verbunt}{Claeys
  et~al.}{2014}]{claeysTheoreticalUncertaintiesType2014}
Claeys J. S.~W.,  Pols O.~R.,  Izzard R.~G.,  Vink J.,   Verbunt F. W.~M.,
  2014, \mn@doi [Astronomy \& Astrophysics] {10.1051/0004-6361/201322714}, 563,
  A83

\bibitem[\protect\citeauthoryear{Collette}{Collette}{2013}]{collettePythonHDF52013}
Collette A.,  2013, Python and {{HDF5}}.
{O'Reilly}

\bibitem[\protect\citeauthoryear{Cooke et~al.,}{Cooke
  et~al.}{2012}]{cookeSuperluminousSupernovaeRedshifts2012}
Cooke J.,  et~al., 2012, \mn@doi [Nature] {10.1038/nature11521}, 491, 228

\bibitem[\protect\citeauthoryear{Costa, Bressan, Mapelli, Marigo, Iorio  \&
  Spera}{Costa et~al.}{2021}]{costaFormationGW190521Stellar2021}
Costa G.,  Bressan A.,  Mapelli M.,  Marigo P.,  Iorio G.,   Spera M.,  2021,
  \mn@doi [Monthly Notices of the Royal Astronomical Society]
  {10.1093/mnras/staa3916}, 501, 4514

\bibitem[\protect\citeauthoryear{Croon, McDermott  \& Sakstein}{Croon
  et~al.}{2020}]{croonMissingActionNew2020}
Croon D.,  McDermott S.~D.,   Sakstein J.,  2020, \mn@doi [Physical Review D]
  {10.1103/PhysRevD.102.115024}, 102, 115024

\bibitem[\protect\citeauthoryear{De~Marco \& Izzard}{De~Marco \&
  Izzard}{2017}]{demarcoImpactCompanionsStellar2017}
De~Marco O.,  Izzard R.~G.,  2017, \mn@doi [Publications of the Astronomical
  Society of Australia] {10.1017/pasa.2016.52}, 34

\bibitem[\protect\citeauthoryear{Disberg \& Nelemans}{Disberg \&
  Nelemans}{2023}]{disbergFailedSupernovaeNatural2023}
Disberg P.,  Nelemans G.,  2023, \mn@doi [Astronomy \& Astrophysics]
  {10.1051/0004-6361/202245693}, 676, A31

\bibitem[\protect\citeauthoryear{Dominik, Belczynski, Fryer, Holz, Berti,
  Bulik, Mandel  \& O'Shaughnessy}{Dominik
  et~al.}{2013}]{dominikDOUBLECOMPACTOBJECTS2013}
Dominik M.,  Belczynski K.,  Fryer C.,  Holz D.~E.,  Berti E.,  Bulik T.,
  Mandel I.,   O'Shaughnessy R.,  2013, \mn@doi [The Astrophysical Journal]
  {10.1088/0004-637X/779/1/72}, 779, 72

\bibitem[\protect\citeauthoryear{Dominik et~al.,}{Dominik
  et~al.}{2015}]{dominikDoubleCompactObjects2015}
Dominik M.,  et~al., 2015, \mn@doi [The Astrophysical Journal]
  {10.1088/0004-637X/806/2/263}, 806, 263

\bibitem[\protect\citeauthoryear{Dray, Dale, Beer, Napiwotzki  \& King}{Dray
  et~al.}{2005}]{drayWolfRayetStar2005}
Dray L.~M.,  Dale J.~E.,  Beer M.~E.,  Napiwotzki R.,   King A.~R.,  2005,
  \mn@doi [Monthly Notices of the Royal Astronomical Society]
  {10.1111/j.1365-2966.2005.09536.x}, 364, 59

\bibitem[\protect\citeauthoryear{Edelman, Doctor, Godfrey  \& Farr}{Edelman
  et~al.}{2022}]{edelmanAinNoMountain2022}
Edelman B.,  Doctor Z.,  Godfrey J.,   Farr B.,  2022, \mn@doi [The
  Astrophysical Journal] {10.3847/1538-4357/ac3667}, 924, 101

\bibitem[\protect\citeauthoryear{Eldridge, Stanway  \& Tang}{Eldridge
  et~al.}{2019}]{eldridgeConsistentEstimateGravitational2019}
Eldridge J.~J.,  Stanway E.~R.,   Tang P.~N.,  2019, \mn@doi [Monthly Notices
  of the Royal Astronomical Society] {10.1093/mnras/sty2714}, 482, 870

\bibitem[\protect\citeauthoryear{Farag, Renzo, Farmer, Chidester  \&
  Timmes}{Farag et~al.}{2022}]{faragResolvingPeakBlack2022}
Farag E.,  Renzo M.,  Farmer R.,  Chidester M.~T.,   Timmes F.~X.,  2022,
  \mn@doi [The Astrophysical Journal] {10.3847/1538-4357/ac8b83}, 937, 112

\bibitem[\protect\citeauthoryear{Farah, Edelman, Zevin, Fishbach,
  Mar{\'i}a~Ezquiaga, Farr  \& Holz}{Farah
  et~al.}{2023}]{farahThingsThatMight2023}
Farah A.~M.,  Edelman B.,  Zevin M.,  Fishbach M.,  Mar{\'i}a~Ezquiaga J.,
  Farr B.,   Holz D.~E.,  2023, Things That Might Go Bump in the Night:
  {{Assessing}} Structure in the Binary Black Hole Mass Spectrum,
  \mn@doi{10.48550/arXiv.2301.00834}, \url
  {https://ui.adsabs.harvard.edu/abs/2023arXiv230100834F}

\bibitem[\protect\citeauthoryear{Farmer, Renzo, {de Mink}, Marchant  \&
  Justham}{Farmer et~al.}{2019}]{farmerMindGapLocation2019}
Farmer R.,  Renzo M.,  {de Mink} S.~E.,  Marchant P.,   Justham S.,  2019,
  \mn@doi [The Astrophysical Journal] {10.3847/1538-4357/ab518b}, 887, 53

\bibitem[\protect\citeauthoryear{Farmer, Renzo, {de Mink}, Fishbach  \&
  Justham}{Farmer et~al.}{2020}]{farmerConstraintsGravitationalWave2020}
Farmer R.,  Renzo M.,  {de Mink} S.,  Fishbach M.,   Justham S.,  2020, \mn@doi
  [The Astrophysical Journal] {10.3847/2041-8213/abbadd}, 902, L36

\bibitem[\protect\citeauthoryear{Farr, Fishbach, Ye  \& Holz}{Farr
  et~al.}{2019}]{farrFuturePercentLevelMeasurement2019}
Farr W.~M.,  Fishbach M.,  Ye J.,   Holz D.,  2019, \mn@doi [The Astrophysical
  Journal] {10.3847/2041-8213/ab4284}, 883, L42

\bibitem[\protect\citeauthoryear{Fenniak, Stamy, {pubpub-zz}, Thoma, Peveler,
  {exiledkingcc}  \& {PyPDF2 Contributors}}{Fenniak
  et~al.}{2022}]{fenniakPyPDF2Library2022}
Fenniak M.,  Stamy M.,  {pubpub-zz} Thoma M.,  Peveler M.,  {exiledkingcc}
  {PyPDF2 Contributors} 2022, The {{PyPDF2}} Library, \url
  {https://pypi.org/project/PyPDF2/}

\bibitem[\protect\citeauthoryear{Fern{\'a}ndez, Quataert, Kashiyama  \&
  Coughlin}{Fern{\'a}ndez et~al.}{2018}]{fernandezMassEjectionFailed2018}
Fern{\'a}ndez R.,  Quataert E.,  Kashiyama K.,   Coughlin E.~R.,  2018, \mn@doi
  [Monthly Notices of the Royal Astronomical Society] {10.1093/mnras/sty306},
  476, 2366

\bibitem[\protect\citeauthoryear{Frohmaier et~al.,}{Frohmaier
  et~al.}{2021}]{frohmaierCoreCollapseSuperluminous2021}
Frohmaier C.,  et~al., 2021, \mn@doi [Monthly Notices of the Royal Astronomical
  Society] {10.1093/mnras/staa3607}, 500, 5142

\bibitem[\protect\citeauthoryear{Fryer}{Fryer}{2004}]{fryerNeutronStarKicks2004}
Fryer C.~L.,  2004, \mn@doi [The Astrophysical Journal] {10.1086/382044}, 601,
  L175

\bibitem[\protect\citeauthoryear{Fryer, Belczynski, Wiktorowicz, Dominik,
  Kalogera  \& Holz}{Fryer et~al.}{2012}]{fryerCompactRemnantMass2012}
Fryer C.~L.,  Belczynski K.,  Wiktorowicz G.,  Dominik M.,  Kalogera V.,   Holz
  D.~E.,  2012, \mn@doi [The Astrophysical Journal]
  {10.1088/0004-637X/749/1/91}, 749, 91

\bibitem[\protect\citeauthoryear{{Gal-Yam}}{{Gal-Yam}}{2019}]{gal-yamMostLuminousSupernovae2019}
{Gal-Yam} A.,  2019, \mn@doi [Annual Review of Astronomy and Astrophysics, vol.
  57, p.305-333] {10.1146/annurev-astro-081817-051819}, 57, 305

\bibitem[\protect\citeauthoryear{{Gallegos-Garcia}, Berry, Marchant  \&
  Kalogera}{{Gallegos-Garcia}
  et~al.}{2021}]{gallegos-garciaBinaryBlackHole2021}
{Gallegos-Garcia} M.,  Berry C. P.~L.,  Marchant P.,   Kalogera V.,  2021,
  \mn@doi [The Astrophysical Journal] {10.3847/1538-4357/ac2610}, 922, 110

\bibitem[\protect\citeauthoryear{Ge, Webbink, Chen  \& Han}{Ge
  et~al.}{2015}]{geAdiabaticMassLoss2015}
Ge H.,  Webbink R.~F.,  Chen X.,   Han Z.,  2015, \mn@doi [The Astrophysical
  Journal] {10.1088/0004-637X/812/1/40}, 812, 40

\bibitem[\protect\citeauthoryear{Ge, Webbink, Chen  \& Han}{Ge
  et~al.}{2020}]{geAdiabaticMassLoss2020}
Ge H.,  Webbink R.~F.,  Chen X.,   Han Z.,  2020, \mn@doi [The Astrophysical
  Journal] {10.3847/1538-4357/aba7b7}, 899, 132

\bibitem[\protect\citeauthoryear{Giacobbo \& Mapelli}{Giacobbo \&
  Mapelli}{2020}]{giacobboRevisingNatalKick2020}
Giacobbo N.,  Mapelli M.,  2020, \mn@doi [The Astrophysical Journal]
  {10.3847/1538-4357/ab7335}, 891, 141

\bibitem[\protect\citeauthoryear{Gilmer, Kozyreva, Hirschi, Fr{\"o}hlich  \&
  Yusof}{Gilmer et~al.}{2017}]{gilmerPairinstabilitySupernovaSimulations2017}
Gilmer M.~S.,  Kozyreva A.,  Hirschi R.,  Fr{\"o}hlich C.,   Yusof N.,  2017,
  \mn@doi [The Astrophysical Journal] {10.3847/1538-4357/aa8461}, 846, 100

\bibitem[\protect\citeauthoryear{Glatzel, Fricke  \& El~Eid}{Glatzel
  et~al.}{1985}]{glatzelFateRotatingPairunstable1985}
Glatzel W.,  Fricke K.~J.,   El~Eid M.~F.,  1985, Astronomy and Astrophysics,
  \href {https://ui.adsabs.harvard.edu/abs/1985A\&A...149..413G} {149, 413}

\bibitem[\protect\citeauthoryear{Gobat}{Gobat}{2022}]{gobatAsymmetricUncertaintyHandling2022}
Gobat C.,  2022, Asymmetric Uncertainty: {{Handling}} Nonstandard Numerical
  Uncertainties, Astrophysics Source Code Library, record ascl:2208.005
  (\mn@eprint {ascl} {2208.005})

\bibitem[\protect\citeauthoryear{Gomez et~al.,}{Gomez
  et~al.}{2019}]{gomezSN2016ietPulsational2019}
Gomez S.,  et~al., 2019, \mn@doi [The Astrophysical Journal]
  {10.3847/1538-4357/ab2f92}, 881, 87

\bibitem[\protect\citeauthoryear{Gomez, Berger, Nicholl, Blanchard  \&
  Hosseinzadeh}{Gomez et~al.}{2022}]{gomezLuminousSupernovaeUnveiling2022}
Gomez S.,  Berger E.,  Nicholl M.,  Blanchard P.~K.,   Hosseinzadeh G.,  2022,
  \mn@doi [The Astrophysical Journal] {10.3847/1538-4357/ac9842}, 941, 107

\bibitem[\protect\citeauthoryear{Grefenstette et~al.,}{Grefenstette
  et~al.}{2016}]{grefenstetteDISTRIBUTIONRADIOACTIVE44Ti2016}
Grefenstette B.~W.,  et~al., 2016, \mn@doi [The Astrophysical Journal]
  {10.3847/1538-4357/834/1/19}, 834, 19

\bibitem[\protect\citeauthoryear{Harris et~al.,}{Harris
  et~al.}{2020}]{harrisArrayProgrammingNumPy2020}
Harris C.~R.,  et~al., 2020, \mn@doi [Nature] {10.1038/s41586-020-2649-2}, 585,
  357

\bibitem[\protect\citeauthoryear{Hendriks \& Izzard}{Hendriks \&
  Izzard}{2023}]{hendriksBinaryCpythonPythonbased2023}
Hendriks D.~D.,  Izzard R.~G.,  2023, \mn@doi [Journal of Open Source Software]
  {10.21105/joss.04642}, 8, 4642

\bibitem[\protect\citeauthoryear{Hobbs, Lorimer, Lyne  \& Kramer}{Hobbs
  et~al.}{2005}]{hobbsStatisticalStudy2332005}
Hobbs G.,  Lorimer D.~R.,  Lyne A.~G.,   Kramer M.,  2005, \mn@doi [Monthly
  Notices of the Royal Astronomical Society]
  {10.1111/j.1365-2966.2005.09087.x}, 360, 974

\bibitem[\protect\citeauthoryear{{Holland-Ashford}, Lopez, Auchettl, Temim  \&
  {Ramirez-Ruiz}}{{Holland-Ashford}
  et~al.}{2017}]{holland-ashfordComparingNeutronStar2017}
{Holland-Ashford} T.,  Lopez L.~A.,  Auchettl K.,  Temim T.,   {Ramirez-Ruiz}
  E.,  2017, \mn@doi [The Astrophysical Journal] {10.3847/1538-4357/aa7a5c},
  844, 84

\bibitem[\protect\citeauthoryear{Hoogendoorn}{Hoogendoorn}{2023}]{hoogendoornEelcoHoogendoornNumpyArraysetops2023}
Hoogendoorn E.,  2023, {{EelcoHoogendoorn}}/{{Numpy}}\_arraysetops\_{{EP}},
  \url {https://github.com/EelcoHoogendoorn/Numpy_arraysetops_EP}

\bibitem[\protect\citeauthoryear{Hummel, Pawlik, Milosavljevi{\'c}  \&
  Bromm}{Hummel et~al.}{2012}]{hummelSOURCEDENSITYOBSERVABILITY2012}
Hummel J.~A.,  Pawlik A.~H.,  Milosavljevi{\'c} M.,   Bromm V.,  2012, \mn@doi
  [The Astrophysical Journal] {10.1088/0004-637X/755/1/72}, 755, 72

\bibitem[\protect\citeauthoryear{Humphreys \& Davidson}{Humphreys \&
  Davidson}{1994}]{humphreysLuminousBlueVariables1994}
Humphreys R.~M.,  Davidson K.,  1994, \mn@doi [Publications of the Astronomical
  Society of the Pacific] {10.1086/133478}, 106, 1025

\bibitem[\protect\citeauthoryear{Hunter}{Hunter}{2007}]{hunterMatplotlib2DGraphics2007}
Hunter J.~D.,  2007, \mn@doi [Computing in Science \& Engineering]
  {10.1109/MCSE.2007.55}, 9, 90

\bibitem[\protect\citeauthoryear{Hurley, Pols  \& Tout}{Hurley
  et~al.}{2000}]{hurleyComprehensiveAnalyticFormulae2000}
Hurley J.~R.,  Pols O.~R.,   Tout C.~A.,  2000, \mn@doi [Monthly Notices of the
  Royal Astronomical Society] {10.1046/j.1365-8711.2000.03426.x}, 315, 543

\bibitem[\protect\citeauthoryear{Hurley, Tout  \& Pols}{Hurley
  et~al.}{2002}]{hurleyEvolutionBinaryStars2002}
Hurley J.~R.,  Tout C.~A.,   Pols O.~R.,  2002, \mn@doi [Monthly Notices of the
  Royal Astronomical Society] {10.1046/j.1365-8711.2002.05038.x}, 329, 897

\bibitem[\protect\citeauthoryear{Ivanov \& Fern{\'a}ndez}{Ivanov \&
  Fern{\'a}ndez}{2021}]{ivanovMassEjectionFailed2021}
Ivanov M.,  Fern{\'a}ndez R.,  2021, \mn@doi [The Astrophysical Journal]
  {10.3847/1538-4357/abe59e}, 911, 6

\bibitem[\protect\citeauthoryear{Izzard \& Halabi}{Izzard \&
  Halabi}{2018}]{izzardPopulationSynthesisBinary2018}
Izzard R.~G.,  Halabi G.~M.,  2018, Population Synthesis of Binary Stars,
  \mn@doi{10.48550/arXiv.1808.06883}, \url
  {https://ui.adsabs.harvard.edu/abs/2018arXiv180806883I}

\bibitem[\protect\citeauthoryear{Izzard \& Jermyn}{Izzard \&
  Jermyn}{2022}]{izzardCircumbinaryDiscsStellar2022}
Izzard R.~G.,  Jermyn A.~S.,  2022, \mn@doi [Monthly Notices of the Royal
  Astronomical Society] {10.1093/mnras/stac2899}

\bibitem[\protect\citeauthoryear{Izzard, Tout, Karakas  \& Pols}{Izzard
  et~al.}{2004}]{izzardNewSyntheticModel2004}
Izzard R.~G.,  Tout C.~A.,  Karakas A.~I.,   Pols O.~R.,  2004, \mn@doi
  [Monthly Notices of the Royal Astronomical Society]
  {10.1111/j.1365-2966.2004.07446.x}, 350, 407

\bibitem[\protect\citeauthoryear{Izzard, Dray, Karakas, Lugaro  \& Tout}{Izzard
  et~al.}{2006}]{izzardPopulationNucleosynthesisSingle2006}
Izzard R.~G.,  Dray L.~M.,  Karakas A.~I.,  Lugaro M.,   Tout C.~A.,  2006,
  \mn@doi [Astronomy \& Astrophysics] {10.1051/0004-6361:20066129}, 460, 565

\bibitem[\protect\citeauthoryear{Izzard, Glebbeek, Stancliffe  \& Pols}{Izzard
  et~al.}{2009}]{izzardPopulationSynthesisBinary2009}
Izzard R.~G.,  Glebbeek E.,  Stancliffe R.~J.,   Pols O.~R.,  2009, \mn@doi
  [Astronomy and Astrophysics] {10.1051/0004-6361/200912827}, 508, 1359

\bibitem[\protect\citeauthoryear{Izzard, Preece, Jofre, Halabi, Masseron  \&
  Tout}{Izzard et~al.}{2018}]{izzardBinaryStarsGalactic2018}
Izzard R.~G.,  Preece H.,  Jofre P.,  Halabi G.~M.,  Masseron T.,   Tout C.~A.,
   2018, \mn@doi [Monthly Notices of the Royal Astronomical Society]
  {10.1093/mnras/stx2355}, 473, 2984

\bibitem[\protect\citeauthoryear{Janka}{Janka}{2013a}]{jankaNatalKicksStellar2013}
Janka H.-T.,  2013a, \mn@doi [Monthly Notices of the Royal Astronomical
  Society] {10.1093/mnras/stt1106}, 434, 1355

\bibitem[\protect\citeauthoryear{Janka}{Janka}{2013b}]{jankaNatalKicksStellarMass2013}
Janka H.-T.,  2013b, \mn@doi [Monthly Notices of the Royal Astronomical
  Society] {10.1093/mnras/stt1106}, 434, 1355

\bibitem[\protect\citeauthoryear{Karathanasis, Mukherjee  \&
  Mastrogiovanni}{Karathanasis et~al.}{2023}]{karathanasisBinaryBlackHoles2023}
Karathanasis C.,  Mukherjee S.,   Mastrogiovanni S.,  2023, \mn@doi [Monthly
  Notices of the Royal Astronomical Society] {10.1093/mnras/stad1373}, 523,
  4539

\bibitem[\protect\citeauthoryear{Kasen, Woosley  \& Heger}{Kasen
  et~al.}{2011}]{kasenPAIRINSTABILITYSUPERNOVAE2011}
Kasen D.,  Woosley S.~E.,   Heger A.,  2011, \mn@doi [The Astrophysical
  Journal] {10.1088/0004-637X/734/2/102}, 734, 102

\bibitem[\protect\citeauthoryear{Katsuda et~al.,}{Katsuda
  et~al.}{2018}]{katsudaIntermediatemassElementsYoung2018}
Katsuda S.,  et~al., 2018, \mn@doi [The Astrophysical Journal]
  {10.3847/1538-4357/aab092}, 856, 18

\bibitem[\protect\citeauthoryear{Klencki, Moe, Gladysz, Chruslinska, Holz  \&
  Belczynski}{Klencki et~al.}{2018}]{klenckiImpactIntercorrelatedInitial2018}
Klencki J.,  Moe M.,  Gladysz W.,  Chruslinska M.,  Holz D.~E.,   Belczynski
  K.,  2018, \mn@doi [Astronomy \& Astrophysics] {10.1051/0004-6361/201833025},
  619, A77

\bibitem[\protect\citeauthoryear{Klencki, Nelemans, Istrate  \&
  Chruslinska}{Klencki et~al.}{2021}]{klenckiItHasBe2021}
Klencki J.,  Nelemans G.,  Istrate A.~G.,   Chruslinska M.,  2021, \mn@doi
  [Astronomy \& Astrophysics] {10.1051/0004-6361/202038707}, 645, A54

\bibitem[\protect\citeauthoryear{Kluyver et~al.,}{Kluyver
  et~al.}{2016}]{kluyverJupyterNotebooksPublishing2016}
Kluyver T.,  et~al., 2016, in Loizides F.,  Schmidt B.,  eds, Positioning and
  Power in Academic Publishing: {{Players}}, Agents and Agendas. {IOS Press},
  pp 87--90

\bibitem[\protect\citeauthoryear{Kobulnicky \& Fryer}{Kobulnicky \&
  Fryer}{2007}]{kobulnickyNewLookBinary2007}
Kobulnicky H.~A.,  Fryer C.~L.,  2007, \mn@doi [The Astrophysical Journal]
  {10.1086/522073}, 670, 747

\bibitem[\protect\citeauthoryear{Kozyreva \& Blinnikov}{Kozyreva \&
  Blinnikov}{2015}]{kozyrevaCanPairinstabilitySupernova2015}
Kozyreva A.,  Blinnikov S.,  2015, \mn@doi [Monthly Notices of the Royal
  Astronomical Society] {10.1093/mnras/stv2287}, 454, 4357

\bibitem[\protect\citeauthoryear{Kozyreva, Blinnikov, Langer  \& Yoon}{Kozyreva
  et~al.}{2014}]{kozyrevaObservationalPropertiesLowredshift2014}
Kozyreva A.,  Blinnikov S.,  Langer N.,   Yoon S.-C.,  2014, \mn@doi [Astronomy
  \& Astrophysics] {10.1051/0004-6361/201423447}, 565, A70

\bibitem[\protect\citeauthoryear{Kroupa}{Kroupa}{2001}]{kroupaVariationInitialMass2001}
Kroupa P.,  2001, \mn@doi [Monthly Notices of the Royal Astronomical Society]
  {10.1046/j.1365-8711.2001.04022.x}, 322, 231

\bibitem[\protect\citeauthoryear{Kuncarayakti et~al.,}{Kuncarayakti
  et~al.}{2023}]{kuncarayaktiBactrianBroadlinedTypeIc2023}
Kuncarayakti H.,  et~al., 2023, The {{Bactrian}}? {{Broad-lined Type-Ic}}
  Supernova {{SN}} 2022xxf with Extraordinary Two-Humped Light Curves,
  \mn@doi{10.48550/arXiv.2303.16925}, \url
  {https://ui.adsabs.harvard.edu/abs/2023arXiv230316925K}

\bibitem[\protect\citeauthoryear{{LIGO Scientific Collaboration and Virgo
  Collaboration} et~al.,}{{LIGO Scientific Collaboration and Virgo
  Collaboration}
  et~al.}{2019}]{ligoscientificcollaborationandvirgocollaborationGWTC1GravitationalWaveTransient2019}
{LIGO Scientific Collaboration and Virgo Collaboration} et~al., 2019, \mn@doi
  [Physical Review X] {10.1103/PhysRevX.9.031040}, 9, 031040

\bibitem[\protect\citeauthoryear{{LIGO Scientific Collaboration}, {Virgo
  Collaboration}  \& {KAGRA Collaboration}}{{LIGO Scientific Collaboration}
  et~al.}{2021}]{ligo_scientific_collaboration_and_virgo_2021_5655785}
{LIGO Scientific Collaboration} {Virgo Collaboration}  {KAGRA Collaboration}
  2021, The Population of Merging Compact Binaries Inferred Using Gravitational
  Waves through {{GWTC-3}} - {{Data}} Release, \mn@doi{10.5281/zenodo.5655785},
  \url {https://doi.org/10.5281/zenodo.5655785}

\bibitem[\protect\citeauthoryear{Langer}{Langer}{2012}]{langerPresupernovaEvolutionMassive2012}
Langer N.,  2012, \mn@doi [Annual Review of Astronomy and Astrophysics]
  {10.1146/annurev-astro-081811-125534}, 50, 107

\bibitem[\protect\citeauthoryear{Laplace, Justham, Renzo, G{\"o}tberg, Farmer,
  Vartanyan  \& de Mink}{Laplace
  et~al.}{2021}]{laplaceDifferentCorePresupernova2021}
Laplace E.,  Justham S.,  Renzo M.,  G{\"o}tberg Y.,  Farmer R.,  Vartanyan D.,
    de Mink S.~E.,  2021, \mn@doi [Astronomy \& Astrophysics]
  {10.1051/0004-6361/202140506}, 656, A58

\bibitem[\protect\citeauthoryear{Li, Wang, Han, Tang, Yuan, Fan  \& Wei}{Li
  et~al.}{2021}]{liFlexibleGaussianProcess2021}
Li Y.-J.,  Wang Y.-Z.,  Han M.-Z.,  Tang S.-P.,  Yuan Q.,  Fan Y.-Z.,   Wei
  D.-M.,  2021, \mn@doi [The Astrophysical Journal] {10.3847/1538-4357/ac0971},
  917, 33

\bibitem[\protect\citeauthoryear{Li, Wang, Tang, Yuan, Fan  \& Wei}{Li
  et~al.}{2022}]{liDivergenceMassRatio2022}
Li Y.-J.,  Wang Y.-Z.,  Tang S.-P.,  Yuan Q.,  Fan Y.-Z.,   Wei D.-M.,  2022,
  \mn@doi [The Astrophysical Journal Letters] {10.3847/2041-8213/ac78dd}, 933,
  L14

\bibitem[\protect\citeauthoryear{Limongi \& Chieffi}{Limongi \&
  Chieffi}{2018}]{limongiPresupernovaEvolutionExplosive2018}
Limongi M.,  Chieffi A.,  2018, \mn@doi [The Astrophysical Journal Supplement
  Series] {10.3847/1538-4365/aacb24}, 237, 13

\bibitem[\protect\citeauthoryear{Lin et~al.,}{Lin
  et~al.}{2023}]{linSuperluminousSupernovaLightened2023}
Lin W.,  et~al., 2023, \mn@doi [Nature Astronomy] {10.1038/s41550-023-01957-3},
  7, 779

\bibitem[\protect\citeauthoryear{Lovegrove \& Woosley}{Lovegrove \&
  Woosley}{2013}]{lovegroveVeryLowEnergy2013}
Lovegrove E.,  Woosley S.~E.,  2013, \mn@doi [The Astrophysical Journal]
  {10.1088/0004-637X/769/2/109}, 769, 109

\bibitem[\protect\citeauthoryear{Lunnan et~al.,}{Lunnan
  et~al.}{2018}]{lunnanUVResonanceLine2018}
Lunnan R.,  et~al., 2018, \mn@doi [Nature Astronomy]
  {10.1038/s41550-018-0568-z}, 2, 887

\bibitem[\protect\citeauthoryear{Maeder \& Meynet}{Maeder \&
  Meynet}{2000}]{maederEvolutionRotatingStars2000}
Maeder A.,  Meynet G.,  2000, \mn@doi [Annual Review of Astronomy and
  Astrophysics] {10.1146/annurev.astro.38.1.143}, 38, 143

\bibitem[\protect\citeauthoryear{Mandel}{Mandel}{2016}]{mandelEstimatesBlackHole2016}
Mandel I.,  2016, \mn@doi [Monthly Notices of the Royal Astronomical Society]
  {10.1093/mnras/stv2733}, 456, 578

\bibitem[\protect\citeauthoryear{Mandel \& {de Mink}}{Mandel \& {de
  Mink}}{2016}]{mandelMergingBinaryBlack2016}
Mandel I.,  {de Mink} S.~E.,  2016, \mn@doi [Monthly Notices of the Royal
  Astronomical Society] {10.1093/mnras/stw379}, 458, 2634

\bibitem[\protect\citeauthoryear{Mapelli, Giacobbo, Santoliquido  \&
  Artale}{Mapelli et~al.}{2019}]{mapelliPropertiesMergingBlack2019}
Mapelli M.,  Giacobbo N.,  Santoliquido F.,   Artale M.~C.,  2019, \mn@doi
  [Monthly Notices of the Royal Astronomical Society] {10.1093/mnras/stz1150},
  487, 2

\bibitem[\protect\citeauthoryear{Mapelli, Bouffanais, Santoliquido, Arca~Sedda
  \& Artale}{Mapelli et~al.}{2022}]{mapelliCosmicEvolutionBinary2022}
Mapelli M.,  Bouffanais Y.,  Santoliquido F.,  Arca~Sedda M.,   Artale M.~C.,
  2022, \mn@doi [Monthly Notices of the Royal Astronomical Society]
  {10.1093/mnras/stac422}, 511, 5797

\bibitem[\protect\citeauthoryear{Marchant \& Moriya}{Marchant \&
  Moriya}{2020}]{marchantImpactStellarRotation2020}
Marchant P.,  Moriya T.~J.,  2020, \mn@doi [Astronomy \& Astrophysics]
  {10.1051/0004-6361/202038902}, 640, L18

\bibitem[\protect\citeauthoryear{Marchant, Langer, Podsiadlowski, Tauris  \&
  Moriya}{Marchant et~al.}{2016}]{marchantNewRouteMerging2016}
Marchant P.,  Langer N.,  Podsiadlowski P.,  Tauris T.~M.,   Moriya T.~J.,
  2016, \mn@doi [Astronomy \& Astrophysics] {10.1051/0004-6361/201628133}, 588,
  A50

\bibitem[\protect\citeauthoryear{Marchant, Renzo, Farmer, Pappas, Taam, {de
  Mink}  \& Kalogera}{Marchant
  et~al.}{2019}]{marchantPulsationalPairinstabilitySupernovae2019}
Marchant P.,  Renzo M.,  Farmer R.,  Pappas K. M.~W.,  Taam R.~E.,  {de Mink}
  S.~E.,   Kalogera V.,  2019, \mn@doi [The Astrophysical Journal]
  {10.3847/1538-4357/ab3426}, 882, 36

\bibitem[\protect\citeauthoryear{Mehta, Buonanno, Gair, Miller, Farag,
  {deBoer}, Wiescher  \& Timmes}{Mehta
  et~al.}{2022}]{mehtaObservingIntermediatemassBlack2022}
Mehta A.~K.,  Buonanno A.,  Gair J.,  Miller M.~C.,  Farag E.,  {deBoer} R.~J.,
   Wiescher M.,   Timmes F.~X.,  2022, \mn@doi [The Astrophysical Journal]
  {10.3847/1538-4357/ac3130}, 924, 39

\bibitem[\protect\citeauthoryear{Moe \& Stefano}{Moe \&
  Stefano}{2017}]{moeMindYourPs2017}
Moe M.,  Stefano R.~D.,  2017, \mn@doi [The Astrophysical Journal Supplement
  Series] {10.3847/1538-4365/aa6fb6}, 230, 15

\bibitem[\protect\citeauthoryear{Moe, Kratter  \& Badenes}{Moe
  et~al.}{2019}]{moeCloseBinaryFraction2019}
Moe M.,  Kratter K.~M.,   Badenes C.,  2019, \mn@doi [The Astrophysical
  Journal] {10.3847/1538-4357/ab0d88}, 875, 61

\bibitem[\protect\citeauthoryear{Mori, Moriya, Takiwaki, Kotake, Horiuchi  \&
  Blinnikov}{Mori et~al.}{2023}]{moriLightCurvesEvent2023}
Mori K.,  Moriya T.~J.,  Takiwaki T.,  Kotake K.,  Horiuchi S.,   Blinnikov
  S.~I.,  2023, \mn@doi [The Astrophysical Journal] {10.3847/1538-4357/acaaff},
  943, 12

\bibitem[\protect\citeauthoryear{Moriya \& Langer}{Moriya \&
  Langer}{2015}]{moriyaPulsationsRedSupergiant2015}
Moriya T.~J.,  Langer N.,  2015, \mn@doi [Astronomy \& Astrophysics]
  {10.1051/0004-6361/201424957}, 573, A18

\bibitem[\protect\citeauthoryear{Moriya, Quimby  \& Robertson}{Moriya
  et~al.}{2022}]{moriyaDiscoveringSupernovaeEpoch2022}
Moriya T.~J.,  Quimby R.~M.,   Robertson B.~E.,  2022, \mn@doi [The
  Astrophysical Journal] {10.3847/1538-4357/ac415e}, 925, 211

\bibitem[\protect\citeauthoryear{Nadezhin}{Nadezhin}{1980}]{nadezhinSecondaryIndicationsGravitational1980}
Nadezhin D.~K.,  1980, \mn@doi [Astrophysics and Space Science]
  {10.1007/BF00638971}, 69, 115

\bibitem[\protect\citeauthoryear{Neijssel et~al.,}{Neijssel
  et~al.}{2019}]{neijsselEffectMetallicityspecificStar2019}
Neijssel C.~J.,  et~al., 2019, \mn@doi [Monthly Notices of the Royal
  Astronomical Society] {10.1093/mnras/stz2840}, 490, 3740

\bibitem[\protect\citeauthoryear{Nicholl et~al.,}{Nicholl
  et~al.}{2013}]{nichollSlowlyFadingSuperluminous2013}
Nicholl M.,  et~al., 2013, \mn@doi [Nature] {10.1038/nature12569}, 502, 346

\bibitem[\protect\citeauthoryear{Nitz et~al.,}{Nitz
  et~al.}{2023}]{nitzGwastroPycbcV22023}
Nitz A.,  et~al., 2023, Gwastro/Pycbc: V2.1.2 Release of {{PyCBC}}, Zenodo,
  \mn@doi{10.5281/zenodo.7885796}, \url
  {https://doi.org/10.5281/zenodo.7885796}

\bibitem[\protect\citeauthoryear{Offner, Moe, Kratter, Sadavoy, Jensen  \&
  Tobin}{Offner et~al.}{2022}]{offnerOriginEvolutionMultiple2022}
Offner S. S.~R.,  Moe M.,  Kratter K.~M.,  Sadavoy S.~I.,  Jensen E. L.~N.,
  Tobin J.~J.,  2022, The {{Origin}} and {{Evolution}} of {{Multiple Star
  Systems}}, \mn@doi{10.48550/arXiv.2203.10066}, \url
  {https://ui.adsabs.harvard.edu/abs/2022arXiv220310066O}

\bibitem[\protect\citeauthoryear{Olejak, Belczynski  \& Ivanova}{Olejak
  et~al.}{2021}]{olejakImpactCommonEnvelope2021}
Olejak A.,  Belczynski K.,   Ivanova N.,  2021, \mn@doi [Astronomy \&
  Astrophysics] {10.1051/0004-6361/202140520}, 651, A100

\bibitem[\protect\citeauthoryear{Perez \& Granger}{Perez \&
  Granger}{2007}]{perezIPythonSystemInteractive2007}
Perez F.,  Granger B.~E.,  2007, \mn@doi [Computing in Science \& Engineering]
  {10.1109/MCSE.2007.53}, 9, 21

\bibitem[\protect\citeauthoryear{Perley et~al.,}{Perley
  et~al.}{2016}]{perleyHostgalaxyProperties322016}
Perley D.~A.,  et~al., 2016, \mn@doi [The Astrophysical Journal]
  {10.3847/0004-637X/830/1/13}, 830, 13

\bibitem[\protect\citeauthoryear{Peters}{Peters}{1964}]{petersGravitationalRadiationMotion1964}
Peters P.~C.,  1964, \mn@doi [Physical Review] {10.1103/PhysRev.136.B1224},
  136, B1224

\bibitem[\protect\citeauthoryear{Petrovi{\'c}}{Petrovi{\'c}}{2020}]{petrovicEvolutionMassiveBinary2020}
Petrovi{\'c} J.,  2020, \mn@doi [Serbian Astronomical Journal]
  {10.2298/SAJ2001001P}, 201, 1

\bibitem[\protect\citeauthoryear{Piro}{Piro}{2013}]{piroTakingOutUnnovae2013}
Piro A.~L.,  2013, \mn@doi [The Astrophysical Journal]
  {10.1088/2041-8205/768/1/L14}, 768, L14

\bibitem[\protect\citeauthoryear{Pols, Schr{\"o}der, Hurley, Tout  \&
  Eggleton}{Pols et~al.}{1998}]{polsStellarEvolutionModels1998}
Pols O.~R.,  Schr{\"o}der K.-P.,  Hurley J.~R.,  Tout C.~A.,   Eggleton P.~P.,
  1998, \mn@doi [Monthly Notices of the Royal Astronomical Society]
  {10.1046/j.1365-8711.1998.01658.x}, 298, 525

\bibitem[\protect\citeauthoryear{Postnov \& Yungelson}{Postnov \&
  Yungelson}{2014}]{postnovEvolutionCompactBinary2014}
Postnov K.,  Yungelson L.,  2014, \mn@doi [Living Reviews in Relativity]
  {10.12942/lrr-2014-3}, 17, 3

\bibitem[\protect\citeauthoryear{Powell, M{\"u}ller  \& Heger}{Powell
  et~al.}{2021}]{powellFinalCoreCollapse2021}
Powell J.,  M{\"u}ller B.,   Heger A.,  2021, \mn@doi [Monthly Notices of the
  Royal Astronomical Society] {10.1093/mnras/stab614}, 503, 2108

\bibitem[\protect\citeauthoryear{{PyTables Developers Team}}{{PyTables
  Developers
  Team}}{2002}]{pytablesdevelopersteamPyTablesHierarchicalDatasets2002}
{PyTables Developers Team} 2002, {{PyTables}}: {{Hierarchical}} Datasets in
  {{Python}}, \url {http://www.pytables.org/}

\bibitem[\protect\citeauthoryear{Rahman, Janka, Stockinger  \& Woosley}{Rahman
  et~al.}{2022}]{rahmanPulsationalPairinstabilitySupernovae2022}
Rahman N.,  Janka H.-T.,  Stockinger G.,   Woosley S.~E.,  2022, \mn@doi
  [Monthly Notices of the Royal Astronomical Society] {10.1093/mnras/stac758},
  512, 4503

\bibitem[\protect\citeauthoryear{Rakavy \& Shaviv}{Rakavy \&
  Shaviv}{1967}]{rakavyInstabilitiesHighlyEvolved1967}
Rakavy G.,  Shaviv G.,  1967, \mn@doi [The Astrophysical Journal]
  {10.1086/149204}, 148, 803

\bibitem[\protect\citeauthoryear{Reimers}{Reimers}{1975}]{reimersCircumstellarAbsorptionLines1975}
Reimers D.,  1975, Memoires of the Societe Royale des Sciences de Liege, \href
  {https://ui.adsabs.harvard.edu/abs/1975MSRSL...8..369R/} {8, 369}

\bibitem[\protect\citeauthoryear{Renzo, Ott, Shore  \& de Mink}{Renzo
  et~al.}{2017}]{renzoSystematicSurveyEffects2017}
Renzo M.,  Ott C.~D.,  Shore S.~N.,   de Mink S.~E.,  2017, \mn@doi [Astronomy
  \& Astrophysics] {10.1051/0004-6361/201730698}, 603, A118

\bibitem[\protect\citeauthoryear{Renzo et~al.,}{Renzo
  et~al.}{2019}]{renzoMassiveRunawayWalkaway2019}
Renzo M.,  et~al., 2019, \mn@doi [Astronomy \& Astrophysics]
  {10.1051/0004-6361/201833297}, 624, A66

\bibitem[\protect\citeauthoryear{Renzo, Farmer, Justham, {de Mink}, G{\"o}tberg
   \& Marchant}{Renzo et~al.}{2020a}]{renzoSensitivityLowerEdge2020}
Renzo M.,  Farmer R.~J.,  Justham S.,  {de Mink} S.~E.,  G{\"o}tberg Y.,
  Marchant P.,  2020a, \mn@doi [Monthly Notices of the Royal Astronomical
  Society] {10.1093/mnras/staa549}, 493, 4333

\bibitem[\protect\citeauthoryear{Renzo, Farmer, Justham, G{\"o}tberg, {de
  Mink}, Zapartas, Marchant  \& Smith}{Renzo
  et~al.}{2020b}]{renzoPredictionsHydrogenfreeEjecta2020}
Renzo M.,  Farmer R.,  Justham S.,  G{\"o}tberg Y.,  {de Mink} S.~E.,  Zapartas
  E.,  Marchant P.,   Smith N.,  2020b, \mn@doi [Astronomy \& Astrophysics]
  {10.1051/0004-6361/202037710}, 640, A56

\bibitem[\protect\citeauthoryear{Renzo, Hendriks, van Son  \& Farmer}{Renzo
  et~al.}{2022}]{renzoPairinstabilityMassLoss2022}
Renzo M.,  Hendriks D.~D.,  van Son L. A.~C.,   Farmer R.,  2022, \mn@doi
  [Research Notes of the AAS] {10.3847/2515-5172/ac503e}, 6, 25

\bibitem[\protect\citeauthoryear{Renzo, Zapartas, Justham, Breivik, Lau,
  Farmer, Cantiello  \& Metzger}{Renzo
  et~al.}{2023}]{renzoRejuvenatedAccretorsHave2023}
Renzo M.,  Zapartas E.,  Justham S.,  Breivik K.,  Lau M.,  Farmer R.,
  Cantiello M.,   Metzger B.~D.,  2023, \mn@doi [The Astrophysical Journal
  Letters] {10.3847/2041-8213/aca4d3}, 942, L32

\bibitem[\protect\citeauthoryear{Riley et~al.,}{Riley
  et~al.}{2022}]{rileyRapidStellarBinary2022}
Riley J.,  et~al., 2022, \mn@doi [The Astrophysical Journal Supplement Series]
  {10.3847/1538-4365/ac416c}, 258, 34

\bibitem[\protect\citeauthoryear{Sadiq, Dent  \& Wysocki}{Sadiq
  et~al.}{2022}]{sadiqFlexibleFastEstimation2022}
Sadiq J.,  Dent T.,   Wysocki D.,  2022, \mn@doi [Physical Review D]
  {10.1103/PhysRevD.105.123014}, 105, 123014

\bibitem[\protect\citeauthoryear{Safarzadeh}{Safarzadeh}{2020}]{safarzadehBranchingRatioLIGO2020}
Safarzadeh M.,  2020, \mn@doi [The Astrophysical Journal]
  {10.3847/2041-8213/ab7cdc}, 892, L8

\bibitem[\protect\citeauthoryear{Sakstein, Croon  \& McDermott}{Sakstein
  et~al.}{2022}]{saksteinAxionInstabilitySupernovae2022}
Sakstein J.,  Croon D.,   McDermott S.~D.,  2022, \mn@doi [Physical Review D]
  {10.1103/PhysRevD.105.095038}, 105, 095038

\bibitem[\protect\citeauthoryear{Sana et~al.,}{Sana
  et~al.}{2012}]{sanaBinaryInteractionDominates2012}
Sana H.,  et~al., 2012, \mn@doi [Science] {10.1126/science.1223344}, 337, 444

\bibitem[\protect\citeauthoryear{Schneider, Podsiadlowski  \&
  M{\"u}ller}{Schneider
  et~al.}{2021}]{schneiderPresupernovaEvolutionCompactobject2021}
Schneider F. R.~N.,  Podsiadlowski P.,   M{\"u}ller B.,  2021, \mn@doi
  [Astronomy \& Astrophysics] {10.1051/0004-6361/202039219}, 645, A5

\bibitem[\protect\citeauthoryear{Schneider, Podsiadlowski  \&
  Laplace}{Schneider et~al.}{2023}]{schneiderBimodalBlackHole2023}
Schneider F. R.~N.,  Podsiadlowski P.,   Laplace E.,  2023, \mn@doi [The
  Astrophysical Journal Letters] {10.3847/2041-8213/acd77a}, 950, L9

\bibitem[\protect\citeauthoryear{Schulze et~al.,}{Schulze
  et~al.}{2023}]{schulze1100DaysLife2023}
Schulze S.,  et~al., 2023, 1100 {{Days}} in the {{Life}} of the {{Supernova}}
  2018ibb \textendash{} the {{Best Pair-Instability Supernova Candidate}}, to
  Date (\mn@eprint {arxiv} {2305.05796}), \mn@doi{10.48550/arXiv.2305.05796}

\bibitem[\protect\citeauthoryear{Shen et~al.,}{Shen
  et~al.}{2023}]{shenNewDetermination12C2023}
Shen Y.,  et~al., 2023, \mn@doi [The Astrophysical Journal]
  {10.3847/1538-4357/acb7de}, 945, 41

\bibitem[\protect\citeauthoryear{Shklovskii}{Shklovskii}{1970}]{shklovskiiPossibleCausesSecular1970}
Shklovskii I.~S.,  1970, Soviet Astronomy, \href
  {https://ui.adsabs.harvard.edu/abs/1970SvA....13..562S} {13, 562}

\bibitem[\protect\citeauthoryear{Siegel, Agarwal, Barnes, Metzger, Renzo  \&
  Villar}{Siegel et~al.}{2022}]{siegelSuperkilonovaeMassiveCollapsars2022}
Siegel D.~M.,  Agarwal A.,  Barnes J.,  Metzger B.~D.,  Renzo M.,   Villar
  V.~A.,  2022, \mn@doi [The Astrophysical Journal] {10.3847/1538-4357/ac8d04},
  941, 100

\bibitem[\protect\citeauthoryear{Soberman, Phinney  \& {van den
  Heuvel}}{Soberman et~al.}{1997}]{sobermanStabilityCriteriaMass1997}
Soberman G.~E.,  Phinney E.~S.,   {van den Heuvel} E. P.~J.,  1997, \mn@doi
  [Astronomy and Astrophysics] {10.48550/arXiv.astro-ph/9703016}, 327, 620

\bibitem[\protect\citeauthoryear{Stevenson, Sampson, Powell, {Vigna-G{\'o}mez},
  Neijssel, Sz{\'e}csi  \& Mandel}{Stevenson
  et~al.}{2019}]{stevensonImpactPairinstabilityMass2019}
Stevenson S.,  Sampson M.,  Powell J.,  {Vigna-G{\'o}mez} A.,  Neijssel C.~J.,
  Sz{\'e}csi D.,   Mandel I.,  2019, \mn@doi [The Astrophysical Journal]
  {10.3847/1538-4357/ab3981}, 882, 121

\bibitem[\protect\citeauthoryear{Talbot \& Thrane}{Talbot \&
  Thrane}{2018}]{talbotMeasuringBinaryBlack2018}
Talbot C.,  Thrane E.,  2018, \mn@doi [The Astrophysical Journal]
  {10.3847/1538-4357/aab34c}, 856, 173

\bibitem[\protect\citeauthoryear{Tanaka, Gilkis, Izzard  \& Tout}{Tanaka
  et~al.}{2023}]{tanakaOptimalEnvelopeEjection2023}
Tanaka A.~M.,  Gilkis A.,  Izzard R.~G.,   Tout C.~A.,  2023, \mn@doi [Monthly
  Notices of the Royal Astronomical Society] {10.1093/mnras/stad971}, 522, 1140

\bibitem[\protect\citeauthoryear{Tanikawa, Kinugawa, Yoshida, Hijikawa  \&
  Umeda}{Tanikawa et~al.}{2021}]{tanikawaPopulationIIIBinary2021}
Tanikawa A.,  Kinugawa T.,  Yoshida T.,  Hijikawa K.,   Umeda H.,  2021,
  \mn@doi [Monthly Notices of the Royal Astronomical Society]
  {10.1093/mnras/stab1421}, 505, 2170

\bibitem[\protect\citeauthoryear{Tanikawa, Moriya, Tominaga  \&
  Yoshida}{Tanikawa et~al.}{2023}]{tanikawaEuclidDetectabilityPair2023}
Tanikawa A.,  Moriya T.~J.,  Tominaga N.,   Yoshida N.,  2023, \mn@doi [Monthly
  Notices of the Royal Astronomical Society: Letters] {10.1093/mnrasl/slac149},
  519, L32

\bibitem[\protect\citeauthoryear{{The Astropy Collaboration} et~al.,}{{The
  Astropy Collaboration}
  et~al.}{2022}]{theastropycollaborationAstropyProjectSustaining2022}
{The Astropy Collaboration} et~al., 2022, \mn@doi [The Astrophysical Journal]
  {10.3847/1538-4357/ac7c74}, 935, 167

\bibitem[\protect\citeauthoryear{{The LIGO Scientific Collaboration}
  et~al.,}{{The LIGO Scientific Collaboration}
  et~al.}{2023}]{theligoscientificcollaborationOpenDataThird2023}
{The LIGO Scientific Collaboration} et~al., 2023, Open Data from the Third
  Observing Run of {{LIGO}}, {{Virgo}}, {{KAGRA}} and {{GEO}} (\mn@eprint
  {arxiv} {2302.03676}), \mn@doi{10.48550/arXiv.2302.03676}, \url
  {http://arxiv.org/abs/2302.03676}

\bibitem[\protect\citeauthoryear{{The pandas development team}}{{The pandas
  development team}}{2020}]{thepandasdevelopmentteamPandasdevPandasPandas2020}
{The pandas development team} 2020, Pandas-Dev/Pandas: {{Pandas}}, Zenodo, \url
  {https://doi.org/10.5281/zenodo.3509134}

\bibitem[\protect\citeauthoryear{Thiele, Breivik, Sanderson  \& Luger}{Thiele
  et~al.}{2023}]{thieleApplyingMetallicitydependentBinary2023}
Thiele S.,  Breivik K.,  Sanderson R.~E.,   Luger R.,  2023, \mn@doi [The
  Astrophysical Journal] {10.3847/1538-4357/aca7be}, 945, 162

\bibitem[\protect\citeauthoryear{Tiwari}{Tiwari}{2022}]{tiwariExploringFeaturesBinary2022}
Tiwari V.,  2022, \mn@doi [The Astrophysical Journal]
  {10.3847/1538-4357/ac589a}, 928, 155

\bibitem[\protect\citeauthoryear{Tolstov, Nomoto, Blinnikov, Sorokina, Quimby
  \& Baklanov}{Tolstov
  et~al.}{2017}]{tolstovPulsationalPairinstabilityModel2017}
Tolstov A.,  Nomoto K.,  Blinnikov S.,  Sorokina E.,  Quimby R.,   Baklanov P.,
   2017, \mn@doi [The Astrophysical Journal] {10.3847/1538-4357/835/2/266},
  835, 266

\bibitem[\protect\citeauthoryear{Tout, Aarseth, Pols  \& Eggleton}{Tout
  et~al.}{1997}]{toutRapidBinaryStar1997}
Tout C.~A.,  Aarseth S.~J.,  Pols O.~R.,   Eggleton P.~P.,  1997, \mn@doi
  [Monthly Notices of the Royal Astronomical Society]
  {10.1093/mnras/291.4.732}, 291, 732

\bibitem[\protect\citeauthoryear{Van~Rossum \& Drake}{Van~Rossum \&
  Drake}{2009}]{vanrossumPythonReferenceManual2009}
Van~Rossum G.,  Drake F.~L.,  2009, Python 3 Reference Manual.
{CreateSpace}, {Scotts Valley, CA}

\bibitem[\protect\citeauthoryear{Vassiliadis \& Wood}{Vassiliadis \&
  Wood}{1993}]{vassiliadisEvolutionLowIntermediatemass1993}
Vassiliadis E.,  Wood P.~R.,  1993, \mn@doi [The Astrophysical Journal]
  {10.1086/173033}, 413, 641

\bibitem[\protect\citeauthoryear{Veske, Bartos, M{\'a}rka  \& M{\'a}rka}{Veske
  et~al.}{2021}]{veskeCharacterizingObservationBias2021}
Veske D.,  Bartos I.,  M{\'a}rka Z.,   M{\'a}rka S.,  2021, \mn@doi [The
  Astrophysical Journal] {10.3847/1538-4357/ac27ac}, 922, 258

\bibitem[\protect\citeauthoryear{Villar, Nicholl  \& Berger}{Villar
  et~al.}{2018}]{villarSuperluminousSupernovaeLSST2018}
Villar V.~A.,  Nicholl M.,   Berger E.,  2018, \mn@doi [The Astrophysical
  Journal] {10.3847/1538-4357/aaee6a}, 869, 166

\bibitem[\protect\citeauthoryear{Vink, {de Koter}  \& Lamers}{Vink
  et~al.}{2000}]{vinkNewTheoreticalMassloss2000}
Vink J.~S.,  {de Koter} A.,   Lamers H. J. G. L.~M.,  2000, \mn@doi [Astronomy
  and Astrophysics] {10.48550/arXiv.astro-ph/0008183}, \href
  {http://adsabs.harvard.edu/abs/2000A\&A...362..295V} {362, 295}

\bibitem[\protect\citeauthoryear{Vink, {de Koter}  \& Lamers}{Vink
  et~al.}{2001}]{vinkMasslossPredictionsStars2001}
Vink J.~S.,  {de Koter} A.,   Lamers H. J. G. L.~M.,  2001, \mn@doi [Astronomy
  and Astrophysics] {10.1051/0004-6361:20010127}, 369, 574

\bibitem[\protect\citeauthoryear{Vink, Higgins, Sander  \& Sabhahit}{Vink
  et~al.}{2021}]{vinkMaximumBlackHole2021}
Vink J.~S.,  Higgins E.~R.,  Sander A. A.~C.,   Sabhahit G.~N.,  2021, \mn@doi
  [Monthly Notices of the Royal Astronomical Society] {10.1093/mnras/stab842},
  504, 146

\bibitem[\protect\citeauthoryear{Virtanen et~al.,}{Virtanen
  et~al.}{2020}]{virtanenSciPyFundamentalAlgorithms2020}
Virtanen P.,  et~al., 2020, \mn@doi [Nature Methods]
  {10.1038/s41592-019-0686-2}, \href {https://rdcu.be/b08Wh} {17, 261}

\bibitem[\protect\citeauthoryear{Wang, Liu, Lin, Wang, Dai, Li  \& Song}{Wang
  et~al.}{2022}]{wangIPTF14hlsCircumstellarMedium2022}
Wang L.-J.,  Liu L.-D.,  Lin W.-L.,  Wang X.-F.,  Dai Z.-G.,  Li B.,   Song
  L.-M.,  2022, \mn@doi [The Astrophysical Journal] {10.3847/1538-4357/ac7564},
  933, 102

\bibitem[\protect\citeauthoryear{{Wes McKinney}}{{Wes
  McKinney}}{2010}]{wesmckinneyDataStructuresStatistical2010}
{Wes McKinney} 2010, in {van der Walt} S.,  {Jarrod Millman} eds, Proceedings
  of the 9th {{Python}} in {{Science Conference}}. pp 56--61,
  \mn@doi{10.25080/Majora-92bf1922-00a}

\bibitem[\protect\citeauthoryear{Wong, Breivik, Kremer  \& Callister}{Wong
  et~al.}{2021}]{wongJointConstraintsFieldcluster2021}
Wong K. W.~K.,  Breivik K.,  Kremer K.,   Callister T.,  2021, \mn@doi
  [Physical Review D] {10.1103/PhysRevD.103.083021}, 103, 083021

\bibitem[\protect\citeauthoryear{Woosley}{Woosley}{2017}]{woosleyPulsationalPairInstabilitySupernovae2017}
Woosley S.~E.,  2017, \mn@doi [The Astrophysical Journal]
  {10.3847/1538-4357/836/2/244}, 836, 244

\bibitem[\protect\citeauthoryear{Woosley}{Woosley}{2019}]{woosleyEvolutionMassiveHelium2019}
Woosley S.~E.,  2019, \mn@doi [The Astrophysical Journal]
  {10.3847/1538-4357/ab1b41}, 878, 49

\bibitem[\protect\citeauthoryear{Woosley \& Heger}{Woosley \&
  Heger}{2021}]{woosleyPairinstabilityMassGap2021}
Woosley S.~E.,  Heger A.,  2021, \mn@doi [The Astrophysical Journal]
  {10.3847/2041-8213/abf2c4}, 912, L31

\bibitem[\protect\citeauthoryear{Woosley \& Smith}{Woosley \&
  Smith}{2022}]{woosleySN1961VPulsational2022}
Woosley S.~E.,  Smith N.,  2022, \mn@doi [The Astrophysical Journal]
  {10.3847/1538-4357/ac8eb3}, 938, 57

\bibitem[\protect\citeauthoryear{Woosley, Heger  \& Weaver}{Woosley
  et~al.}{2002}]{woosleyEvolutionExplosionMassive2002}
Woosley S.~E.,  Heger A.,   Weaver T.~A.,  2002, \mn@doi [Reviews of Modern
  Physics] {10.1103/RevModPhys.74.1015}, 74, 1015

\bibitem[\protect\citeauthoryear{Woosley, Blinnikov  \& Heger}{Woosley
  et~al.}{2007}]{woosleyPulsationalPairInstability2007}
Woosley S.~E.,  Blinnikov S.,   Heger A.,  2007, \mn@doi [Nature]
  {10.1038/nature06333}, 450, 390

\bibitem[\protect\citeauthoryear{Yoon}{Yoon}{2017}]{yoonBetterUnderstandingEvolution2017}
Yoon S.-C.,  2017, \mn@doi [Monthly Notices of the Royal Astronomical Society]
  {10.1093/mnras/stx1496}, 470, 3970

\bibitem[\protect\citeauthoryear{Zevin, Spera, Berry  \& Kalogera}{Zevin
  et~al.}{2020}]{zevinExploringLowerMass2020}
Zevin M.,  Spera M.,  Berry C. P.~L.,   Kalogera V.,  2020, \mn@doi [The
  Astrophysical Journal] {10.3847/2041-8213/aba74e}, 899, L1

\bibitem[\protect\citeauthoryear{Ziegler \& Freese}{Ziegler \&
  Freese}{2021}]{zieglerFillingBlackHole2021}
Ziegler J.,  Freese K.,  2021, \mn@doi [Physical Review D]
  {10.1103/PhysRevD.104.043015}, 104, 043015

\bibitem[\protect\citeauthoryear{Ziegler \& Freese}{Ziegler \&
  Freese}{2022}]{zieglerGapNoMore2022}
Ziegler J.,  Freese K.,  2022, \mn@doi [arXiv e-prints]
  {10.48550/arXiv.2212.13903}, \href
  {https://ui.adsabs.harvard.edu/abs/2022arXiv221213903Z} {}

\bibitem[\protect\citeauthoryear{de Mink \& Belczynski}{de~Mink \&
  Belczynski}{2015}]{minkMERGERRATESDOUBLE2015}
de Mink S.~E.,  Belczynski K.,  2015, \mn@doi [The Astrophysical Journal]
  {10.1088/0004-637X/814/1/58}, 814, 58

\bibitem[\protect\citeauthoryear{{deBoer} et~al.,}{{deBoer}
  et~al.}{2017}]{deboer12C16OReaction2017}
{deBoer} R.~J.,  et~al., 2017, \mn@doi [Reviews of Modern Physics]
  {10.1103/RevModPhys.89.035007}, 89, 035007

\bibitem[\protect\citeauthoryear{{du~Buisson} et~al.,}{{du~Buisson}
  et~al.}{2020}]{dubuissonCosmicRatesBlack2020}
{du~Buisson} L.,  et~al., 2020, \mn@doi [Monthly Notices of the Royal
  Astronomical Society] {10.1093/mnras/staa3225}, 499, 5941

\bibitem[\protect\citeauthoryear{van Son et~al.,}{van Son
  et~al.}{2022a}]{sonRedshiftEvolutionBinary2022}
van Son L. A.~C.,  et~al., 2022a, \mn@doi [The Astrophysical Journal]
  {10.3847/1538-4357/ac64a3}, 931, 17

\bibitem[\protect\citeauthoryear{van Son et~al.,}{van Son
  et~al.}{2022b}]{sonNoPeaksValleys2022}
van Son L. A.~C.,  et~al., 2022b, \mn@doi [The Astrophysical Journal]
  {10.3847/1538-4357/ac9b0a}, 940, 184

\bibitem[\protect\citeauthoryear{van Son, de Mink, Chru{\'s}li{\'n}ska, Conroy,
  Pakmor  \& Hernquist}{van Son et~al.}{2023}]{sonLocationsFeaturesMass2023}
van Son L. A.~C.,  de Mink S.~E.,  Chru{\'s}li{\'n}ska M.,  Conroy C.,  Pakmor
  R.,   Hernquist L.,  2023, \mn@doi [The Astrophysical Journal]
  {10.3847/1538-4357/acbf51}, 948, 105

\makeatother
\end{thebibliography}

\begin{appendices}

\section{Remnant masses of single stars}
\label{sec:effect-single-stars}
\begin{figure*}
  \centering
  \includegraphics[width=\textwidth]{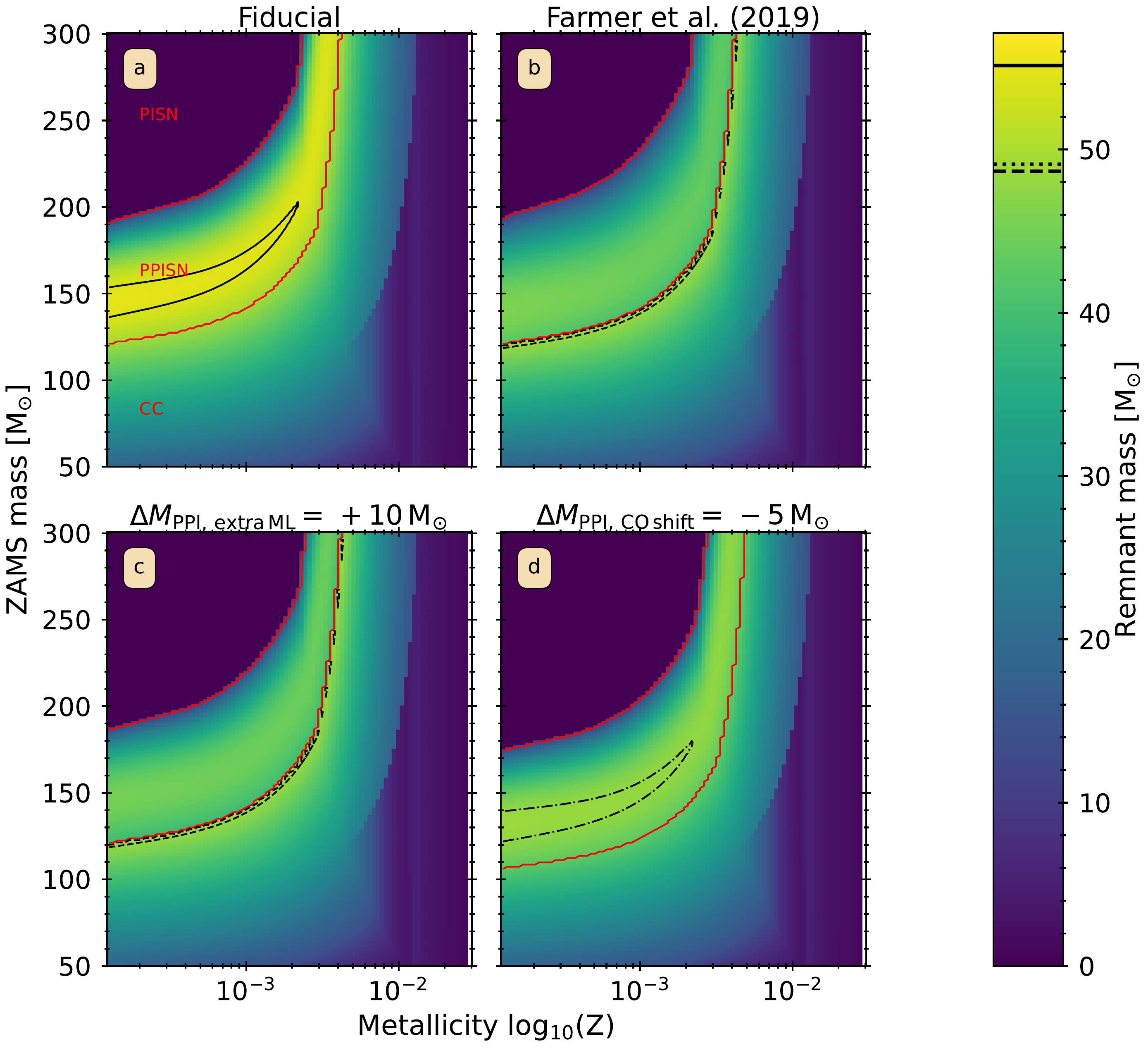}
  \caption[Remnant-mass distribution of single-star evolution as a function of
  ZAMS mass and metallicity for our fiducial,
  \citetalias{farmerMindGapLocation2019}, $\extraML = +10\,\solarmass$ and
  $\COshift = -5\,\solarmass$ models.]{Remnant-mass distribution in our single
	star grid as a function of ZAMS masses (abscissa) and metallicities
	(ordinate). The colour-scale indicates the final remnant mass. The four
	panels show our fiducial {\ppisn} model (a), the
	\citetalias{farmerMindGapLocation2019} model (b) and two examples of our
	variation on the {\ppisn} prescription, the first being the extra mass loss
	variation (c) and the second the core-mass shift variation (d). The thin red
	line indicates the region of ZAMS mass and metallicity where the single star
	undergoes {\ppisn} + CC. The region in the top left corner is where stars
	undergo {\pisn} and leave no remnant. The remaining region indicates stars
	that undergo CC only. The black lines of various line styles indicate
	regions with remnant masses within $1\,\solarmass$ of the maximum mass in
	each distribution. For the models \citetalias{farmerMindGapLocation2019} and
	$\extraML = +10\,\solarmass$ these regions overlap and we indicate them with
	the same line style.}
  \label{fig:three_panel_variations}
\end{figure*}

In \Figref{fig:three_panel_variations} we show how our fiducial prescription for
{\ppisn} mass loss, the \citetalias{farmerMindGapLocation2019} model for
{\ppisne} mass loss, and examples of our two variations affect the remnant-mass
prediction as a function of initial stellar (ZAMS) mass and metallicity. The
final mass is determined by a combination of wind mass loss, compact-object
formation by {\ccsne}, black-hole formation by {\ppisne} and total disruption by
{\pisne}.

In \Figref{fig:three_panel_variations} (a) we show the fiducial remnant masses
as a function of initial mass and metallicity, in
\Figref{fig:three_panel_variations} (b) we show the remnant masses as predicted
with the \citetalias{farmerMindGapLocation2019} variation, in
\Figref{fig:three_panel_variations} (c) we show an example of variation with an
extra mass loss of $\extraML = 10\,\solarmass$, and in the
\Figref{fig:three_panel_variations} (d) we show the final mass predictions for
the CO core-mass shift variation of $\COshift = -5\,\solarmass$.

In the \citetalias{farmerMindGapLocation2019} variation,
\Figref{fig:three_panel_variations} (b), we see that the region of ZAMS masses
that undergo {\ppisne} is not strongly affected, but the supernova type that
leads to the most massive BH is now a {\ccsn}. Moreover, the remnant masses
\Figref{fig:three_panel_variations} (c) match well with those in
\Figref{fig:three_panel_variations} (b), which indicates that in \binaryc\ the
\citetalias{farmerMindGapLocation2019} prescription results in a discontinuity
between the \textit{last} {\ccsn} and the \textit{first} {\ppisn} of about
$10\,\solarmass$.

In the CO core-mass shift variation, \Figref{fig:three_panel_variations} (d),
the ZAMS mass and metallicity region is shifted to lower mass and higher
metallicity compared to our fiducial model, \Figref{fig:three_panel_variations}
(a), with for the lowest metallicity, a downward shift of the ZAMS mass for the
{\ppisne} region by about $10\,\solarmass$. The lower ZAMS-mass requirement
leads to an increase in the highest metallicity for which the {\ppisne} occur,
in this case from $3\,\times\,10^{-3}$ to $5\,\times\,10^{-3}$.

\section{Fiducial primary-mass distribution including systems that undergo and
  survive CE}
\label{app:primary-additional}
In this study we calculate the properties of merging {\bhbh} systems from
simulations with varying assumptions on the remnant mass predictions from
{\ppisne}, and compare them to observed distributions. In the results in
Sections~\ref{sec:primary-mass-distribution-variations}
and~\ref{sec:rate-dens-supern} we exclude systems that undergo one or more CE
episodes. Here we show which mass ranges are most affected by this.

\begin{figure*}
  \centering \includegraphics[width=0.75\textwidth]{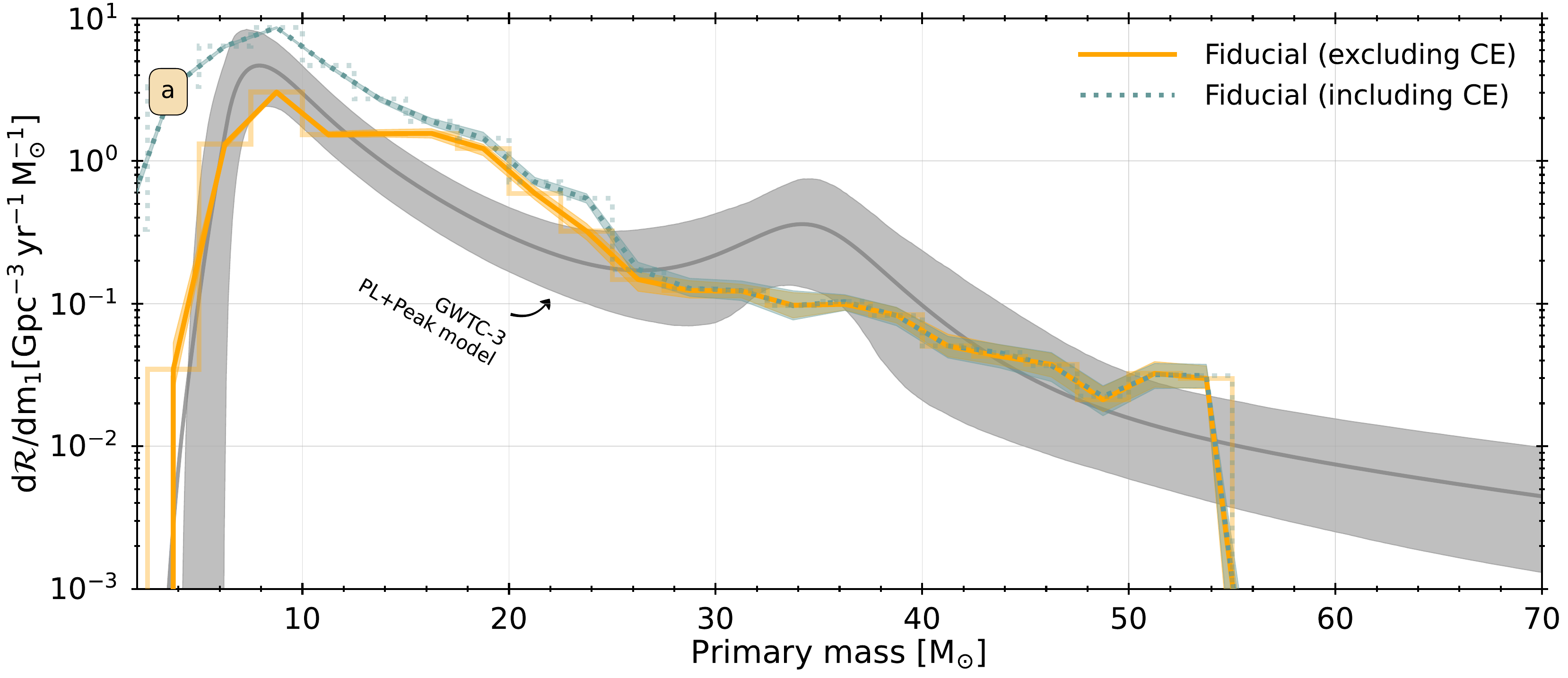}
  \caption[Merger rate at $z = \ligoredshift$ as a function of primary mass for
  our fiducial model including CE systems and excluding CE systems.]{Panel (a):
	Distribution of primary masses for {\bhbh} systems merging at
	$z = \ligoredshift$. The dark-grey line indicates the mean of the
	{\powerlawpeak} model $z = \ligoredshift$ and the grey shaded region
	indicates the $90$ per cent confidence interval. The orange line and region
	shows the mean and $90$ per cent confidence interval of our fiducial model
	results that exclude CE systems. The blue-dotted line and region shows the
	mean and $90$ per cent confidence interval of our fiducial model results
	that include CE results.  }
  \label{app:fiducial_distribution_highlight}
\end{figure*}

In \Figref{app:fiducial_distribution_highlight} we show a comparison between our
fiducial model where CE systems are excluded and those where CE systems are
included. Below $14\,\solarmass$ systems that undergo and survive CE systems
contribute significantly to the total merger rate ($>40$ per cent), especially
around the peak at $10\,\solarmass$. Between $14\,\solarmass$ and
$25\,\solarmass$ {\bhbh} mergers that undergo and survive CE contribute, but not
as much as the other channels ($<40$ per cent).

\section{Supernova-kick scaling}
\label{app:supern-kick-scal}
In this study we sample the BH kick speed from a Maxwellian velocity
distribution \citep{hobbsStatisticalStudy2332005} and we scale this kick by
$(1-f_{\mathrm{fallback}})$, which measures how much of the initially ejected
mass actually is lost from the system (\Secref{sec:supernova-natal-kick}). There
is, however, some debate on whether BHs do receive kicks and of what amplitude
\citep[e.g.,][]{drayWolfRayetStar2005, renzoMassiveRunawayWalkaway2019,
  atriPotentialKickVelocity2019a, callisterStateFieldBinary2021}

Because the velocity distribution of supernova kicks of
\citet{hobbsStatisticalStudy2332005} is based on observed pulsars, i.e. neutron
stars, a different scaling approach is based on that the kick distribution
should be regarded as a momentum kick distribution, and that the velocity kicks
for BHs should be scaled by the ratio of the BH remnant mass to the neutron star
mass (taken as $1.4\,\solarmass$) to conserve momentum, as,
\begin{equation}
  \label{eq:kick_scaling_masses}
  V_{\mathrm{scaled\,kick}} = V_{\mathrm{sampled\,kick}} \frac{M_{\mathrm{NS}}}{M_{\mathrm{remnant}}},
\end{equation}
where $M_{\mathrm{NS}}$ is the NS mass and $M_{\mathrm{remnant}}$ is the remnant BH
mass.

There are indications that BH kick speeds are indeed lower than those for NSs
\citep[e.g.][]{mandelEstimatesBlackHole2016, atriPotentialKickVelocity2019a}.
There are, however, also studies that show that black holes can still attain
high-velocity kicks, based on theory and simulations
\citep[e.g.][]{jankaNatalKicksStellarMass2013, chanImpactFallbackCompact2020}
motivated by strong asymmetries in the ejecta and the 'tug-boat' mechanism.

Moreover, there does not seem to be a consensus in the population-synthesis
studies on which scaling should be used. For example
\citet{sonRedshiftEvolutionBinary2022} uses fallback-only scaling, while
\citet{brielUnderstandingHighmassBinary2023} only scales by the NS-mass to
remnant-mass ratio. \citet{giacobboRevisingNatalKick2020} compares different
(combinations of) scaling factors and finds that these different prescriptions
do not affect the {\bhbh}-merger rate significantly.

\begin{figure}
  \centering \includegraphics[width=\columnwidth]{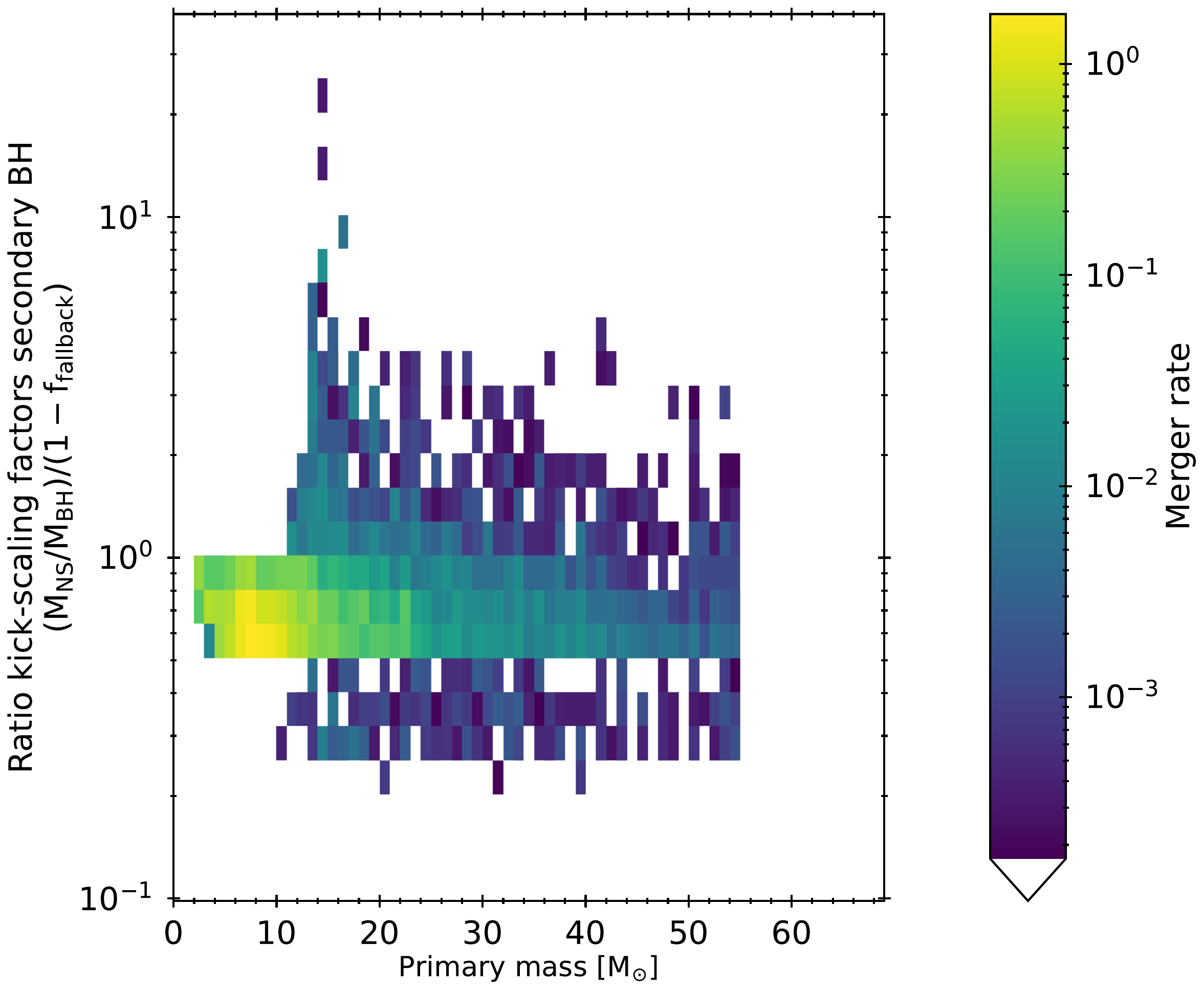}
  \caption[Ratio kick-scaling factors for secondary BH's in merging systems for
  fiducial models.]{Merger rate of {\bhbh} systems in our fiducials models as a
	function of primary mass (ordinate) and the ratio of kick-scaling factors of
	secondary BHs (abcissa).
	Overall, the fallback scaling factors are larger than the
	$(M_{\mathrm{NS}}/M_{\mathrm{remnant}})$ for most of the merging {\bhbh}
	systems. Around $M_{\mathrm{primary}}\,\sim 15\,\solarmass$ there are
	secondary BHs that would get significantly lower kicks, but this is only the
	case for a few systems.}
  \label{fig:kick_scaling_plots}
\end{figure}

For the mass range in primary BH mass $M_\mathrm{BH,1}$ relevant here, the
merging {\bhbh} systems in our fiducial models achieve to merge thanks to the
eccentricity induced by the kick onto the lower mass companion. We calculate the
ratio between the $(M_{\mathrm{NS}}/M_{\mathrm{remnant}})$ scaling factor and
the $(1-f_{\mathrm{fallback}})$ scaling factor to estimate the how this would
affect our results.
The colour in \Figref{fig:kick_scaling_plots} shows the merger rate of {\bhbh}
systems in our fiducials models as a function of primary mass and the ratio of
kick-scaling factors of secondary BHs in systems where the secondary BH recieves
a kick.
The plot shows that for the majority of mergers the ratio is below $1$,
indicating that the $(1-f_{\mathrm{fallback}})$ scaling factor is larger than
with the $(M_{\mathrm{NS}}/M_{\mathrm{remnant}})$ scaling. Thus, for the
majority of mergers that recieve a kick, the kick is lower than what it would be
with the $(M_{\mathrm{NS}}/M_{\mathrm{remnant}})$ scaling. Over the entire range
we do find a spread in the ratio of the two scaling factors, mostly between
$0.2$ and $4$, except for some systems with primary BH masses around
$15\,\solarmass$, indicating that in those systems the secondary BH kick would
be scaled down more.
Using the $(M_{\mathrm{NS}}/M_{\mathrm{remnant}})$ scaling leads to higher kicks
(so presumably higher post-CC eccentricities, but also higher disruption rates)
at the formation of the secondary BH for the majority of the merging systems. A
combination of both scalings may be appropriate
\citep{giacobboRevisingNatalKick2020}, but our fallback scaling already leads to
a dampening of many of the kicks comparable to what the
$(M_{\mathrm{NS}}/M_{\mathrm{remnant}})$ scaling would do.

\section{Merger-rate calculation}
\label{app:merger-rate-calculation}
In our population-synthesis framework {\binaryc} we take a grid-based approach
to sample the properties of the systems in our populations. We use the
distributions of the birth properties primary mass ($M_{1}$), secondary mass
($M_{2}$), period ($P$) (\Secref{sec:simulated-populations}) to span a hypercube
which we split up into equally sized volumes of phase-space. Each of these
volumes represents a system with a set of birth properties, $\zeta_{i}$, and a
probability $p_{i}$. We use this probability to calculate the number of systems
this system represents ($N_{i}$) in a population of $N_{\mathrm{bin}}$
binaries. We combine this probability with the binary fraction,
$f_{\mathrm{bin}}$, to take into account the fact that not all systems are
binary systems, which allows us to calculate $N_{i}$ given a population of
single and binary systems $N_{\mathrm{systems}}$,
\begin{equation}
  \label{eq:calculation-numbers}
  N_{i} = p_{i}\,\times\,f_{\mathrm{bin}}\,\times\,N_{\mathrm{systems}}.
\end{equation}
In this study we are, however, more interested in the rate of formation and
merging of systems than in their total number. To calculate the rate of
formation of system $i$, $R_{i}$, we combine \Eqref{eq:calculation-numbers} with
a star formation rate (SFR) and the average mass of a system,
\begin{equation}
  \label{eq:calculation-rate}
    R_{i} = p_{i} \times\,f_{\mathrm{bin}} \times\,\frac{\mathrm{SFR}}{\left<M\right>},
\end{equation}
where $\frac{\mathrm{SFR}}{\left<M\right>}$ is the rate of formation of all
systems and $\left<M\right>$ is the average mass of all systems,
$\left<M\right>$. We calculate this average mass by integrating the mass
distributions between their global bounds, in both single and binary systems
(see \Secref{sec:simulated-populations}) and obtain an average system mass of
$\left<M\right> = 0.689\,\solarmass$.

To calculate the merger-rate density, $R_{\mathrm{merge},\,i}$, of a system with
a metallicity, $Z_{i}$, and a delay time, $t_{\mathrm{delay},\,i}$, at a given
merger redshift, $z_{\mathrm{merge}}$, we modify \Eqref{eq:calculation-rate} to
calculate the SFR density at the birth redshift, $z_{\mathrm{birth}}$, of
system, $i$, given that the system merges at redshift $z_{\mathrm{merge}}$,
\begin{equation}
  \label{eq:merger_rate_calculation}
  R_{\mathrm{merge},\,i},\ z_{\mathrm{merge}} = p_{i}\,\times\,f_{\mathrm{bin}}\,\times\,\frac{\mathrm{SFR}(z_{\mathrm{birth},\,i},\ Z_{i})\,\times\,dZ_{i}}{\left<M\right>},
\end{equation}
where $Z_{i}$ is the metallicity of the system and $dZ_{i}$ is the width of the
metallicity bin in which the system lies. $z_{\mathrm{birth},\,i}$ corresponds
to the birth-redshift of the system, which is determined by calculating the
birth lookback time
$t^{*}_{\mathrm{birth},\,i} = t^{*}_{\mathrm{merge}} + t_{\mathrm{delay},\,i}$
and converting it to the corresponding redshift using the cosmology defined in
\Secref{sec:cosmo-convo}. We repeat this for all systems at all metallicities.

\begin{figure}
\centering
\includegraphics[width=\columnwidth]{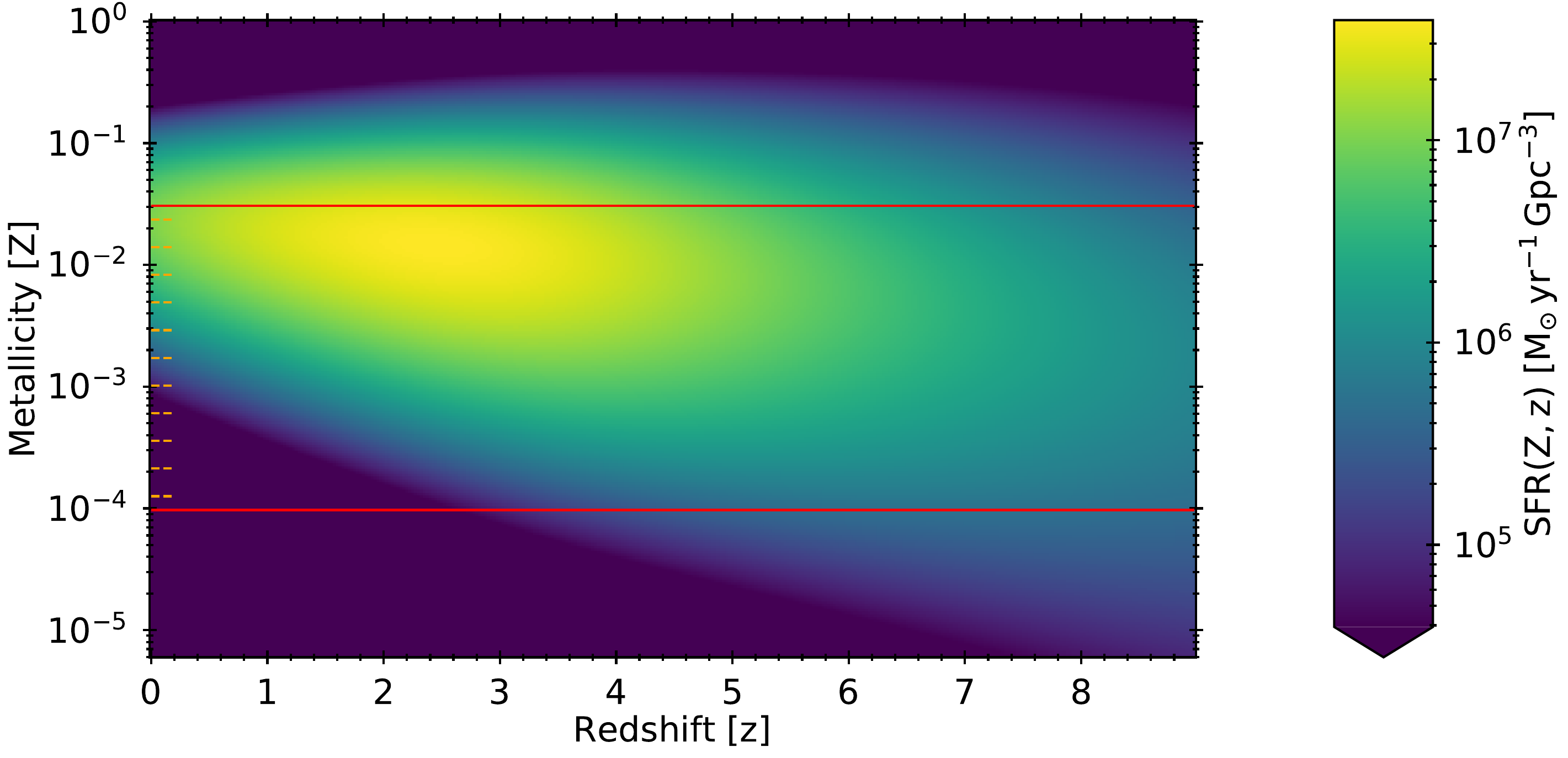}
\caption[Star-formation rate density $\mathrm{SFR}(z,Z)$ used in this
study.]{Star-formation rate density, $\mathrm{SFR}(z,Z)$ used to convolve our
  {\bhbh} systems \citep{sonRedshiftEvolutionBinary2022}, as a function of
  redshift (ordinate) and metallicity (abscissa). The colour indicates the
  star-formation rate density, the red lines indicate the extent of our
  metallicity grid, and the orange dashed ticks on the left are the
  metallicities of the binary populations we simulate.}
  \label{fig:mssfr}
\end{figure}

\Figref{fig:mssfr} shows the SFR distribution, along with the extent of the
metallicities we use, in our input populations and the individual metallicities.

\end{appendices}

\bsp	
\label{lastpage}
\end{document}